\documentclass[a4paper,11pt]{article}

\usepackage{amsmath,amsfonts,amssymb,caption,graphicx,natbib}
\usepackage{appendix}
\usepackage{ushort}
\usepackage{makeidx}
\usepackage{float}
\usepackage{accents}
\usepackage{color}
\usepackage{framed}
\usepackage{versions}
\usepackage{xcolor}
\usepackage{emptypage}
\usepackage{mathbbol}
\usepackage{array}
\usepackage{multirow}
\usepackage{lipsum}

\def\hmath$#1${\texorpdfstring{{\rmfamily\textit{#1}}}{#1}}
\usepackage{tikz}
\usetikzlibrary{trees}

\usepackage[bookmarks]{hyperref}
\hypersetup{
    colorlinks=true,   	
    linkcolor=red,      
    citecolor = [rgb]{0 0.7 0},   	
    filecolor=magenta, 	
    urlcolor=blue
}
 
\input{epsf}

\newcommand\blfootnote[1]{%
  \begingroup
  \renewcommand\thefootnote{}\footnote{#1}%
  \addtocounter{footnote}{-1}%
  \endgroup
}

\newcommand{\IND}{{\mathbb I}}

\newcommand{\maxi}{\mathop{{\rm max}^{(i)}}}
\newcommand{\maxip}{\mathop{{\rm max}^{(i')}}}
\newcommand{\maxij}{\mathop{{\rm max}^{(i_j)}}}
\newcommand{\maxiz}{\mathop{{\rm max}^{(i_0)}}}
\newcommand{\maxio}{\mathop{{\rm max}^{(i_1)}}}
\newcommand{\maxim}{\mathop{{\rm max}^{(i_{m-1})}}}

\newcommand{\Qhati}{\mbox{$\hat{Q}_i$}}

\newcommand{\TMAX}{T_{\rm MAX}}

\def\ba{\begin{align}}
\def\ea{\end{align}}
\def\ban{\begin{align*}}
\def\ean{\end{align*}}

\def\be{\begin{eqnarray}}
\def\ee{\end{eqnarray}}
\def\ben{\begin{eqnarray*}}
\def\een{\end{eqnarray*}}

\def\bqq{\begin{equation}}
\def\eqq{\end{equation}}
\def\bqqn{\begin{equation*}}
\def\eqqn{\end{equation*}}

\def\CTW{{\sf CTW}}

\def\BCT{{\sf BCT}}
\def\BCTs{{\sf BCT}s}
\def\BCTk{{\sf $k$-BCT}}

\def\elabel#1{\label{e:#1}}

\def\sq{$\Box$}

\def\qed{\ifmmode\sq\else{\unskip\nobreak\hfil
\penalty50\hskip1em\null\nobreak\hfil\sq
\parfillskip=0pt\finalhyphendemerits=0\endgraf}\fi\par\medbreak}

\newsavebox{\junk}
\savebox{\junk}[1.6mm]{\hbox{$|\!|\!|$}}

\def\til={{\widetilde =}}

\def\clQ{{\cal Q}}

\def\clT{{\cal T}}

 \def\eq#1/{(\ref{#1})}

\newtheorem{theorem}{Theorem}[section]
\newtheorem{corollary}[theorem]{Corollary}

\newtheorem{lemma}[theorem]{Lemma}

\newtheorem{example}[theorem]{Example}

\def\eq#1/{(\ref{e:#1})}

\newcommand{\beqn}[1]{\notes{#1}%
\begin{eqnarray} \elabel{#1}}

\newcommand{\eeqn}{\end{eqnarray} } 

\newcommand{\beq}[1]{\notes{#1}%
\begin{equation}\elabel{#1}}

\newcommand{\eeq}{\end{equation}} 

\def\bdes{\begin{description}}
\def\edes{\end{description}}

\DeclareMathOperator*{\B7cup}{\text{\raisebox{0.25ex}%
	{\scalebox{0.7}{$\bigcup$}}}}

\def\notes#1{}

\definecolor{mag}{rgb}{0.7,0,0.3}
\definecolor{dgreen}{rgb}{0.1,0.5,0.1}
\definecolor{dred}{rgb}{.8,0,0}
\definecolor{gray}{rgb}{.8,.8,.8}
\definecolor{brown}{rgb}{0.6451,0.3706,0.1745}

\begin{document}

\pagestyle{empty}


\title{\vspace{-1.5cm}%
Bayesian context trees:
modelling and exact inference\\ for discrete time series
}

\author
{
        Ioannis Kontoyiannis
    \thanks{Statistical Laboratory,
	Centre for Mathematical Sciences,
        University of Cambridge,
	Wilberforce Road,
	Cambridge CB3 0WB, UK.
                Email: \texttt{\href{mailto:yiannis@maths.cam.ac.uk}%
			{yiannis@maths.cam.ac.uk}}.
	}
\and
 	Lambros Mertzanis
    \thanks{Department of Electrical and Computer Engineering,
        University of Maryland,
	College Part, MD, USA.
                Email: \texttt{\href{mailto:lambros@umd.edu}%
			{lambros@umd.edu}}.
 	}
\and
 	Athina Panotopoulou
    \thanks{Department of Computer Science,
        Dartmouth College, Hanover, HN, USA.
                Email: \texttt{\href{mailto:ath1na@bu.edu}%
			{ath1na@bu.edu}}.
 	}
\and
 	Ioannis Papageorgiou
    \thanks{Department of Engineering,
        University of Cambridge,
        Trumpington Street, Cambridge CB2 1PZ, UK.
                Email: \texttt{\href{mailto:ip307@cam.ac.uk}%
			{ip307@cam.ac.uk}}.
 	}
\and
 	Maria Skoularidou
    \thanks{MRC-BSU, University of Cambridge,
	Cambridge, UK.
                Email: \texttt{\href{mailto:ms2407@cam.ac.uk}%
			{ms2407@cam.ac.uk}}.
 	}
}

\date{\today}

\thispagestyle{empty}

\maketitle

\thispagestyle{empty}

\setcounter{page}{-1}

\bigskip

\noindent
{\bf Summary.} 
We develop a new Bayesian modelling framework for the 
class of higher-order, variable-memory Markov chains,
and introduce an associated collection 
of methodological tools for exact
inference with discrete time series.
We show that a version of
the context tree weighting algorithm 
can compute
the prior predictive likelihood exactly
(averaged over both models and parameters), 
and two related algorithms are introduced,
which identify the {\em a posteriori} most 
likely models and compute their exact posterior probabilities. 
All three algorithms are deterministic and have linear-time complexity.
A family of variable-dimension Markov chain Monte Carlo
samplers is also provided, facilitating further exploration 
of the posterior.
The performance of the proposed methods 
in model selection, Markov order 
estimation and prediction is 
illustrated through simulation experiments 
and real-world applications with 
data from finance, genetics, 
neuroscience, and animal communication. 
The associated algorithms are implemented in the
{\sf R} package {\sf BCT}.

\bigskip

\noindent
{\bf Keywords.} 
Discrete time series; 
Bayesian context tree;
Model selection;
Prediction;
Exact Bayesian inference; 
Markov order estimation;
Bayes factors;
Markov chain Monte Carlo;
Context tree weighting

\blfootnote{
	Preliminary versions of some
	of the results in this work were presented
	in \cite{ctw-itw:12}, \cite{ctw-spawc:18}
	and \cite{ctw-isit:21}.
	}

\newpage

\tableofcontents

\newpage

\pagestyle{plain}
\setcounter{page}{1}

\section{Introduction}
\label{s:intro}

Higher-order Markov chains are frequently the first and most natural modelling 
choice for discrete time series with significant -- apparent or suspected -- 
temporal structure, especially when no specific underlying data generating 
mechanism can be assumed. But the description of a full Markov chain of 
order $d$ with values in a set of size $m$,
say, requires the specification 
of $m^d(m-1)$ parameters, which makes the use of full Markov chains 
problematic in practice: As has been often 
noted \citep{raftery:85,buhlmann:99,sarkar:16},
the dimension of the parameter space grows exponentially with the memory 
length, and the resulting model class lacks modelling wealth and flexibility. 
This lack of flexibility severely hinders, among other things, 
the important goal of balancing the bias-variance tradeoff
between more complex models that fit the data closely, and simpler models
that generalise well.

To address these issues and to offer better solutions to a wealth
of related scientific and engineering problems that arise in connection
with discrete time series, numerous approaches have been developed 
since the mid-1980s.

The mixture transition distribution (MTD) models introduced
by \cite{raftery:85} and later generalised in
\cite{raftery-tavare:94,berchtold:02},
allow for more parsimonious parametrisations of the 
transition distribution of some $d$th order Markov chains,
as mixtures of first order transition
matrices corresponding to different lags. 
MTD models make it possible to consider longer
memory lengths $d$ and to quantify the relative importance
of different lags, but the resulting model class is still 
structurally poor.

A much more flexible class of Markov chain models with 
{\em variable memory} are the {\em tree sources}
introduced in the celebrated 
work of \cite{rissanen:83b,rissanen:83,rissanen:86}.
The gist of this approach is that the length
of the memory that determines the transition probability
of the chain can depend on the exact pattern of the most 
recently occurring symbols. Initially tree
sources received a lot of attention 
in the information-theoretic literature
in connection with data compression
\citep{weinbergeretal:94,willems-shtarkov-tjalkens:95}.
One of their first applications
outside information theory was by
\cite{ron-et-al:96}, who introduced the 
notion of a probabilistic suffix tree (PST)
as an effective structure for 
representing variable-memory chains. The PST point 
of view, along with the associated model selection 
technique {\em Learn-PSA},
have been used for
bioinformatics problems and other machine learning 
tasks \citep{bejerano:01,gabadinho:16}.

In the statistics literature, tree-structured models 
were examined by \cite{buhlmann:99,buhlmann:00}.
Their variable-length Markov chains (VLMCs) 
and the associated model selection tools are based, 
in part, on Rissanen's tree sources and his {\sc context} 
algorithm. The VLMC approach has been successful 
in applications \citep{buhlmann:04,busch-et-al:09}
that include
DNA modelling \citep{ben-gal:05,browning:06}
and linguistics \citep{galves-et-al:12,abakuks:12}.

Tree sources and VLMC models group together 
certain patterns of past symbols that lead to the 
same conditional distribution for the chain,
under the constraint that each such group consists
of all patterns of length $d$ that share a common
suffix. A more general class of parsimonious Markov 
models, known as sparse Markov chains (SMC), 
arises when this constraint is removed.
Originally introduced
as ``minimal Markov models'' by
\cite{garcia:11}, they were later examined
in more detail in \cite{jaaskinen:14,xiong:16,garcia:17}.
But the lack of structure of this vast model class
makes it difficult to identify appropriate models
in practice.

\newpage

A different interesting class of higher-order Markov models
was more recently introduced by \cite{sarkar:16},
who used conditional tensor factorisation (CTF) to 
give parsimonious representations of the full 
transition probability distribution
(viewed as a high-dimensional tensor) 
of a Markov chain.
These representations are based on extensions of
earlier ideas on tensor factorisation 
for categorical regression \citep{yang:16}.
CTF effectively shrinks the high-dimensional 
transition probability tensor to a lower-dimensional structure
that can still capture high-order 
dependence. Unlike MTD, CTF accommodates complex 
interactions between the lags, and is accompanied 
by computational tools that allow 
for rich Bayesian inference.


In this work we revisit the class of variable-memory 
Markov models. We introduce a new Bayesian framework 
for a version of these models,
and we develop algorithmic tools that lead to
very effective and efficient {\em exact} inference.
Although our 
methods open the door to a wide range of statistical 
and machine learning applications --
including
anomaly detection, change-point estimation,
and pattern analysis -- here we focus primarily
on the more fundamental tasks of model selection, 
estimation, and sequential prediction.

\subsection{Outline of contributions}
\label{s:summary}

In Section~\ref{s:VLMC} we define a class of 
models for variable-memory Markov chains
that admit natural representations as context trees.
Given a finite set~$A$ and a maximal memory length
$D\geq 0$, the class $\clT(D)$ contains 
all variable-memory models of Markov chains with
values in $A$ and memory no longer
than~$D$. A new family of discrete prior 
distributions $\pi_D(T;\beta)$ on models $T\in\clT(D)$
is introduced, indexed by a hyperparameter $\beta\in(0,1)$.
Roughly speaking, $\pi_D$ penalises larger and more complex models
by an exponential amount.
Given a model~$T$, we place independent Dirichlet
priors $\pi(\theta|T)$ on the associated parameters $\theta$.
We refer to the models in $\clT(D)$ equipped with this 
prior structure as {\em Bayesian context trees} (\BCT).

Sections~\ref{s:MMLA}--\ref{s:kMAPT} contain our 
core methodological
results, in the form of three exact inference algorithms
for \BCTs. First we show that a version of the context tree
weighting (\CTW) algorithm 
\citep{ctw-multi:94,willems-shtarkov-tjalkens:95}
can be used to not only evaluate the 
marginal likelihoods $P(x|T)=\int P(x|\theta,T)\pi(\theta|T)d\theta$
of observations~$x$ with respect to models~$T$, which are 
easy to obtain, but also the 
{\em prior predictive likelihood} $P_D^*(x)$,
averaged over all models, 
\be
P_D^*(x)
:=\sum_{T\in\clT(D)}\pi_D(T;\beta)P(x|T)
=\sum_{T\in\clT(D)} \pi_D(T;\beta) \int P(x|\theta,T)\pi(\theta|T)d\theta.
\label{eq:PPL}
\ee
The \CTW\ algorithm computes $P_D^*(x)$ exactly, 
and its complexity
is only linear in the length of the observed time series $x$.
Since the most basic obstacle to
performing effective Bayesian inference
is the difficulty to either sample from or
obtain expectations with respect to
the posterior distribution, typically
stemming from the impossibility
of computing its normalizing factor $P_D^*(x)$,
it is clear that the exact nature of the 
results produced 
by the \CTW\ algorithm
should facilitate the development 
of efficient methods for numerous core statistical
tasks and related applications.

In Section~\ref{s:MAPT} we describe the Bayesian context tree
(\BCT) algorithm and prove that 
it identifies 
the maximum {\em a posteriori} probability (MAP) model.
This is a generalisation of the ``context tree maximizing''
algorithm of
\cite{volf-willems:94}.
And in Section~\ref{s:kMAPT} we show that a new
algorithm, the $k$-\BCT\ algorithm, can be used
to identify the $k$ {\em a posteriori} most likely
tree models, for any $k\geq 1$.
Despite the fact that the class $\clT(D)$ is vast,
consisting of doubly-exponentially many models in the
memory length $D$, the complexity of both the \BCT\ 
and $k$-\BCT\ algorithms is only linear in 
$D$ and in the length of the observations $x$. 
But as a function of $k$,
the complexity of $k$-\BCT\ grows faster than
linearly in $k$; in fact, in its naive implementation,
it grows like $k^m$, 
where $m$ is the number
of possible values of the time series $x$.
So its practical applicability is limited
to relatively small values of $k$.

In order to enable broader exploration
of the posterior distributions
$\pi(T|x)$ and $\pi(\theta,T|x)$,
in Section~\ref{s:MH} we develop
a new family of variable-dimension Markov chain
Monte Carlo (MCMC) algorithms
that obtain samples
from $\pi(\theta,T|x)$. Their performance
is illustrated in Section~\ref{s:MCMC} on model selection,
parameter estimation, and Markov order
estimation problems, on 
simulated and real data examples.

In Section~\ref{s:compare} we present extensive
model selection results, 
comparing the performance of the
\BCT\ framework with that of the corresponding
VLMC and MTD methods,
on both real and simulated
data.
We find that the \BCT\
algorithm consistently performs at least
as well as VLMC and MTD and usually
gives a better fit on
simulated data. Moreover, the $k$-\BCT\ algorithm
in combination with the MCMC samplers of 
Section~\ref{s:MH} identify a number of candidate
models for the observed data, also providing
a quantitative measure of uncertainty for the 
selected models in the form of posterior probabilities.
Using the \CTW\ algorithm, these posterior probabilities
can be computed exactly, as can the relevant Bayes factors
and posterior odds for a variety of hypotheses of interest.
In terms of complexity, the \BCT\ algorithm
is found to be computationally
much more effective than MTD and VLMC.
In fact, the linear complexity of \CTW\
and \BCT\ facilitates their use
in big-data applications, as illustrated,
e.g.,
in the analysis of a neural spike train data
set of $\approx 4\times 10^6$ samples,
with memory lengths up to $D=1500$.

In Section~\ref{s:prediction} we compare
the natural predictor
induced by the \BCT\ framework
with the predictors provided
by the MTD, VLMC, SMC and CTF 
methodologies. The \BCT\ predictor
is seen to have two
significant advantages,
which lead to superior performance.
The first
is that the {\em posterior predictive distribution}
can be computed exactly, as
$P_D^*(x_{n+1}|x_1,\ldots,x_n)=
P_D^*(x_1,\ldots,x_{n+1})/
P_D^*(x_1,\ldots,x_n)$,
via the \CTW\
algorithm.
This way, the induced predictor
is obtained by implicitly averaging
over all models with respect to their
exact posterior probabilities, thus 
avoiding the need to perform approximate 
model averaging via simulation or other
numerical integration methods.
The second advantage is that, because the 
\CTW\ algorithm can be updated sequentially,
so can the \BCT\ predictor,
so that it continues
to ``learn'' from
the data
even past the training phase.
Results on both simulated and real time 
series illustrate the
performance of the \BCT\ predictor,
confirming these observations.

\subsection{Further connections and comments}
\label{s:connections}

Variable-memory models, like the
Bayesian context trees
considered in
this work, describe a flexible and rich class of 
higher-order Markov chains that admit parsimonious
parametrisations and allow for natural 
graphical representations of important
structural dependencies. The shape
of the context tree can be easily interpreted
and provides useful information about
the regularities present in the data
\citep{bejerano:01,buhlmann:04}.
Because \BCTs\ are a vast model class,
global model selection techniques based, e.g.,
on criteria
like AIC and BIC, cannot be applied directly
via, say, exhaustive search.
But efficient tools
like the \BCT\ algorithm presented here
make it possible to describe
complex sequential data in a way
that offers an effective
balance for the 
simplicity-expressivity 
tradeoff
\citep{garivier:11}.


Another point of view which naturally relates
to the present development
is Rissanen's celebrated
Minimum Description Length (MDL) 
principle
\citep{rissanen:87,rissanen:book,grunwald:book}.
The MDL principle
provides a broad operational foundation
for statistical inference,
as well as constructive tools and 
appealing metaphors 
for selecting prior distributions
\citep{chipman:01}.
In particular, MDL considerations
underpin much of the original 
work on the
\CTW\ algorithm \citep{ctw-multi:94,ctw:02}
and our own choice of priors
in Section~\ref{s:VLMC}.

\newpage

A method commonly used for model comparison is
the Bayesian information criterion 
(BIC), derived by
\cite{schwarz:78}
as an asymptotic approximation to 
twice the logarithm of the Bayes factor 
$P(x|T)/P(x|T')$ between two models $T,T'$ 
\citep{kass:95}.
The form of the BIC and its
familiar ``$(1/2)\log n$-per-degree-of-freedom'' 
log-likelihood penalty also shares
deep connections with the MDL principle,
see, e.g., the discussions by
\cite{barron-rissanen-yu:98,csiszar:00}.
Further comments on this are given
in Section~\ref{s:prediction}
in connection with Theorem~\ref{thm:redundancy}.

Finally, we note that there exist
a number of alternative
approaches to modelling 
discrete time series.
An important collection of tools
is provided by hidden Markov models.
HMMs are a general
and very broadly used 
model class, with a wide range
of applications and a variety of associated
methodological
procedures for learning and inference
\citep{bishop:book06,cappe:book06}.
A more classical approach to discrete time 
series modelling is via discrete analogs 
of linear models, often using
multinomial logit or probit regression
\citep{yee:10}.
Such a linear-predictor approach is described 
in~\cite{zeger-liang:86},
and a different parsimonious class of models
with an emphasis on binary time series
is given in~\cite{fahrmeir:1987}.
A treatment of
partial likelihood inference
on generalised discrete linear models is
presented in~\cite{fokianos:03},
and an extension of the traditional ARMA methodology
to integer autoregressive models for count time series 
is developed in~\cite{fokianos:12}.

\newpage

\section{Bayesian context trees} 
\label{s:VLMC}

\begin{figure}[ht!]
\centerline{\includegraphics[width=4.4in]{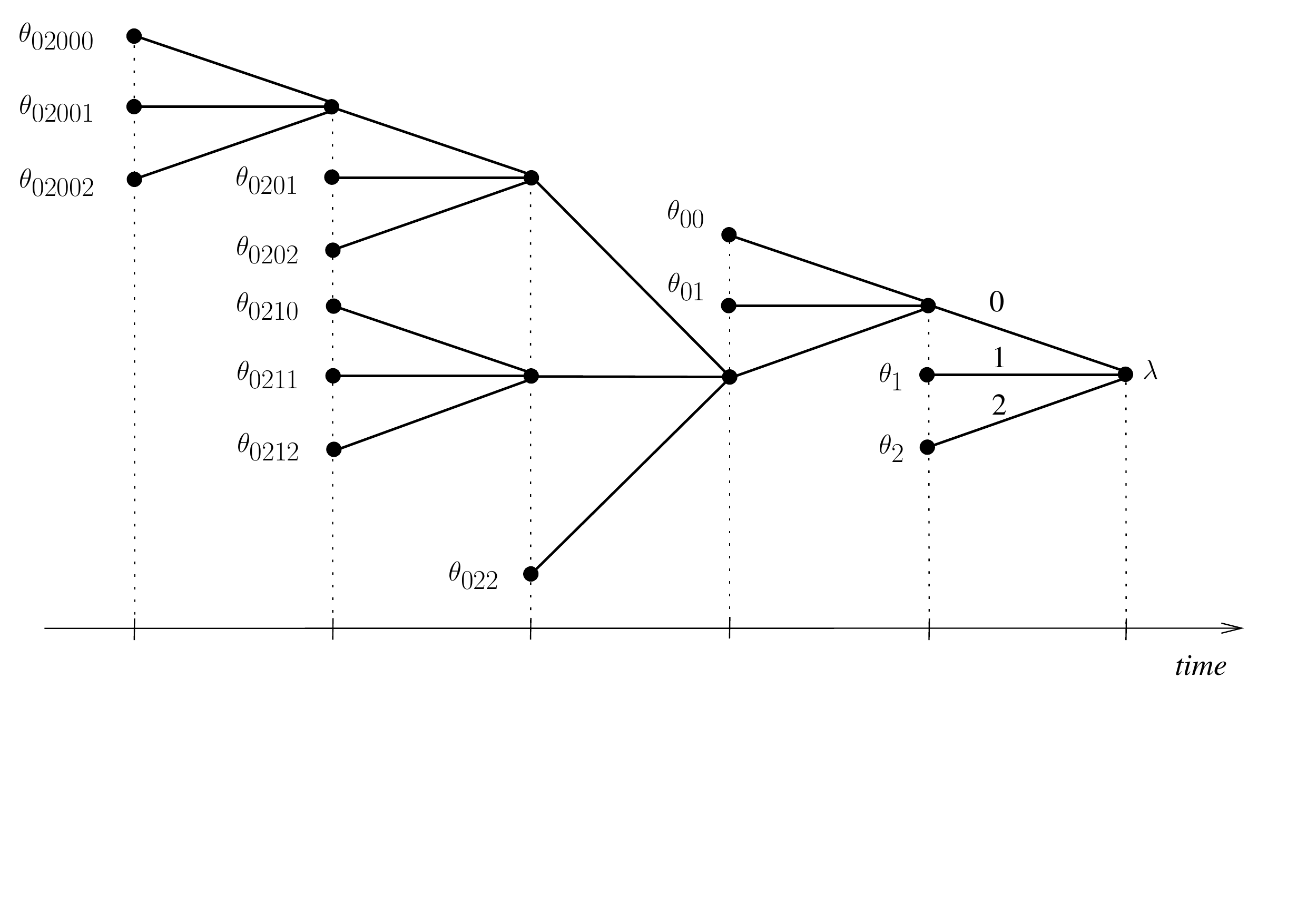}}
\vspace{-0.8in}
\caption{The tree model and parameters 
of a 5th order variable-memory chain.}
\label{fig:running}
\end{figure}

The distribution of a full
$d$th order Markov chain $\{X_n\}$
with values in the finite state space, or {\em alphabet}, $A$,
is identified by its conditional
distributions,
\ben
\theta_s(a):=\Pr(X_{n}=a|X_{n-1}=x_{n-1},X_{n-2}=x_{n-2},\ldots,
X_{n-d}=x_{n-d}),
\qquad a\in A,
\een
for every {\em context} 
$s=(x_{n-1},x_{n-2},\ldots,x_{n-d})\in A^d$
of length $d$. If the alphabet $A$ has size
$m$, the description of these conditional
distributions requires the specification
of $(m-1)m^d$ parameters. 
But suppose,
for example, that $A=\{0,1,2\}$,
and that when the most recent symbol $x_{n-1}$ is a ``1''
the distribution of the next symbol is independent
of the remaining values $x_{n-2},...,x_{n-d}$,
that is, the conditional probabilities,
$$\Pr(X_{n}=a|X_{n-1}=1,X_{n-2}=x_{n-2},\ldots,
X_{n-d}=x_{n-d}),$$
only depend on $a$.
An example of how the distribution of such 
a 5th order, {\em variable-memory} Markov chain may
be represented by
a labeled tree is shown
in Figure~\ref{fig:running}.

\noindent
Each leaf of the tree corresponds to a string $s$ containing
between one and five symbols from~$A$,
determined by the labels along the path from the root
$\lambda$ to that leaf.
Whenever the string~$s$ consists of the same sequence
of symbols as the most recent $x_i$'s, 
the probability that the next value of the chain will 
be ``$a$'' is 
$\theta_s(a)$, given by the distribution $\theta_s$
marked on that leaf, e.g.,
$$\Pr(X_n=2|X_{n-1}=0,X_{n-2}=1,X_{n-3}=1,\ldots)
=\Pr(X_n=2|X_{n-1}=0,X_{n-2}=1)=\theta_{01}(2).$$
Note that, since the tree model for this chain has 13 leaves,
instead of the full $2\times 3^5=486$ parameters,
it suffices to only specify $2\times 13=26$. 

\subsection{Variable-memory Markov chains}
\label{s:VLMC-1}

Let $\{X_n\}$ 
be a $d$th order Markov chain,
for some $d\geq 0$, with values in an alphabet 
$A$ of size $m\geq 2$. Without
loss of generality, we take $A=\{0,1,\ldots,m-1\}$
throughout. The {\em model} describing $\{X_n\}$ as
a variable-memory chain will always be represented by
a tree as in the example above.

Let $T$ be an $m$-ary tree of depth no greater than~$d$,
which is {\em proper}, in that, if a node in $T$ 
is not a leaf, then it has exactly $m$ children.
This means that every length-$d$ context
$s:=(x_{-1},x_{-2},\ldots,x_{-d})\in A^d$
has a unique suffix $(x_{-1},\ldots,x_{-j})$
which is a leaf of $T$,
for some $j\leq d$.

Let $C$ be the 
function that maps each context $s\in A^d$
to a leaf of $T$. 
Viewing $T$ as the collection of its leaves,
it is the range of $C$,
$$C:A^d\to T\subset\B7cup_{i=0}^d A^i,$$
with the 
convention
that $A^0=\{\lambda\}$ contains only 
the {\em empty string $\lambda$}.

For indices $ i\leq j$,
we write $X_i^j$ for the vector
of random variables
$(X_i,X_{i+1},\ldots,X_j)$ and similarly
$x_i^j \in A^{j-i+1}$ for
a string $(x_i,x_{i+1},\ldots,x_j)$ representing
a particular realisation of these random variables.
Then the 
Markov property for a variable-memory chain $\{X_n\}$ 
with model $T$ takes
the form:
\be
P(x_1^n|x_{-d+1}^0)
:=
\Pr(X_1^n=x_1^n|X_{-d+1}^0=x_{-d+1}^0)
=
\prod_{i=1}^n P(x_i|x_{i-d}^{i-1})
=
\prod_{i=1}^n P(x_i|C(x_{i-d}^{i-1})).
\label{eq:markovP}
\ee
The complete description
of the distribution of $\{X_n\}$,
in addition to the model $T$,
requires the specification
of a set of {\em parameters}
$\theta=\{\theta_s\;;\;s\in T\}$:
To every $s\in T$
we associate a probability vector,
$$\theta_s:=(\theta_s(0),\theta_s(1),\ldots,\theta_{s}(m-1)),$$
where the $\theta_s(j)$ are nonnegative and sum to one.
Then~(\ref{eq:markovP}) can be written,
\ben
P(x_1^n|x_{-d+1}^0)
=
\prod_{i=1}^n \theta_{C(x_{i-d}^{i-1})}(x_i),
\een
or, alternatively,
\be
P(x_1^n|x_{-d+1}^0)
=
\prod_{s\in T} \prod_{j\in A}
\theta_s(j)^{a_s(j)},
\label{eq:likelihood2}
\ee
where each element $a_s(j)$ of
the {\em count vector} $a_s=(a_s(0),a_s(1),\ldots,a_{s}(m-1))$
is,
\be
a_s(j)\;:=\;\mbox{\# times symbol $j\in A$ follows 
context $s$ in $x_1^n$}.
\label{eq:vector-a}
\ee

\subsection{Bayesian context trees}
\label{s:prior}

Throughout the paper, we consider
every 
$d$th order, variable-memory chain
$\{X_n\}$ as described by a (proper) tree model $T$
with
associated parameters
$\theta=\{\theta_s\;;\;s\in T\}$.
Here we define a family of prior distributions
on $(T,\theta)$. 

\medskip

\noindent
{\bf Model prior.}
For a maximal depth $D\geq 0$
and an alphabet $A=\{0,1,\ldots,m-1\}$
of size $m\geq 2$,
let $\clT(D)$ denote the collection of all
(proper) tree models $T$ on $A$ with depth no
greater than $D$.
Given an arbitrary 
$\beta\in(0,1)$, we define the prior distribution,
\be
\pi(T):=\pi_D(T;\beta):=\alpha^{|T|-1}\beta^{|T|-L_D(T)},
\qquad T\in\clT(D),
\label{eq:tree-prior}
\ee
where $\alpha:=(1-\beta)^{1/(m-1)}$,
$|T|$ denotes the number of leaves of $T$,
and $L_D(T)$ denotes the number of leaves 
$T$ has at depth $D$. The following lemma,
proved in Section~\ref{s:proofs} of the
supplementary material,
states that~(\ref{eq:tree-prior}) does indeed define
a probability distribution. 

\begin{lemma}
\label{lem:prior}
For any $D\geq 0$ and any $\beta\in(0,1)$:
$
\sum_{T\in{\cal T}(D)}\pi_D(T;\beta) =1.
$
\end{lemma}

\noindent
{\bf Prior on $\theta$.}
Given a model $T\in\clT(D)$, 
we place an independent Dirichlet prior
with parameters $(1/2,1/2,\ldots,1/2)$ 
on each $\theta_s$ so that,
$\pi(\theta|T)=\prod_{s\in T}\pi(\theta_s)$,
where,
\be
\pi(\theta_s)
= \pi(\theta_s(0), \theta_s(1), \ldots, \theta_s({m-1}))
= \frac{\Gamma(m/2)}{\pi^{m/2}}\prod_{j=0}^{m-1}\theta_s(j)^{-\frac{1}{2}}
\propto \prod_{j=0}^{m-1}\theta_s(j)^{-\frac{1}{2}}.
\label{eq:theta-prior}
\ee 
Although we take the parameter vector of the
Dirichlet prior for all $\theta_s$
to be $(1/2,\ldots,1/2)$, corresponding
to the Jeffreys prior,
the extension of all our results
to arbitrary Dirichlet parameters
is straightforward as outlined in 
Section~\ref{s:dirichlet} of the supplementary material.

Finally, given the model $T$ and associated
parameters $\theta=\{\theta_s\;;\;s\in T\}$,
from~(\ref{eq:likelihood2}) we have,
\be
P(x_1^n|x_{-d+1}^0,\theta,T)
=
\prod_{s\in T} \prod_{j=0}^{m-1} \theta_s(j)^{a_s(j)},
\label{eq:likelihood3}
\ee
where $a_s$ are the count vectors 
defined in~(\ref{eq:vector-a}).
By convention, when we write $\sum_{s\in T}$
or $\prod_{s\in T}$, we take the corresponding
sum or product over all the {\em leaves} $s$ of the
tree, not all its nodes. Also, in order to avoid
cumbersome notation, in what follows we often
write $x$ for the entire time series $x_1^n$ and 
suppress the dependence on its initial 
context $x_{-d+1}^0$, so that, for example, 
we denote,
$$
P(x,\theta|T)=P(x_1^n,\theta|x_{-d+1}^0,T)
=P(x_1^n|x_{-d+1}^0,\theta,T)\pi(\theta|T).
$$

\smallskip

\noindent
{\bf Choice of $\beta$.}
The first factor, $\alpha^{|T|-1}$, in the definition
of the model prior $\pi_D(T;\beta)$
is the more important
and easier one to interpret,
showing that larger models are 
penalised by an exponential amount, while the 
second factor, $\beta^{|T|-L_D(T)}$, 
adds a less intuitive penalty to the 
leaves at depth strictly 
smaller than $D$. 

Consider two models $S,T\in\clT(D)$
such that $S$ is a subtree of $T$. If
$T$ is produced by adding a single branch of $m$ 
nodes to one of the leaves, $s$, say, of $S$,
and $s$ is at depth $D-2$ or smaller, then
$|T|=|S|+m-1$ and $L_D(T)=L_D(S)$, so that,
\ben
\frac{\pi_D(T;\beta)}{\pi_D(S;\beta)}=(1-\beta)\beta^{m-1},
\een
which, as desired, is always less than one. 
But if $T$ is produced from $S$ by adding a branch
of $m$ nodes to a node $s$ at depth $D-1$, then
$L_D(T)=L_D(S)+m$, and,
$$\frac{\pi_D(T;\beta)}{\pi_D(S;\beta)}=\frac{1-\beta}{\beta}.$$
This is strictly decreasing in $\beta$, and for
$\beta<1/2$ it is greater than one, while for 
$\beta>1/2$ it is smaller than 1. 

Therefore,
$\pi_D(T;\beta)$ penalises larger trees 
by an exponential amount as long as $\beta\geq1/2$,
and larger values of $\beta$ make the
penalisation more severe. Also, for larger alphabet 
sizes,
$\mbox{$\alpha=(1-\beta)^{1/(m-1)}$}$ becomes very close to~1 and
the second factor dominates, an effect 
which is unintuitive and less desirable.
Therefore, in practice we will always
take $\beta\approx 1-2^{-m+1}$,
so that $\alpha\approx 1/2$, unless there 
are specific reasons for a different
choice. 

\subsection{Marginal likelihood and prior predictive likelihood}
\label{s:mml}

A useful property of the \BCT~framework 
is that the parameters $\theta$ can
easily be integrated out, so that the 
{\em marginal likelihoods} $P(x|T)$ can be expressed in closed form.
This result, stated without proof in Lemma~\ref{lem:integrate},
is based on a standard computation.


\begin{lemma}
\label{lem:integrate}
The marginal likelihood $P(x|T)$ 
of the observations
$x$ given a model $T$ is,
$$P(x|T)
=\int P(x,\theta|T)d\theta
=\int P(x|\theta,T)\pi(\theta|T)d\theta
=\prod_{s\in T}P_e(a_s),$$
where the count vectors 
$a_s=(a_s(0),a_s(1),\ldots,a_s(m-1))$
are defined
in {\em (\ref{eq:vector-a})} 
and 
the {\em estimated probabilities} $P_e(a_s)$ are defined by,
\be
P_{e}(a_s):= 
\frac{\prod_{j=0}^{m-1} 
[(1/2)(3/2)\cdots 
(a_s(j)- 1/2)]}
{(m/2)(m/2+1) 
\cdots (m/2+M_s-1)},
\label{eq:Pe}
\ee
where $M_s:=a_s(0)+a_s(1)+\cdots+a_s(m-1)$, 
with the convention that any empty 
product is taken to be equal to 1.
\end{lemma}

In terms of inference, one of the main objects of
interest is the model posterior distribution,
$$\pi(T|x)=\frac{P(x|T)\pi(T)}{P(x)},$$
where the denominator is
the {\em prior predictive likelihood} 
$P(x)=P_D^*(x)$ in~(\ref{eq:PPL}).
The difficulty in computing
$P^*_D(x)$
comes, in part, from the fact that the class $\clT(D)$
of variable-memory models is enormously rich, 
even for moderate (or even small) alphabet sizes $m$ 
and maximal depths~$D$: The size of
$\clT(D)$ 
grows doubly exponentially in $D$;
a simple computation shows that,
\be
|\clT(D)|\geq \sum_{d=0}^{D-1}2^{m^d}\,-D.
\label{eq:doublyE}
\ee

In Sections~\ref{s:MMLA}--\ref{s:kMAPT} we describe
three exact inference algorithms for the
class of Bayesian context trees.
We show that, perhaps somewhat surprisingly,
$P_D^*(x)$ can be computed {\em exactly}
and in a very efficient manner,
and that the mode
of the posterior $\pi(T|x)$ as well
as the next few most likely models
can be explicitly identified.

\newpage

\section{Methodology}
\label{s:algorithms}

This section contains our core methodological 
contributions, which form the basis for most of
the statistical and learning tasks that can
be performed within the \BCT\ framework.

The \CTW\ algorithm is described in Section~\ref{s:MMLA},
and it is proved that it indeed computes the prior predictive
likelihood $P_D^*(x)$ of a time series $x$. 
Section~\ref{s:MAPT} introduces
the \BCT\ algorithm, which identifies the 
maximum {\em a posteriori} probability (MAP) tree model
$T_1^*$ given $x$.
A more general algorithm, 
$k$-\BCT, is described in
Section~\ref{s:kMAPT};
the $k$-\BCT\ algorithm
identifies not just the mode $T_1^*$ of the posterior $\pi(T|x)$
on model space, but the $k$
{\em a posteriori} most likely tree models,
$T_1^*,T_2^*,\ldots,T_k^*$, for any $k\geq 1$.
All three algorithms are implemented
in the publicly available
{\sf R} package {\sf BCT}~\citep{BCT:Rv1.1}.

In Section~\ref{s:additional} we discuss how the
posterior probability $\pi(T|x)$ of any model $T$
can be easily computed, and we identify 
the full conditional density of the parameters,
$\pi(\theta|x,T)$. In order to explore the posterior
distribution further, a family of variable-dimension
MCMC samplers are
developed in Section~\ref{s:MH}.
Finally, in Section~\ref{s:sequential}
we describe how the \CTW, \BCT\ and \BCTk\
algorithms can all be updated sequentially,
and show that their complexity 
is only linear in the length of the time series $x$.
In particular, the fact that the \CTW\ algorithm can
be implemented in a sequential fashion
means
that it can be effectively used for online prediction;
cf.~Section~\ref{s:prediction}.

\subsection{\CTW: The context tree weighting algorithm}
\label{s:MMLA}

The \CTW\ algorithm takes as input:
The size of the alphabet $m\geq 2$,
the maximum context depth $D\geq 0$,
a time series of observations $x_1^n$ together with
the initial context $x_{-D+1}^0$, all taking values
in the alphabet $A=\{0,1,\ldots,m-1\}$,
and the value of the prior parameter $\beta\in(0,1)$.
It executes the following steps:

\begin{enumerate}
\item[$(i)$]
Build an $m$-ary tree, $\TMAX$, whose leaves are 
all the contexts $x_{i-D}^{i-1}$,
$i=1,2\ldots,n$,
that appear 
in the observations $x_{-D+1}^n$.
If some node $s$ of $\TMAX$ is at depth $d<D$ 
and some but not all of its children are in
$\TMAX$, then add all its remaining children
as well, so that 
$\TMAX$ is a proper tree.
\item[$(ii)$]
Compute the count vector $a_s$ 
as in~(\ref{eq:vector-a}),
at {\em every} node $s$ of the tree $\TMAX$ 
(not only at the leaves), so that 
$a_s$ will be the all-zero vector for 
the additional leaves included in the last
step of~$(i)$.
\item[$(iii)$]
Compute the {\em estimated probability} $P_{e,s}:=P_e(a_s)$ 
given by~(\ref{eq:Pe}),
at each node $s$ of the tree $\TMAX$,
with the convention that $P_e(a_s)=1$
when $a_s$ is the all-zero count vector.
\item[$(iv)$]
Write $sj$ for the concatenation
of context $s$ and symbol $j$, corresponding
to the $j$th child of node $s$.
Starting at the leaves and proceeding
recursively towards the root,
at each node $s$ of the tree $\TMAX$
compute the {\em mixture} (or {\em weighted}) {\em probabilities}:
\ben
P_{w,s}:=
\left\{
\begin{array}{ll}
P_{e,s}, &\;\;\mbox{if $s$ is a leaf,}\\
\beta P_{e,s}+(1-\beta)\prod_{j=0}^{m-1} P_{w,sj}, &\;\;\mbox{otherwise.}
\end{array}
\right.
\een
\item [$(v)$]
Output the mixture probability $P_{w,\lambda}$ 
at the root $\lambda$.
\end{enumerate}

\noindent 
Theorem~\ref{thm:MMLA} is proved in Section~\ref{s:proofs}
of the supplementary material.

\begin{theorem} 
\label{thm:MMLA}
The mixture probability $P_{w,\lambda}$ 
at the root $\lambda$ computed by \CTW~is exactly
the prior predictive likelihood of the observations,
$$
P_{w,\lambda}
=P^*_D(x)=
\sum_{T\in{\cal T}(D)}\pi_D(T;\beta)
\int_\theta P(x_1^n|x_{-D+1}^0,\theta,T)\pi(\theta|T)d\theta,
$$
where the sum is over all 
proper context tree models of depth no greater 
than $D$.
\end{theorem}

\subsection{\BCT: The Bayesian context tree algorithm}
\label{s:MAPT}

The \BCT~algorithm takes the same input
as \CTW\ and
executes the following steps:
\begin{enumerate}
\item[$(i)$]
Build the $m$-ary tree $\TMAX$ 
and 
compute the count vectors $a_s$ 
and the estimated probabilities $P_{e,s}=P_e(a_s)$ 
at all nodes $s$ of $\TMAX$,
as in steps~$(i)$--$(iii)$ of \CTW.
\item[$(ii)$]
Starting at the leaves and proceeding
recursively towards the root,
at each node $s$ of the tree $\TMAX$
compute the {\em maximal probabilities},
\be
P_{m,s}:=
\left\{
\begin{array}{ll}
P_{e,s}, &\;\;\mbox{if $s$ is a leaf at depth $D$,}\\
\beta, &\;\;\mbox{if $s$ is a leaf at depth $<D$,}\\
\max\Big\{\beta P_{e,s},
\;(1-\beta)\prod_{j=0}^{m-1} P_{m,sj}\Big\}, &\;\;\mbox{otherwise.}
\end{array}
\right.
\label{eq:max-prob}
\ee
\item [$(iii)$]
Starting at the root and proceeding
recursively with its descendants, 
for each node $s$:
If the maximum in~(\ref{eq:max-prob})
can be achieved by the first term, 
then prune all its descendants from the 
tree $\TMAX$; otherwise, repeat the same
process at each of the $m$ children
of node $s$.
\item [$(iv)$]
After all nodes have been exhausted in~$(iii)$,
output the resulting tree $T^*_1$ and the
maximal probability at the root $P_{m,\lambda}$.
\end{enumerate}

\noindent
The following theorem is proved in Section~\ref{s:proofs}
of the supplementary material.


\begin{theorem} 
\label{thm:MAPT}
For all $\beta\geq 1/2$, the
tree $T^*_1$ produced by 
the \BCT~algorithm is the MAP tree model
(or one of the MAP tree models, in case the 
maximum below is not uniquely achieved),
\be
\pi(T^*_1|x)=\max_{T\in\clT(D)}\pi(T|x),
\label{eq:thm-MAPT}
\ee
and the maximal probability at the root satisfies,
\be
P_{m,\lambda}=P(x|T_1^*)\pi_D(T_1^*;\beta)=P(x,T_1^*).
\label{eq:max-root}
\ee
\end{theorem}

\subsection{{\hmath $k$}-{\sf BCT}: The top-{\hmath $k$} Bayesian context
trees algorithm}
\label{s:kMAPT}

The \BCTk\ is one of the main novel 
contributions of this work. 
Although it is conceptually a 
natural generalisation 
of \BCT, the precise description of its most efficient implementation 
is quite lengthy. So, for the sake of both clarity and brevity, 
we first describe here a less practical, 
idealised version of \BCTk, which is conceptually identical with 
its the practical version and only differs from it in 
the initialisation step of the leaves at depth $d<D$. 
The actual practical 
algorithm is given in Section~\ref{s:SMkBCT} of the supplementary 
material, 
along with a discussion of its implementation complexity.

The idealised \BCTk\ algorithm takes the same input as \CTW\
together with the number $k\geq 1$ of the 
{\em a posteriori} most likely models to be determined,
and it executes the following steps:
\begin{enumerate}
\item[$(i)$]
Let $\TMAX$ be 
the {\em complete} $m$-ary tree at depth $D$
(this is the ``idealised'' part),
and compute the count vectors $a_s$ 
and the estimated probabilities $P_{e,s}=P_e(a_s)$ 
at all nodes $s$ of $\TMAX$
as in steps~$(ii)$ and $(iii)$ of \CTW.

\item[$(ii)$]
Starting at the leaves 
and proceeding towards the
root, at each
node~$s$ compute a list of~$k$
maximal probabilities $P_{m,s}^{(i)}$
and~$k$ position vectors 
$c^{(i)}_s=(c^{(i)}_s(0),c^{(i)}_s(1),\ldots,c^{(i)}_s(m-1))$,
for $i=1,2,\ldots,k$, where
each $c^{(i)}_s(j)$ is an integer
between $0$ and $k$, recursively, as follows.
\begin{itemize}
\item[$(iia)$]
At each leaf $s$, we 
let $P^{(1)}_{m,s}=P_{e,s}$
and $c_s^{(1)}=(0,0,\ldots,0)$, where the all-zero
vector~$c_s^{(1)}$ indicates
that $P^{(1)}_{m,s}$ corresponds to the value of $P_{e,s}$
and does not depend on the children of $s$ (since there 
are none). 
For $i=2,3,\ldots,k$, we leave $P^{(i)}$
and $c_s^{(i)}$ undefined.
\item[$(iib)$]
At each node $s$ having only $m$ descendants (which
are necessarily leaves), we compute
the two probability-position vector pairs 
$\beta P_{e,s},$ $(0,0,\ldots,0)$ 
and
$(1-\beta)\prod_{j=0}^{m-1}P^{(1)}_{m,sj},$ $(1,1,\ldots,1)$
(where the all-1 vector indicates that the latter
probability only depends on the first maximal probability of
each of the children),
and sort them as
$P^{(1)}_{m,s}$, $c_s^{(1)}$ and
$P^{(2)}_{m,s}$, $c_s^{(2)}$ in order
of decreasing probability.
For $i=3,4,\ldots,k$, we leave $P^{(i)}$
and $c_s^{(i)}$ undefined.
\item[$(iic)$]
A general internal node $s$ has $m$ children,
where each child $sj$ has a list of $k_j$ (for
some $1\leq k_j\leq k$)
probability-vector pairs 
$P^{(i)}_{m,sj},c_{sj}^{(i)}$, $1\leq i\leq k_j$. We compute
the probability $\beta P_{e,s}$ with associated
position vector $(0,0,\ldots,0)$, and all possible
probability-position vector pairs,
\be
(1-\beta)\prod_{j=0}^{m-1}P^{(i_j)}_{m,sj},\;\;\;(i_0,i_1,\ldots,i_{m-1}),
\label{eq:probimax}
\ee
for all possible combinations of indices
$1\leq i_j\leq k_j$ for $0\leq j\leq m-1$.
We then sort these $k'=1+k_0\times k_1\times\cdots\times k_{m-1}$
probabilities in order of decreasing probability,
and rename the top $k$ of them as $P_{m,s}^{(i)}$,
for $i=1,2,\ldots,k$, together with their associated
position vectors $c^{(i)}_{s}$. [Of course, if
$k'<k$, after sorting we leave
the remaining $k-k'$ probability-position vector pairs
undefined.]
\end{itemize}

\item [$(iii)$]
Having determined all maximal probabilities 
$P_{m,s}^{(i)}$ for all nodes $s$ and $1\leq i\leq k$, we 
now determine the ``top $k$'' trees $T_1^*,T_2^*,\ldots,T_k^*$
from the corresponding position vectors $c_s^{(i)}$.
For each $i$ we repeat the following process,
starting at the root and proceeding 
until all available nodes of the tree $\TMAX$
have been exhausted.
\begin{itemize}
\item[$(iiia)$]
{\em Depth $d=0$.}
At the root node $\lambda$, we examine 
$c_\lambda^{(i)}$. If it is the all-zero vector,
then $T_i^*$ is the tree consisting of the root
node only. Otherwise, we add to $T_i^*$ the branch
of $m$ children starting at the root, and 
proceed to examine each of the nodes corresponding
to the $m$ children recursively. 
\item[$(iiib)$]
{\em Depth $d=1$.}
Reaching node $s=j$ corresponding to the 
$j$th child of the root, means
that $t=c_\lambda^{(i)}(j)$ is nonzero.
We examine $c_s^{(t)}$: If it is the all-zero vector,
then we prune from $T_i^*$ all the descendants of $s$
and move to the next unexamined node;
otherwise, we add to $T_i^*$ the branch
of $m$ children starting at $s$, and 
proceed to examine each of the nodes 
corresponding to the $m$ children recursively. 
\item[$(iiic)$]
{\em General depth $1\leq d\leq D-1$.}
Reaching a node $sj$ at depth $d$ from its parent 
node $s$ means that we decided to visit
$sj$ because $t=c_s^{(u)}(j)$ is nonzero
for the appropriate index $u$ (corresponding
to the position vector $c_s^{(u)}$ that was
examined at node $s$).
We examine $c_{sj}^{(t)}$: If it is the all-zero vector,
then we prune from $T_i^*$ all the descendants of $sj$
and move to the next unexamined node;
otherwise, we add to $T_i^*$ the branch
of $m$ children starting at $sj$, and 
proceed to examine each of the nodes 
corresponding to the $m$ children recursively. 
\item[$(iiid)$]
{\em Depth $d=D$.} Reaching a node $s$ at
depth $D$ means we have reached a leaf of
$\TMAX$, so we simply add $s$ to 
$T_i^*$ and proceed to the next unexamined
node.
\end{itemize}
\item [$(iv)$]
Output the $k$ resulting trees $T^*_i$
and the $k$ maximal probabilities at the root,
$P^{(i)}_{m,\lambda}$, $i=1,2,\ldots,k$.
\end{enumerate}

\noindent
Theorem~\ref{thm:MAPT-k}
is proved in Section~\ref{s:proofs2}
of the supplementary material.

\begin{theorem} 
\label{thm:MAPT-k}
For any $\beta\geq 1/2$, the trees 
$T^*_1,T^*_2,\ldots,T^*_k$ produced 
by the \BCTk~algorithm are the $k$
{\em a posteriori} most likely tree models: 
For each
$i=1,2,\ldots,k$,
\be
\pi(T^*_i|x)=\maxi_{T\in\clT(D)}\pi(T|x),
\label{eq:thm-MAPT-k}
\ee
where $\max_{t\in\clT}^{(i)}f(t)$ of a function
$f$ defined on a discrete set of arguments $t\in \clT$
denotes the $i$th largest value of $f$. Moreover, 
the $i$th maximal probability at the root 
satisfies,
\ben
P^{(i)}_{m,\lambda}
=P(x|T_i^*)\pi_D(T_1^*;\beta)
=P(x,T_i^*),\;\;\;i=1,2,\ldots,k.
\een
\end{theorem}

\noindent
Once again we note that, if some of the maxima in~(\ref{eq:thm-MAPT})
are not uniquely achieved, then $T^*_1,\ldots,T_k^*$ are one of the
equivalent collections of $k$ {\em a posteriori} most likely models.

\subsection{Computation of posterior probabilities}
\label{s:additional}

In addition to the prior predictive likelihood 
$P_D^*(x)=P_{w,\lambda}$ computed by \CTW,
and the {\em a~posteriori} most likely models 
identified by the \BCT~and \BCTk~algorithms, 
there are several other useful quantities that can 
easily be obtained through the results
of these algorithms.

For a fixed maximal depth $D\geq 0$ and a fixed
$\beta\in[1/2,1)$, let $x$ denote a
time series $x_1^n$ with 
initial context 
$x_{-D+1}^0$.

\medskip

\noindent
{\em Model posterior probabilities.} 
For any model $T\in\clT(D)$, 
\be
\pi(T|x)=\frac{P(x|T)\pi(T)}{P_D^*(x)}
=\frac{\pi(T)\prod_{s\in T}P_e(a_s)}{P_{w,\lambda}},
\label{eq:mpost}
\ee
where $P_D^*(x)=P_{w,\lambda}$ is the prior predictive likelihood 
computed by \CTW,
and the numerator is given as in 
Lemma~\ref{lem:integrate}.
If a model $T$ has a leaf
$s$ that is 
not included in
the tree $\TMAX$ generated by \CTW,
so that no corresponding count-vector $a_s$
is available,
we follow the convention of setting
$a_s=(0,0,\ldots,0)$ and $P_e(a_s)=1$.

%


\medskip

\noindent
{\em Full conditional density of $\theta$.}
Conditional on a model $T$ and observations $x$, 
the density of the parameters
$\theta=\{\theta_s\;;\;s\in T\}$ is,
$$\pi(\theta|x,T)
\propto
P(x|\theta,T)\pi(\theta|T)
\propto
\left(
\prod_{s\in T}
\prod_{j=0}^{m-1} \theta_s(j)^{a_s(j)}\right)
\left(\prod_{s\in T}
\prod_{j=0}^{m-1}\theta_s(j)^{-\frac{1}{2}}\right),$$
where we have used the definition
of the prior on $\theta$ in~(\ref{eq:theta-prior}) and
the expression for the likelihood~(\ref{eq:likelihood3}).
Therefore, the full conditional density
$\pi(\theta|x,T)$ is the product of
Dirichlet densities,
\be
\pi(\theta|x,T)\sim
\prod_{s\in T}\mbox{Dir}
(a_s(0)+1/2,a_s(1)+1/2,\ldots,a_s(m-1)+1/2).
\label{eq:full-cond}
\ee

\subsection{MCMC samplers}
\label{s:MH}

The \BCTk\ algorithm can identify
the $k$ {\em a posteriori} most likely models,
including in cases where the posterior $\pi(T|x)$ is multimodal,
but, as discussed in Section~\ref{s:sequential},
its complexity grows faster than linearly in $k$,
which makes it impractical
for large values of $k$. 
Next, we describe a family of effective variable-dimension
MCMC
samplers that 
make it possible to explore
$\pi(T|x)$ further,
and to sample from the posterior $\pi(\theta,T|x)$ jointly
on models and parameters.

The random walk (RW) sampler for $\pi(T|x)$ is
a Metropolis-Hastings
algorithm with a proposal distribution 
that, at each step, either adds or removes an $m$-tuple of leaves 
from the current tree. 

\medskip

\noindent
{\bf RW sampler.}
It takes as input the same parameters
as \BCT, and also:
An initial model $T^{(0)}\in\clT(D)$,
the required number $N\geq 1$ 
of MCMC iterations, and the
tree $\TMAX$ together with
the estimated probabilities $P_{e,s}$ at all nodes 
of $\TMAX$, 
as computed by \CTW.
It executes the following steps at 
each iteration $t=1,2,\ldots,N-1$:

\vspace{-0.1in} 

\begin{itemize}
\item[$(i)$]
Given $T^{(t)}$, propose a new tree $T'$ as follows:
\begin{enumerate}
\item[$(a)$]
If $T^{(t)}$ is the empty tree $\Lambda:=\{\lambda\}$ consisting
only of the root node, let $T'$ be the
complete tree at depth 1, $T_c(1)$.
\item[$(b)$]
If $T^{(t)}$ is the complete $m$-ary tree at depth $D$,
$T_c(D)$,
choose uniformly at random one of the internal
nodes $s$ at depth $D-1$ and let $T'$ be 
the same as $T^{(t)}$ but with the $m$ children 
of $s$ removed.
\item[$(c)$]
Otherwise, with probability 1/2 decide 
to propose a larger tree $T'$, formed by 
choosing uniformly at random one of the
leaves of $T^{(t)}$ at depths $\leq D-1$ and
adding its $m$ children to form $T'$;
\item[$(d)$]
Or, with probability 1/2
decide 
to propose a smaller tree $T'$, formed 
by choosing uniformly at random one of the
internal nodes $s$ of $T^{(t)}$ that only have
$m$ descendants and removing the $m$
leaves that stem from $s$, to form $T'$.
\end{enumerate}
\item[$(ii)$] 
Either 
accept $T'$
and set $T^{(t+1)}=T'$, or reject
it and set $T^{(t+1)}=T^{(t)}$, with
corresponding probabilities
$\alpha(T^{(t)},T')=\min\{1,r(T^{(t)},T')\}$ 
and $1-\alpha(T^{(t)},T')$, respectively; 
explicit expressions for the
ratios $r(T,T')$ are given in Section~\ref{s:explicit}
of the supplementary material.
\end{itemize}

The jump sampler for $\pi(T|x)$ 
is a modification of the RW sampler, 
which, in addition to nearest neighbour moves,
also allows for jumps to any one of the $k$
most likely models.
This way we overcome the common difficulty of RW 
samplers to move between separated modes of 
multimodal posterior distributions.

\medskip

\noindent
{\bf Jump sampler.}
It takes as input the same 
parameters as the RW sampler, and also
the value of a jump parameter $p\in(0,1)$
and the collection $\clT^*=\{T_1^*,\ldots,T_k^*\}$
of the top $k$ trees computed by \BCTk.
It executes the following steps at 
each iteration $t=1,2,\ldots,N-1$:

\vspace{-0.05in} 

\begin{itemize}
\item[$(i)$]
Given $T^{(t)}$, propose a new tree $T'$ as follows:
\begin{enumerate}
\item[$(a)$]
With probability $(1-p)$, propose a new tree $T'$ as in
steps $(ia)$--$(id)$ of the RW sampler;
\item[$(b)$]
Or, with probability $p$,
propose a jump move:
Let $T'$ be one of the top $k$ trees
$T_i^*$, uniformly chosen from $\clT^*$.
\end{enumerate}

\item[$(ii)$]
Either accept $T'$
and set $T^{(t+1)}=T'$, or reject
it and set $T^{(t+1)}=T^{(t)}$, with
corresponding probabilities
$\alpha(T^{(t)},T')=\min\{1,r(T^{(t)},T')\}$ 
and $1-\alpha(T^{(t)},T')$, 
where the ratios $r(T,T')$ are
given explicitly in Section~\ref{s:explicit}
of the supplementary material.
\end{itemize}

\noindent
Note that a jump move to one of the top $k$
models $T_i^*$ may be proposed from any state
$T^{(t)}$ of the sampler, but it only has a nonzero
probability of being 
accepted if $T^{(t)}$ itself is either one of the
$T_i^*$ or a neighbour of one of them.
This suggests that the jump parameter
should be chosen so that $(1-p)$ is not
too small; in all of our experiments below
we take $p=1/2$. 

\medskip

\noindent
{\bf MCMC convergence.} The target distribution of
both the RW and jump samplers is the posterior
$\pi(T|x)$, $T\in\clT(D)$, on model space. Unlike
with most MCMC samplers used in Bayesian inference,
here we can in fact compute the value of the target
distribution $\pi(T|x)$ exactly, for any specific
$T\in\clT(D)$,
as noted in Section~\ref{s:additional}.
Nevertheless, because of the enormity of the
space $\clT(D)$ and the fact that it does not
possess any easily exploitable structure, it is still
practically impossible to sample from $\pi(T|x)$
directly, hence we resort to MCMC. On the other hand,
knowing the posterior probabilities $\pi(T|x)$
precisely (not only up to a multiplicative constant)
means that it is easy to obtain a good first indication
of whether the sampler has converged, or at least whether 
it has spent the ``right'' amount of time in 
the important areas of the support of $\pi(T|x)$ near its
mode(s): Simply compute the
frequency of each of the top $k$ models $T_i^*$
in the MCMC sample, and compare it with its actual
posterior probability $\pi(T_i^*|x)$.

Although we have not observed convergence issues
in our experiments on simulated or real data, we note
that more sophisticated MCMC methods can also be used,
e.g., tempering the likelihood \citep{robert:book}
or using a Wang-Landau-style algorithm to force
the sampler to spend a specified proportion of time
at models of each depth \citep{atchade:04}.

\medskip

\noindent
{\bf Joint sampler.}
Being able to obtain MCMC samples $\{T^{(t)}\}$ for $\pi(T|x)$,
and knowing the full conditional
density $\pi(\theta|x,T)$ of the parameters
explicitly as in~(\ref{eq:full-cond}),
it is simple to obtain a corresponding sequence 
of samples $\{(\theta^{(t)},T^{(t)})\}$ 
for the posterior $\pi(\theta,T|x)$
jointly on models and parameters.
This can be done by drawing
a conditionally independent sample 
$\theta^{(t)}\sim \pi(\theta|x,T^{(t)})$
(a Gibbs-type step)
at each MCMC iteration.

\subsection{Sequential updates, prediction, and complexity}
\label{s:sequential}

As more observations become available,
the results of all three exact inference
algorithms in 
Sections~\ref{s:MMLA}--\ref{s:kMAPT}
can be updated sequentially.
This facilitates their online use 
in applications where
it is important that data be processed
sequentially rather than in large blocks. 

For \CTW, having computed 
$P^*_D(x_1^n|x_{-D+1}^0)$ 
and given an additional sample $x_{n+1}$,
the new prior predictive likelihood
$P_D^*(x_1^{n+1}|x_{-D+1}^0)$ can be obtained
in $O(D)$ operations
by updating steps~$(ii)$--$(iv)$ of
\CTW~as follows.
Let $s_D,s_{D-1},\ldots,s_0$ be the contexts
of lengths $D,D-1,\ldots,0$, respectively,
immediately preceding $x_{n+1}$, 
so that in particular 
$s_D=x_{n-D+1}^n$ and $s_0=\lambda$. 
Suppose $x_{n+1}=j$; at each
of the nodes $s_D,s_{D-1},\ldots,s_0$
in the tree $\TMAX$ already constructed
(in that order, and {\em only} there):
\begin{itemize}
\item[$(ii')$]
Update $a_s(j)$ and
$M_s$
by increasing
each of their values by $1$.
\item[$(iii')$]
Update $P_{e,s}=P_e(a_s)$
by multiplying its earlier value by
$(a_s(j)-1/2)/
(m/2+M_s-1)$, for the updated
values of $a_s(j)$ and $M_s$.
\item[$(iv')$]
Re-compute the  probability
$P_{w,s}$.
\end{itemize}
The required result $P_D^*(x_1^{n+1}|x_{-D+1}^0)$
is the (updated) mixture probability
$P_{w,\lambda}$ at the root.


The corresponding update rules 
for \BCT\ and \BCTk\ are 
analogous and easy to determine
and implement.


The ability
to compute the prior predictive
likelihood  sequentially
makes it easy to perform
online prediction.
The canonical Bayesian rule
for predicting the next 
observation $x_{n+1}$ given the past $x_1^n$,
is given by the {\em posterior predictive distribution},
\be
P^*_D(x_{n+1}|x_1^n)=\sum_T\int_\theta P(x_{n+1}|x_1^n,\theta,T)
\pi(\theta,T|x_1^n)\,d\theta
\label{eq:defn},
\ee
where we suppress the dependence
on the initial context $x_{-D+1}^0$
for brevity.
Although it is common that
$P^*_D(x_{n+1}|x_1^n)$
can only be estimated 
(e.g., using MCMC sampling or approximate 
model averaging),
here its value
can be computed exactly and
sequentially by \CTW, via,
\be
P^*_D(x_{n+1}|x_1^n)=\frac{P^*_D(x_1^{n+1})}{P^*_D(x_1^n)}.
\label{eq:defn2}
\ee
Applications of this methodology 
on both simulated and real data are 
given in Section~\ref{s:prediction}.

Next we briefly discuss the implementation complexity
of the three algorithms in 
Sections~\ref{s:MMLA}--\ref{s:kMAPT},
as a function of the parameters $n,m,D$, and $k$.
The complexity of \CTW\ is linear in each of 
$n,m$ and $D$, and in fact it is of order 
$O(nmD)$. To see this, observe that 
in step~$(i)$, for each~$x_i$, $1\leq i\leq n$, 
a new node is created for each of the
$(D+1)$ contexts of $x_i$, of lengths $0,1,\ldots,D$. 
This requires $O(nD)$ operations and produces 
the tree $\TMAX$ which has no more 
than $nD+1$ nodes. 
The second and third steps, 
where the count-vectors $a_s$ 
and the probabilities $P_e(a_s)$ 
are computed, can be integrated 
into the first one. For each $i$,
when we visit each of the $(D+1)$ contexts
$s$ of $x_i$, we increase the 
corresponding counts $a_s(x_i)$ 
by one and update the values
of $P_e(a_s)$, using a constant
number of operations.
Therefore,
the additional complexity of
steps~$(ii)$ and~$(iii)$ is again
$O(nD)$. Lastly, to compute
the mixture probabilities
in step~$(iv)$, at each of
the (no more than $nD+1$) nodes
of $\TMAX$ we perform 
$O(m)$ operations, so that the overall 
complexity is $O(nD)+O((nD+1)m)=O(nmD)$ 
operations.

A similar argument shows that,
as a function of $n$ and $D$, the
complexity of both the \BCT\
and \BCTk\ algorithms is also $O(nD)$.
But as a function of $k$
the complexity of \BCTk\
grows faster than linearly
and increases very substantially
as we require more 
information about the area near the 
mode of the posterior $\pi(T|x)$ in model
space; more details are given in
the relevant discussion in Section~\ref{s:SMkBCT}
of the supplementary material.
This, in part, was the motivation
for the MCMC
algorithms described in Section~\ref{s:MH}.

From the above discussion and the 
description of the algorithms
it is also easy to see that 
the memory requirements of the
\CTW~and \BCT~algorithms are
of order $O(nmD)$, and for 
\BCTk~of order $O(nmDk)$.

\subsection{Bibliographical remarks}
\label{s:CTWhistory}

The \CTW\ algorithm~was first introduced 
for data compression in 1993. 
It was originally described 
for binary observations ($m=2$) and only in the 
special case of $\beta=1/2$
by \cite{ctw-1:93}.
The general version of \CTW\ for
chains on non-binary alphabets
and arbitrary $\beta$ was introduced by
\cite{ctw-multi:94}, 
without 
reference to Bayesian 
inference. 
The connection between \CTW~and
the prior predictive likelihood $P_D^*(x)$ established
in Theorem~\ref{thm:MMLA} was given
in the restricted setting $m=2,\beta=1/2$
in the unpublished work~\cite{willems-shtarkov-tjalkens:pre},
and the outline of a corresponding 
argument in the case of general $\beta$ 
(still only for binary time series) was 
later described in~\cite{ctw:02}.
The result of Theorem~\ref{thm:MMLA}
at the level of generality stated here
is new, as is the class of
prior distributions $\pi_D(T;\beta)$.

A special case of the \BCT~algorithm,
termed the
context tree maximizing (CTM) algorithm,
was first introduced, again in the context
of data compression,
by \cite{willems-shtarkov-tjalkens:pre,volf-willems:94,volf-willems:95b},
for binary observations 
($m=2$) and only in the special case of 
$\beta=1/2$. An extension
for arbitrary $\beta$ (still only for $m=2$) 
was later given
in \cite{ctw:02}; the general version
of the \BCT\ algorithm~presented here is 
new. 
The fact that the \BCT~algorithm identifies the 
MAP tree model
was established
in the restricted setting $m=2$, $\beta=1/2$
in~\cite{willems-shtarkov-tjalkens:pre,ctw:00},
and some generalisations
(still restricted to $m=2$) are discussed
in~\cite{ctw:02}.
Theorem~\ref{thm:MAPT} at the level
of generality stated here is new.

The \BCTk~algorithm,
Theorem~\ref{thm:MAPT-k},
and the MCMC samplers in Section~\ref{s:MH} are new.

\newpage

\section{Model selection}
\label{s:compare}

Here, we compare
the model selection performance of the \BCT\ algorithms
of Section~\ref{s:algorithms} 
with the VLMC and MTD approaches
described in the Introduction.
In the rest of this section we give
brief descriptions of how these three
different techniques will be used.
In Section~\ref{s:simulated}
we compare them on 
simulated data and discuss some
preliminary conclusions.
Section~\ref{s:realdata} contains
corresponding results on real data.

\medskip

\noindent
{\bf Bayesian context trees.}
The \BCT~framework
provides
a consistent foundation for learning
and evaluating appropriate models for
a given data set $x$. 
The \BCT~and \BCTk~algorithms
can be used to 
identify the $k$ {\em a posteriori}
most likely models $T_i^*$, $i=1,2,\ldots,k$,
for some reasonable $k$,
and we can further explore
$\pi(T|x)$ using the MCMC 
samplers of Section~\ref{s:MH}.
With the interpretation of $\pi(T|x)$
as a measure of the ``truth'' of a model $T$ \citep{chipman:01},
the posterior probability provides
a quantitative confidence measure 
for the resulting models.

Let $\{X_n\}$ be a variable-memory chain with model 
$T\in\clT(D)$. 
The specific model $T$ that describes the chain is 
in general
not unique. E.g., every 
independent and identically distributed 
(i.i.d.)
sequence $\{X_n\}$
can also trivially be described as a first order
Markov chain, and adding $m$ children to any leaf
of $T$ which is not at maximal depth and giving each of them
the same parameters as their parent, leaves the distribution
of the chain unchanged.
Naturally, the goal in model selection is to identify
the smallest model that can
fully describe the distribution of the chain.
We call a model $T\in\clT(D)$ {\em minimal} with respect
to the parameter vector $\theta=\{\theta_s\;;\;s\in T\}$, 
if $T$ is either equal to $\Lambda=\{\lambda\}$ 
or, if $T\neq\Lambda$,
then every $m$-tuple of leaves $\{sj\;;\;j=0,1,\ldots,m-1\}$
in $T$
contains at least two with non-identical parameters,
i.e., there are $j\neq j'$ such that $\theta_{sj}\neq\theta_{sj'}$.
It is easy to see 
that every $D$th order 
Markov chain $\{X_n\}$ has a unique minimal model 
$T^*\in\clT(D)$.

The following result, established by
\cite{willems-shtarkov-tjalkens:pre,ctw-4:93}
in the context of the MDL principle,
offers a partial
frequentist justification
for the Bayesian \BCT\ framework.
It says that, for
binary chains, and in the special
case $\beta=1/2$, the MAP model $T^*(n)$ is 
eventually equal to~$T^*$, with probability~1.

\begin{theorem}
\label{thm:modelC}
Let $\{X_n\}$ be an ergodic, variable-memory
chain, with alphabet size $m=2$ and minimal model $T^*\in\clT(D)$.
For $\beta=1/2$, the MAP model $T^*(n)$ based on the random
sample $X_{-D+1}^n$ is eventually unique with probability~1,
and in fact:
$$T^*(n)=T^*,\;\;\;\mbox{eventually, with probability~1.}$$
\end{theorem}

\noindent
{\bf Variable-length Markov chains.}
The VLMC class~\citep{buhlmann:99,buhlmann:00}
consists of tree models
very similar to those in the \BCT\ class $\clT(D)$,
except for the fact that VLMC trees
are not necessarily proper.
The associated VLMC model 
selection algorithm also has similarities
with the pruning procedure of the
\BCT~algorithm: First a version of the
tree $\TMAX$ is constructed and the 
count vectors $a_s$ are computed,
as in the \BCT,
and then
$\TMAX$ is pruned,
based on a {\em cut-off parameter}~$K$
that plays a role analogous to $\beta$,
producing the final model. 
Theoretical justifications
for the resulting VLMC model
are provided in~\cite{buhlmann:99},
where general conditions
for asymptotic consistency 
are established.

The VLMC results in our experiments below 
are obtained using the implementation
in the {\sf R} package {\sf VLMC}.
Since the algorithm that uses
the default value of the cut-off parameter~$K$
(``default-VLMC'') generally gives significantly
inferior results,
we also examine the results obtained by
optimizing the choice of~$K$ in order to minimise
the BIC or the AIC score (``best-BIC-VLMC'' and
``best-AIC-VLMC''). But these parameter optimisations
are computationally very costly, as we point out
in more detail at the end of Section~\ref{s:simulated}.

\newpage


\noindent
{\bf Mixture transition distribution models.}
The original `single-matrix' version of the 
MTD model \citep{raftery:85} is
based on a different way of parsimoniously
representing the $D$th order transition probabilities
of a Markov chain, as a mixture of the form,
$$P(x_0|x_{-D}^{-1})=\sum_{d=1}^D\lambda_dQ(x_0|x_{-d}),$$
where $(\lambda_1,\lambda_2,\ldots,\lambda_D)$ 
are lag parameters 
and $Q=(Q(j|i))$ is a 
stochastic matrix. 
Numerical methods for fitting
this model via approximate
maximum likelihood were developed
by \cite{raftery-tavare:94},
and a generalisation, the {\em multi-matrix
MTD model}, or MTDg, was introduced by
\cite{berchtold:96,berchtold:98}.
The MTDg model is based on the more
general representation, 
$$P(x_0|x_{-D}^{-1})=\sum_{d=1}^D\lambda_dQ^{(d)}(x_0|x_{-d}),$$
where a different stochastic matrix
$Q^{(d)}= (Q^{(d)}(j|i))$
is used for each lag $d=1,2,\ldots,D$.

In our experiments below we use the MTD implementation in the 
{\sf R} package {\sf march}.
For each data set we run the MTD algorithm 
for a range of possible depths $D$, and choose the 
value of $D$ that maximises the corresponding BIC or AIC score.
We refer to the resulting models as the best-BIC-MTD 
and best-AIC-MTD models.
Similarly we obtain 
the best-BIC-MTDg and best-AIC-MTDg models
from the multi-matrix version.

\medskip

\noindent
{\bf Model comparison.}
A natural and logically consistent
way to compare different models from
$\clT(D)$ is to compare their
posterior probabilities $\pi(T|x)$,
but this is not possible for
models produced by different methods.
In such cases, we follow the common practice
of comparing their 
AIC \citep{akaike:73} and BIC \citep{schwarz:78}
scores.
BIC is asymptotically consistent
when dealing with a fixed number of finite-dimensional 
parametric models \citep{hannan:79,wei:92}
whereas AIC is typically asymptotically efficient 
and minimax optimal in infinite-dimensional,
nonparametric 
settings \citep{shibata:80,barron:99}; see 
also \cite{dziak:19}.
This dichotomy suggests that, for our purposes,
BIC is the more relevant criterion,
as confirmed by our findings
in Section~\ref{s:simulated} and Section~\ref{s:scores2}
of the supplementary material.

\subsection{Simulated data}
\label{s:simulated}

We compare the performance of the three different 
model selection approaches described above
on a relatively short synthetic time series.
Analogous comparisons on simulated data from two 
more chains -- one from the VLMC paper \citep{buhlmann:00}
and one from the MTD paper \citep{berchtold:02} --
are carried out in Section~\ref{s:scores2} 
of the supplementary material.

Consider the 5th order variable-memory chain $\{X_n\}$
on the alphabet $A=\{0,1,2\}$ of $m=3$ letters,
with model given by the tree $T$
shown in Figure~\ref{fig:running}
of Section~\ref{s:prior}; the associated
parameter vector $\theta=\{\theta_s\;;\;s\in T\}$
is given in Section~\ref{s:explicit} of the
supplementary material.

With $x_1^n$ consisting of
$n=10,000$ simulated observations $(x_1,x_2,\ldots,x_n)$
and with
a given initial context $x_{-D+1}^0=(x_{-D+1},\ldots,x_0)$, 
the MAP model $T_1^*$
obtained by the \BCT~algorithm
with depth $D=10$ and
$\beta=1-2^{-m+1}=3/4$
is the true underlying model
with respect to which the data were generated.
Its posterior probability is
$\pi(T_1^*|x)\approx 0.3946$ 
while its prior probability is
$\pi(T_1^*)\approx 5.8\times 10^{-6}$.
Although it may seem quite unremarkable
that the ``correct'' model is identified based on a 
series of 10,000 samples,
it is worth noting that, with $D=10$, there are
more than $10^{5900}$ different models in $\clT(D)$.


\newpage

\begin{figure}[ht!]
\vspace{-0.2in}
\centerline{\includegraphics[width=4.9in]{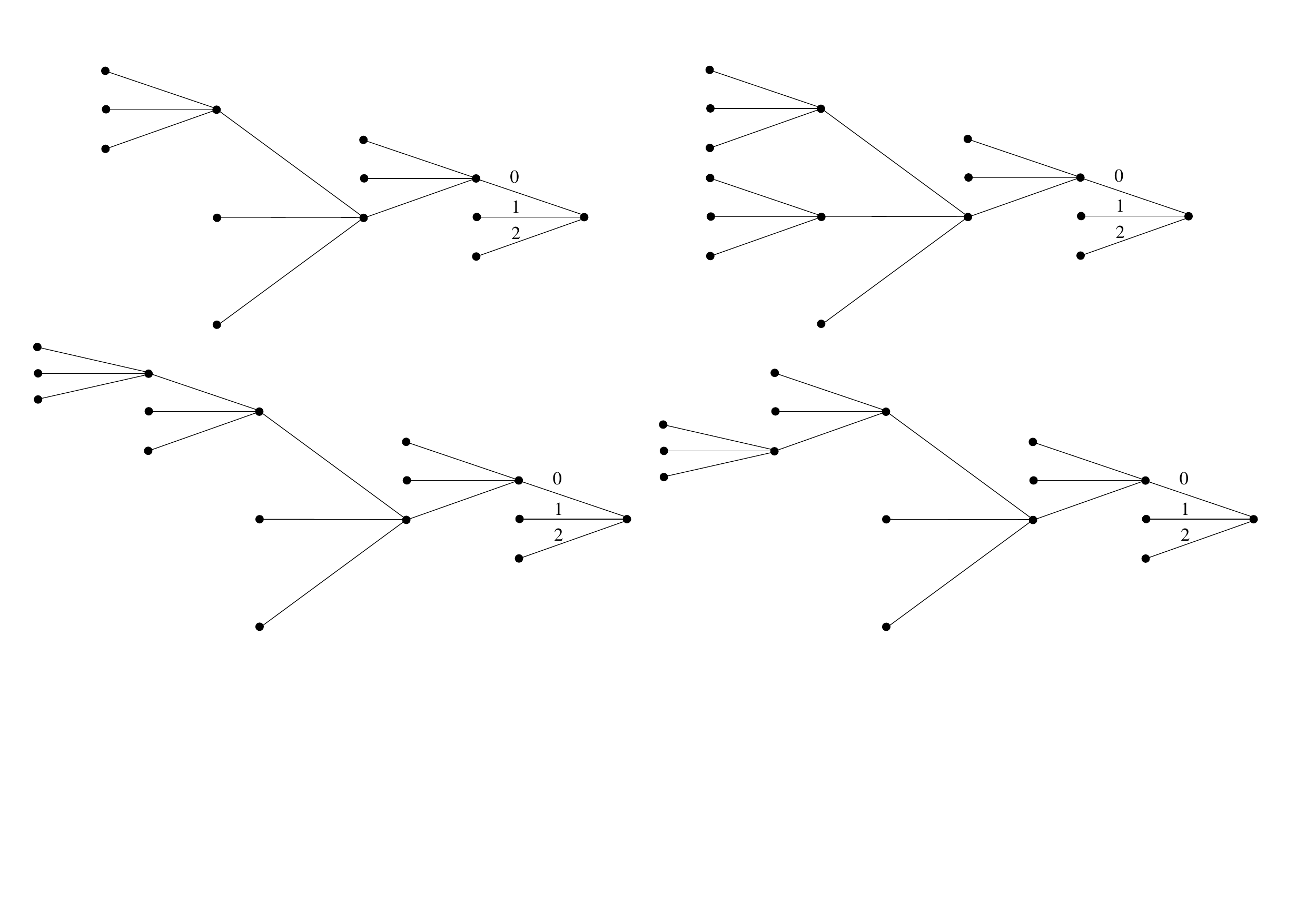}}
\vspace{-1.1in}
\caption{The first, second, third, and fifth {\em a posteriori}
most likely models $T_1^*,T_2^*,T_3^*,T_5^*$ produced by 
\BCTk~with 
$n=1,000$ samples.
The posterior odds $\pi(T_1^*|x)/\pi(T_i^*|x)$ 
with respect
to the MAP model $T_1^*$ are approximately
2.369, 5.358, 12.69, and 15.24, for $i=2,4,5$, respectively.}
\label{fig:scores1b}
\end{figure}

With $n=1,000$ samples, $D=10$, 
and $\beta=1-2^{-m+1}=3/4$, the five
{\em a posteriori} most likely models 
produced by the \BCT~algorithm 
are shown in Figure~\ref{fig:scores1b}.
The MAP model is a depth-4 subtree 
of the true underlying model,
with prior probability $\pi(T_1^*)\approx 2.9\times 10^{-4}$
and posterior
$\pi(T_1^*|x)\approx 0.2702$.
The true model appears as the 4th most 
likely tree, $T_4^*$,
with posterior 
probability $\pi(T_4^*|x)\approx0.0213$.
The 
sum of the posterior probabilities
of the top $5$ models is approximately
0.4737.

\begin{figure}[ht!]
\includegraphics[width=6.6in]{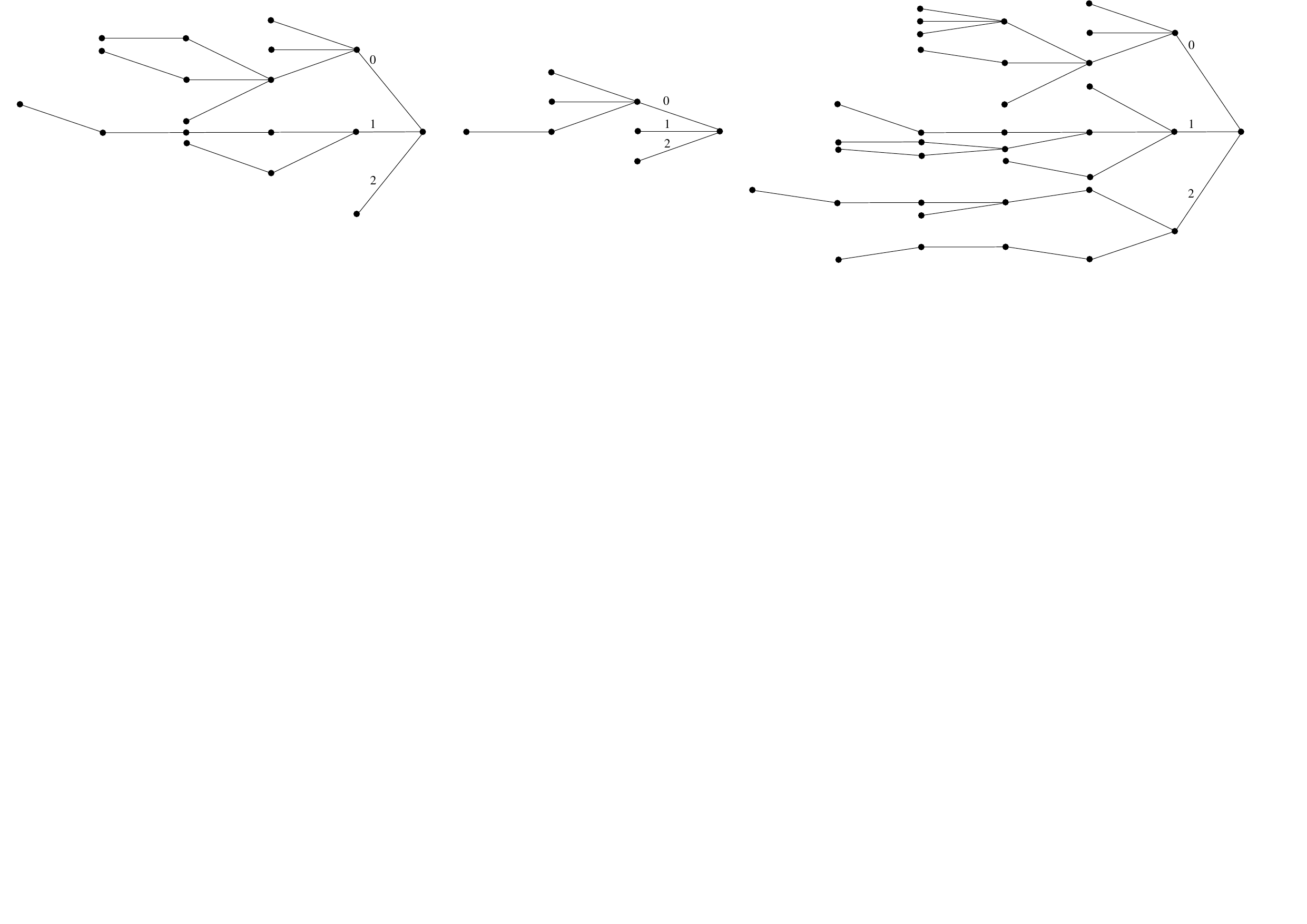}
\vspace{-3.5in}
\caption{The models produced by the default-VLMC,
the best-BIC-VLMC and the best-AIC-VLMC methods,
with $n=1,000$ samples.}
\label{fig:scores2}
\end{figure}

The default-VLMC model is the first tree shown 
in Figure~\ref{fig:scores2};
only about half
of its nodes appear in 
the true underlying model. It has a worse BIC score and 
a better AIC score than the MAP model.
The best-BIC-VLMC produces the small tree 
of depth~3 shown second
in Figure~\ref{fig:scores2},
which is a subtree of the true model;
it has a good BIC score and a poor AIC score. 
In sharp contrast, the best-AIC-VLMC produces a clearly
overfitted model of depth 6, shown 
third in Figure~\ref{fig:scores2}.
Although it has a poor BIC score, it has a very 
good AIC score, as expected.

Finally, the best-BIC-MTD and the best-BIC-MTDg both 
give $D=0$, whereas the best-AIC-MTD gives $D=3$
and the best-AIC-MTDg gives $D=2$. 
Their scores (both BIC and AIC) are generally 
quite a bit worse than those of the MAP model 
or the models produced by VLMC.

Overall, the \BCT\ and the best-BIC-VLMC algorithms achieve the
best performance in terms of scores, and they learn part of the 
true underlying tree. VLMC has a marginally better BIC score whereas 
the \BCT\ algorithm has a slightly better AIC score (by approximately 
1\% in both cases). More importantly, the \BCT\ MAP model $T_1^*$ has 
an additional full branch at depth 4 that reveals more of the 
true underlying structure,
and the \BCTk\ identifies the true model as $T_4^*$.


\newpage

\noindent
{\bf Discussion.}
In the results on simulated data 
above and
in the two examples in Section~\ref{s:scores2} 
of the supplementary material,
the \BCT~and \BCTk~algorithms 
consistently give the most accurate model fit,
with the best-BIC version of VLMC often 
giving similar results.
The \BCT~framework has the 
advantage of identifying not just
a single model, but the top $k$ {\em a posteriori}
most likely models, together with their
exact prior and posterior probabilities.
And, importantly, in
terms of complexity,
the \BCT~algorithm is more efficient than either
VLMC or MTD,
typically by at least two orders of magnitude,
since it does not require any tuning.

{\sl VLMC.} The default and best-AIC versions of VLMC
generally gave similar or identical models,
usually much larger than the true underlying model, 
in rather typical examples of overfitting.
The best-BIC-VLMC was found to be much more accurate
in revealing significant parts 
of the true model,
as expected in view of the earlier 
AIC-vs-BIC discussion.
The resulting models were smaller,
which is consistent with the 
observation that optimizing the BIC score
led to larger values for the cut-off parameter $K$
and more frequent pruning.
For the best-AIC and best-BIC versions,
we executed VLMC approximately 500 times with different
values of $K\in[1,10]$.
Although a smaller range 
($2.8\leq K\leq 6$) was used by \cite{buhlmann:04},
we found it often necessary to look further;
e.g., for the financial data
in Section~\ref{s:FB} of the
supplementary material
the best value was found to be close to 10,
and for the genetic data in Section~\ref{s:realdata}
it was larger than~$20$.
In most cases, using fewer runs 
resulted in different models
giving poor fits.

{\sl BCT.}
The MAP models produced by the \BCT~algorithm were usually
similar to those produced by the best-BIC-VLMC, they had good AIC and
BIC scores, and they generally learned the most accurate 
approximations of the underlying model among all methods considered. 
The additional models obtained by \BCTk~offered further
indications of the type of structure present in the
data, and they were accompanied by posterior probabilities, 
indicating the level of ``posterior confidence'' we may have in these models. 
Also, \BCT~and \BCTk\ 
only require a single run with the default value of the 
hyperparameter $\beta$.

{\sl MTD.}
None of the MTD methods performed as well 
as the \BCT\ or VLMC algorithms,
partly because they gave (by design) less flexible models.
As their only output in terms of the model is the  
Markov order $D$ and the number of nonzero lag parameters
$\lambda_d$,
which can only take a few discrete values,
the best-BIC and best-AIC results were often same.
MTDg in most cases had worse scores than MTD,
but even the MTD's overall performance was not 
competitive with that of \BCT\ and VLMC.  
Finally, MTD had very high complexity, 
since the implicit maximum likelihood computation
is over a non-convex space
defined by a large number of non-linear
constraints \citep{raftery-tavare:94}. For this reason,
it appears infeasible in practice to use
values for $D$ significantly larger that $D=10$.

\subsection{Real data}
\label{s:realdata}

In view of the above discussion,
for the comparisons in the real-world data examples
we only consider the best-BIC versions of VLMC, MTD
and MTDg. One more example with a financial time series
is given in Section~\ref{s:FB} 
of the supplementary material.

\medskip

\noindent
{\bf SARS-CoV-2 genome.}
The severe acute respiratory syndrome coronavirus 2, 
$\mbox{SARS-CoV-2}$, 
is the novel coronavirus responsible for the Covid-19 global
pandemic in 2019-20. Here we examine the
SARS-CoV-2 genome, available in the GenBank
database \citep{genbank:16}
as the sequence~MN908947.3 identified in \cite{wu:20}.
It consists of $n=29,903$ base pairs, 
and we translate the four-letter DNA alphabet 
to $\{0,1,2,3\}$ via the map
(A,C,G,T)\;$\mapsto(0,1,2,3).$

\begin{figure}[ht!]
\vspace{-0.05in}
\includegraphics[width=6.5in]{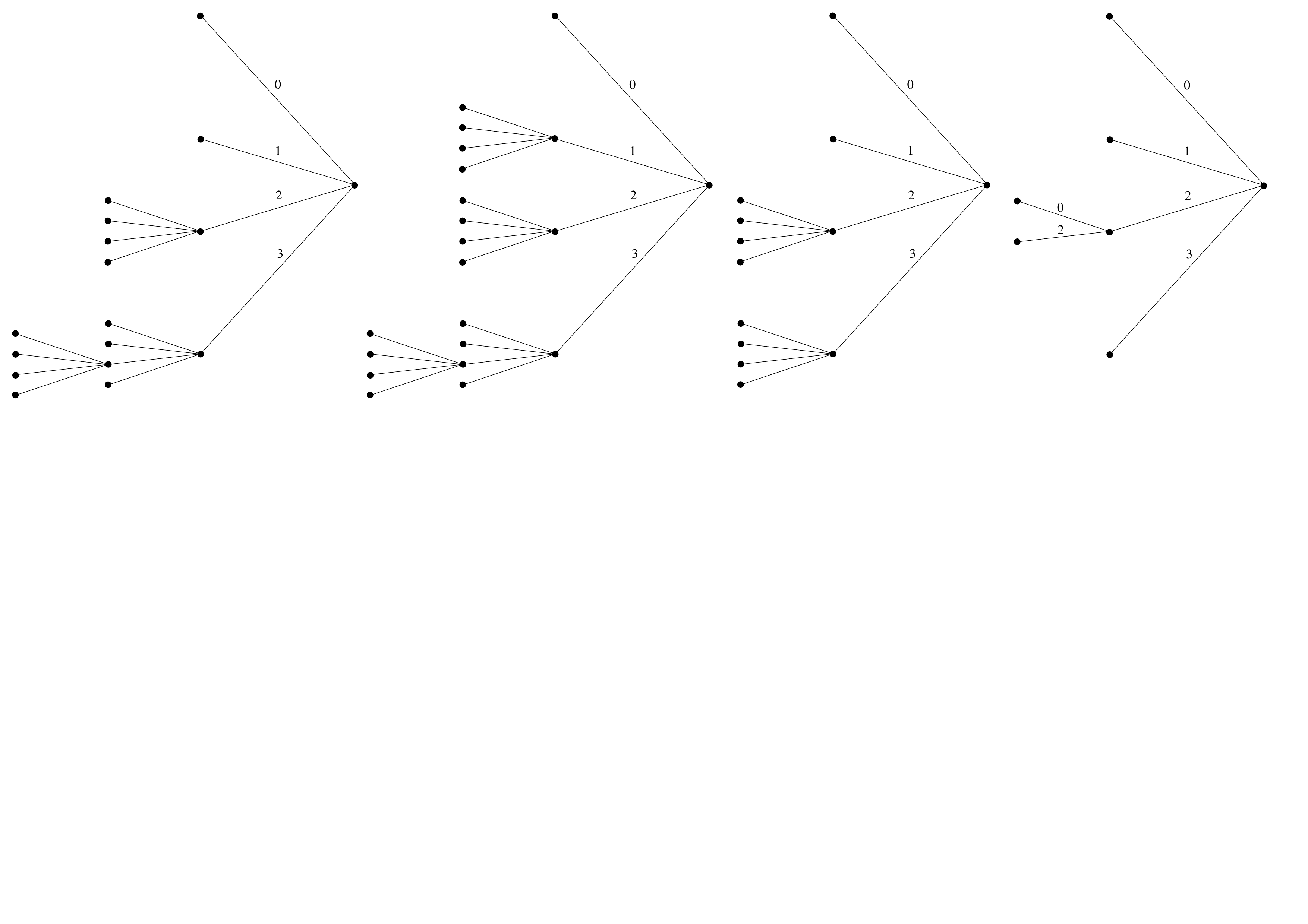}
\vspace{-2.85in}
\caption{First three trees:
The {\em a posteriori} most likely models $T_1^*,T_2^*,T_3^*$
obtained by the \BCTk~algorithm 
on the 
SARS-CoV-2 genome.
The posterior odds 
$\pi(T_1^*|x)/\pi(T_2^*|x)$ 
and $\pi(T_1^*|x)/\pi(T_3^*|x)$ are
approximately 
35.75 and 101.4, respectively.
Last tree:
The best-BIC-VLMC model for 
the SARS-CoV-2 genome; its
AIC and BIC scores are both within $0.1\%$ of the
corresponding scores of the \BCT\ model $T_1^*$.
}
\label{fig:covid}
\end{figure}

The top 3 models obtained by the $k$-\BCT\ algorithm
with $D=10$, $\beta=1-2^{-m+1}=7/8$ and $k=3$ are shown 
as the first three trees in 
Figure~\ref{fig:covid}. The MAP model
has posterior probability
$\pi(T_1^*|x)\approx 0.963$
and prior $\pi(T_1^*)\approx 4.3\times 10^{-5}$. 
The sum of the posterior probabilities 
of these three models is $\approx 0.9994$.

The VLMC model is the depth-2 subtree of $T_1^*$
shown last in Figure~\ref{fig:covid}. The optimisation
of the cut-off parameter $K$ required to find the
best-BIC model, involved a rather laborious search, 
resulting in the choice $K=22.14.$
Both MTD and MTDg give $D=1$ as the optimal
depth, corresponding to a simple first order
Markov chain.

The AIC and BIC scores
of all models generated by all three approaches
are within 0.3\% of each other.
It is interesting that the MAP model
has such high posterior probability,
and that 
\BCTk\ gives models of depth~3 with
very high confidence.
Although it is not possible to otherwise verify 
the significance of the bigger depth of the \BCT\ models
$T_i^*$
compared to those
produced by the other methods,
it may be that the \BCT\ finds 
evidence of the fact that DNA naturally 
gets encoded into triplets 
of bases to form codons that specify particular amino acids. 

\medskip

\noindent
{\bf Pewee birdsong.}
The twilight song of the wood pewee bird, studied extensively
in \cite{craig:43book}, can be described as a 
sequence consisting of an arrangement of musical phrases 
taken from an alphabet of three specific, distinct phrases.
The data in this section consist of a single contiguous song 
by a wood pewee, of length $n=1327$ phrases,
first recorded by~\cite{craig:43book}.
It was analysed by \cite{raftery-tavare:94,berchtold:01}
using MTD models,
by \cite{kharin:11} using a different MTD-type
model representation,
and by \cite{sarkar:16} using CTF models;
it is also contained in
the {\sf R} package {\sf march}.

It is well known that the pewee birdsong contains significant
variability but it is also fairly structured, with specific
repeating patterns. The most common of these,
described by \cite{saunders:44} 
as ``the commonest and most pleasing sentence,'' is the 
string $s^*=0201$, which dominates the data set,
appearing 266 times and occupying 1064 positions
or a little over 80\% of the data. 

\begin{figure}[ht!]
\vspace{-0.2in}
\hspace{0.3in}
\includegraphics[width=6.6in]{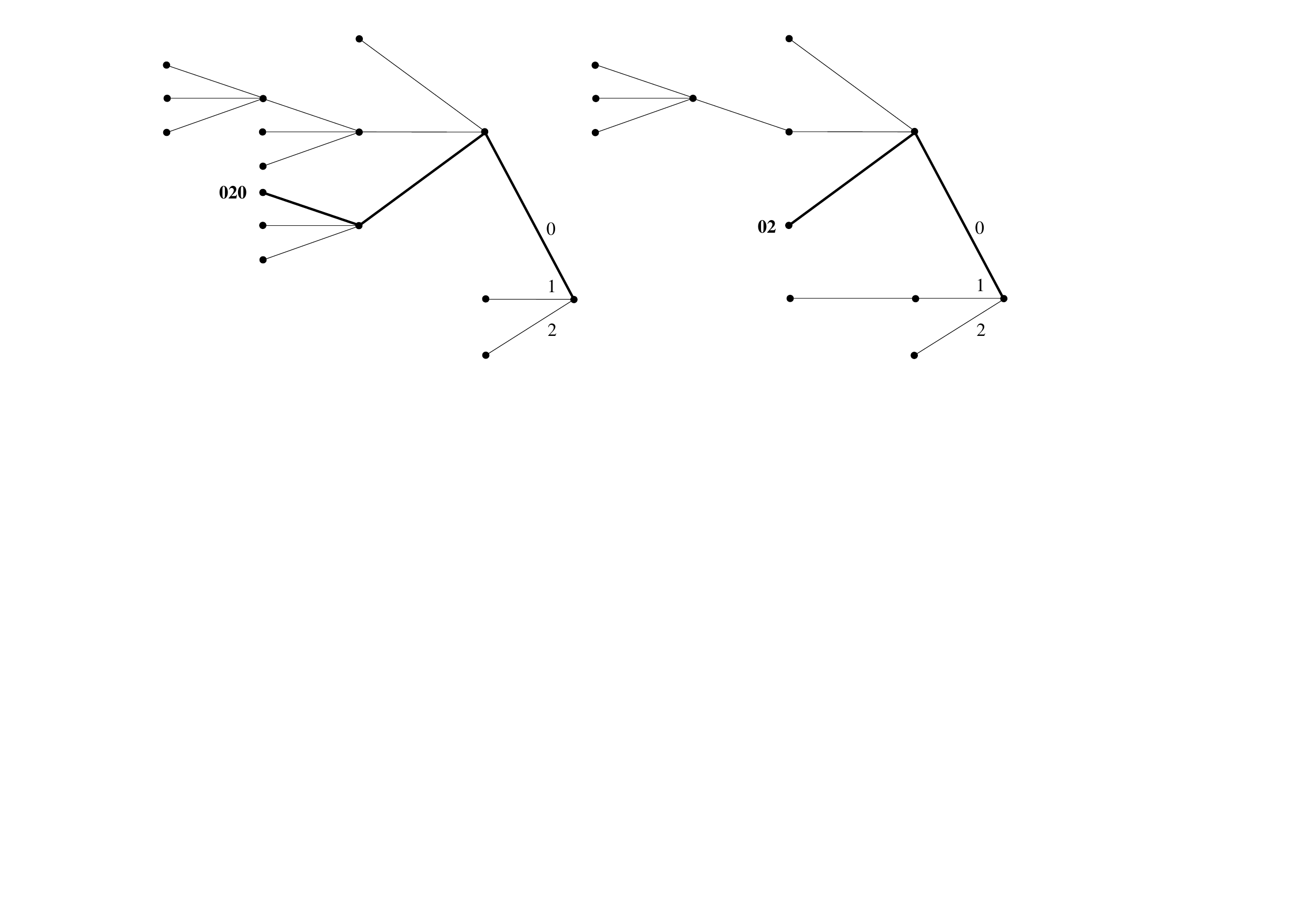}

\vspace{-2.9in}

\caption{Left: The MAP model $T_1^*$ for the pewee birdsong
data. Right:
The corresponding VLMC model.}
\label{fig:bird}
\end{figure}

The MAP tree $T_1^*$ obtained by \BCTk~with $m=3$,
$\beta=1-2^{-m+1}=3/4$, $D=10$ and $k=5$ is shown in Figure~\ref{fig:bird}.
Its prior probability is 
$\pi(T_1^*)\approx 4.1\times 10^{-5}$, and its posterior
$\pi(T_1^*|x)\approx 0.1244$.

Clearly the importance of $s^*=0201$
is captured in the MAP tree $T_1^*$,
which includes the entire prefix $020$ 
necessary to predict the appearance of $s^*=0201$.
The VLMC produces an interesting model, 
also shown in Figure~\ref{fig:bird}, which is 
quite similar to $T_1^*$.
Although it does not include the
$020$ context, it does include the
context $02$,
which contains most
of the predictive power regarding $s^*=0201$:
While $s^*$ appears 266 times in the data,
the context $s=201$ appears 267 times, 
therefore, we can ``statistically''
almost identify $s^*$ with $s$.
Compared to the VLMC model, the 
MAP tree $T_1^*$ has a marginally better AIC 
score and a slightly worse BIC score,
but it is clear that the two methods 
learn much of the same structure from the data.



\begin{figure}[ht!]
\centerline{\includegraphics[width=4.70in]{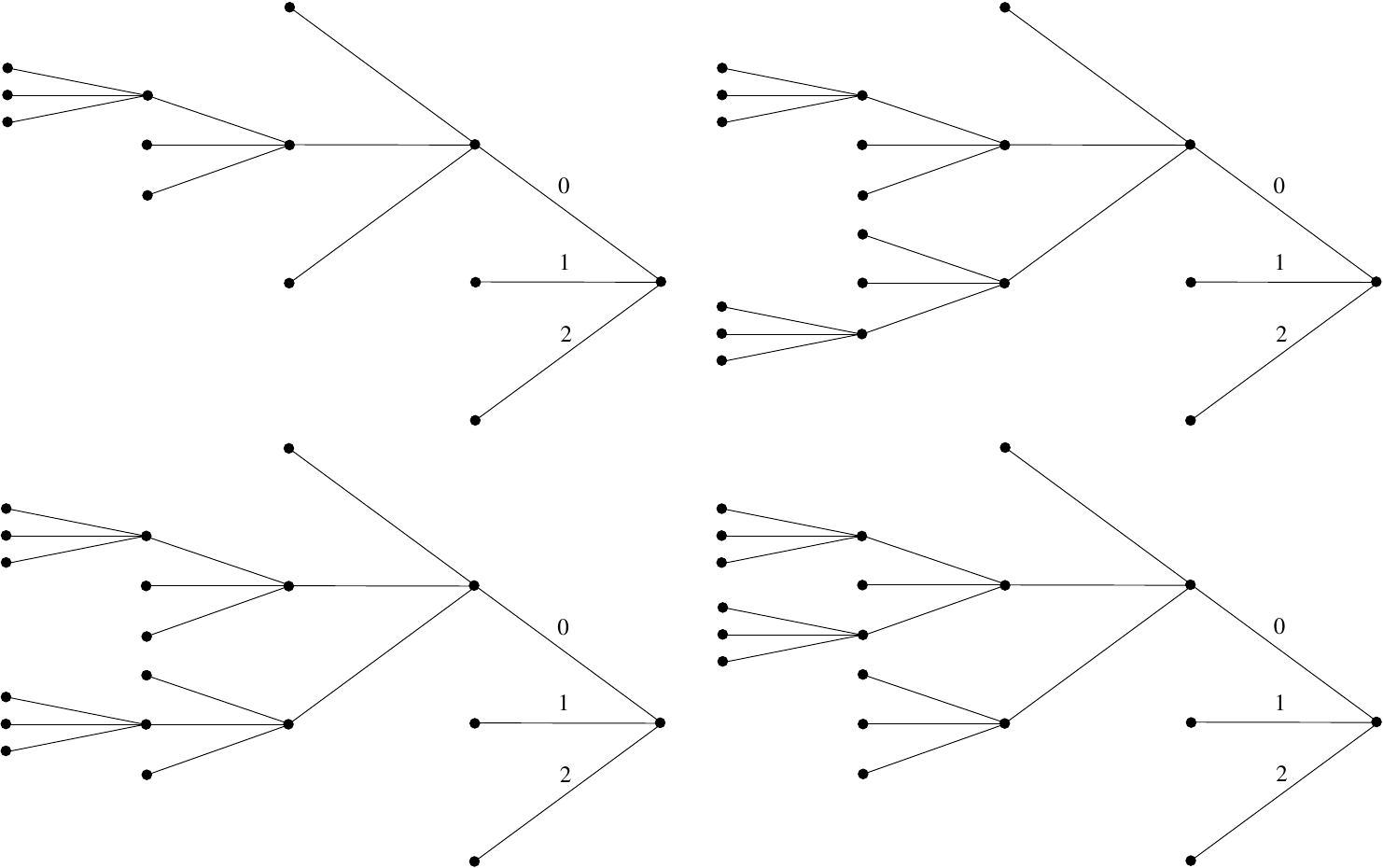}}
\caption{The four {\em a posteriori} most likely models $T_i^*$,
$i=2,3,4,5$, after $T_1^*$,
obtained by the \BCTk~algorithm from 
the pewee birdsong data.
The corresponding
posterior odds are 
$\pi(T_1^*|x)/\pi(T_2^*|x)\approx 5.727$
and
$\pi(T_1^*|x)/\pi(T_i^*|x)\approx 7.111$,
for $i=3,4,5$.
The sum of the posteriors of the top 5 models is $\approx 0.1985$.}
\label{fig:wood}
\end{figure}

The four {\em a posteriori} most likely models after $T_1^*$
are shown in Figure~\ref{fig:wood}. They are all quite similar 
to the MAP tree, each one differing from $T_1^*$ in only 
a single branch.

MTD gives $D=6$ as the optimal depth, and MTDg gives
$D=3$. The resulting MTDg model has better AIC and BIC
scores than the one from MTD, but they are still much
worse (by 39\% for AIC and 27\% for  BIC) than those
of the MAP model. This confirms the findings of 
both~\cite{raftery-tavare:94}
and~\cite{berchtold:01}, where it was 
noted that the MTD family of models is not
appropriate for this data set,
in large part due to the high significance of individual
patterns like $s^*$.

\begin{figure}[ht!]
\centerline{\includegraphics[width=5.6in]{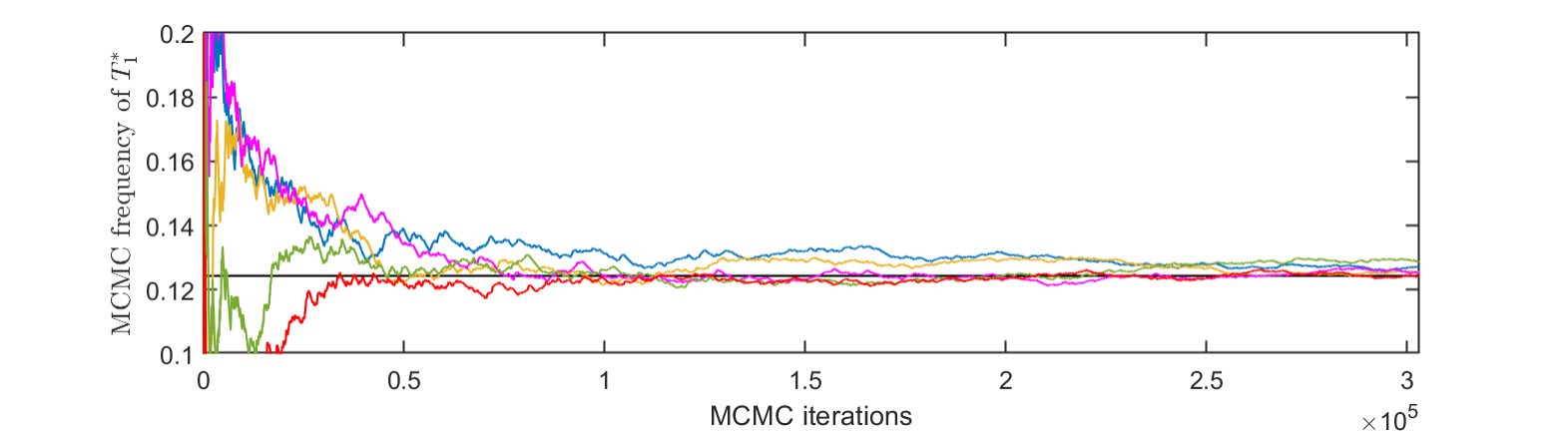}}
\caption{MCMC frequency of 
$T_1^*$.
The five graphs plotted correspond 
to five independent repetitions of the experiment
with $N=3\times 10^5$ iterations;
the horizontal line is the limiting frequency,
$\pi(T_1^*|x)$.}
\label{fig:pw_T1}
\end{figure}

The top five models here carry less
than $20\%$ of the total posterior mass.
Therefore, in order to get a better
sense of the posterior distribution on model space,
we employed the RW MCMC sampler
described in Section~\ref{s:MH}
to produce $N=10^6$ samples from $\pi(T|x)$.
The acceptance rate was 57.8\%, a total 
of $274,721$ unique trees were visited,
and the sum of their posterior probabilities
was 61.2\%. The MCMC frequency of $T_1^*$
in Figure~\ref{fig:pw_T1} indicates
that the sampler converged quite quickly.

The 100 most visited models have a total 
posterior probability of $38.3\%$.
They all have depths between 4 and 7,
with 48 of them having depth 5.
Together with the results of \BCTk\
and VLMC, this suggests that there is
significant fourth- and fifth-order
structure in the data.

\newpage

\noindent
{\bf Neural spike trains.}
Next we consider a long binary time series $x$
that consists of $n = 3,919,361$ bits, describing 
the spike train recorded from a single neuron in 
the V4 region of a monkey's brain. The recording
was made during the experiment described 
in~\cite{gregoriou:09,gregoriou:12}, while
the monkey was performing an attention task.
The recorded signal was discretised into one-millisecond
bins (with $x_i=1$ if there was a spike in
bin $i$ and $x_i=0$ otherwise), corresponding
to a trial lasting a little over 65 minutes.
As we have not been able to find implementations
of VLMC or MTD that can operate on data sets
of this length, in this section we
only present the results obtained by 
the \BCT~and \BCTk~algorithms.

\begin{figure}[ht!]
\centerline{\includegraphics[width=6.0in]{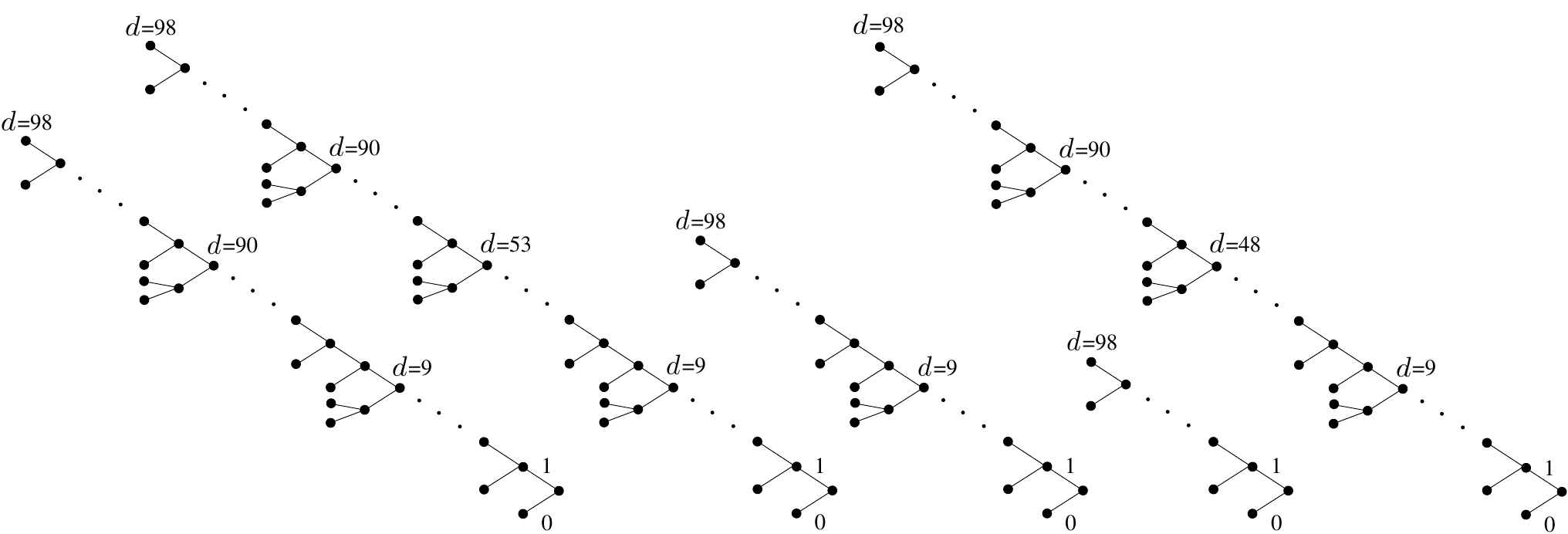}}
\caption{
The top $k=5$ models obtained by \BCTk~on
the spike train data.
The posterior 
odds 
$\pi(T_1^*|x)/\pi(T_i^*|x)$
for $i=2,3,4,5$, are
$\approx 1.2$, 
$1.43$,
$1.52$ and
$1.53$, respectively.
}
\label{fig:spikes}
\end{figure}

With $D=100$, $\beta=1-2^{-m+1}=1/2$ and $k=5$, the MAP
model $T_1^*$ is shown 
in Figure~\ref{fig:spikes}. It has depth~98
and, with the exception of two additional branches at
depths 9 and 90, it is very similar to 
a renewal model, qualitatively similar 
to the first chain in Section~\ref{s:scores2}
of the supplementary material.
Its prior probability is $\pi(T_1^*)\approx 3.1\times 10^{-61}$
and its posterior $\pi(T_1^*|x)\approx 2.1\times 10^{-8}$.
Although this probability is small, 
we note that there 
are more than $79^{10^{29}}$ possible models of depth
no more than~100, cf.~(\ref{eq:doublyE}),
and that this posterior is still larger than the corresponding
prior probability by more than 50 orders of magnitude. 
Therefore, the observations $x$ offer significant
support for this model.

The next four {\em a posteriori} most likely models 
shown in Figure~\ref{fig:spikes} are also very similar
to simple renewal models, offering a partial justification
to the biological intuition behind the 
elementary Poisson/renewal models commonly used in
neuroscience \citep{spikes:book,dayan:book},
and confirming relevant earlier findings 
\citep{neuro-isit,gao:08}.
In other words, we have learned from the data
that the statistically most significant factor
in determining whether a neuron will fire,
given its past history, is the time when it most recently 
fired before.

Although the sum of the posterior probabilities of the top
five models is less than $10^{-7}$, in this case it would
not make sense to employ an MCMC sampler to explore the posterior
further. The reason is that, given the very small values of the
posterior probabilities of the top 5 models, even with $10^6$
MCMC samples visiting $10^6$ distinct models, we would still only
visit around $1\%$ of the support of the posterior, at best.
On the other hand, increasing the value of $k$ can give a 
better idea of the shape of the posterior near its mode $T_1^*$.
With $k=50$, \BCTk~produced 50 trees, all of depth 98, and
all of them being small variations of the renewal model $T_4^*$:
All the resulting models $T_i^*$ had between 
one and five additional branches
at various depths.
The sum of their posterior probabilities is $\approx 4.1 \times 10^{-7}$. 

As a final test of the scalability of the \BCT~algorithm on a large data
set, we obtained the MAP tree with maximum depth $D=1500$. 
Interestingly, it is the same as the MAP tree for $D=100$,
and with only a slightly smaller
posterior probability of $\approx 1.6\times 10^{-8}$.

\newpage

\section{Posterior exploration and estimation}
\label{s:MCMC}

We present two examples that 
illustrate the utility of the MCMC samplers
in Section~\ref{s:MH} for posterior exploration,
and the application of the \BCT\ methodology to
parameter estimation and Markov order estimation.

\vspace{-0.05in}

\begin{example}[Daily changes in S\&P 500]
\label{ex:SP} {\em
We consider the daily changes in Standard \& Poor's index,
from January 2, 1928 until October 7, 
2016 (available 
at \texttt{\url{https://finance.yahoo.com/quote/^GSPC/}}),
quantised to $m=7$ values:
If the change between two successive
trading days, day $(i-1)$ and day $i$, is smaller than $-3\%$, 
$x_i$ is set equal to 0; 
for changes in the intervals 
$(-3\%,-2\%]$,
$(-2\%,-1\%]$, $(-1\%,+1\%]$, $(1\%,2\%]$, and $(2\%,3\%],$
$x_i$ is set equal to $1,2,3,4$ and~5, respectively;
and for changes greater than $3\%$, $x_i=6$.
%
%
%

Based on 
the resulting $n=22900$ points $x_i$,
the top $k=5$ $\mbox{\em a posteriori}$ most likely models
obtained by
the \BCTk~algorithm 
with maximum tree depth $D=260$ (corresponding
to approximately one calendar year's trading days),
are described in
Figure~\ref{fig:SPXtree}.

\begin{figure}[ht!]
\vspace{-0.05in}
\hspace{1.4in}
\includegraphics[width=4.8in]{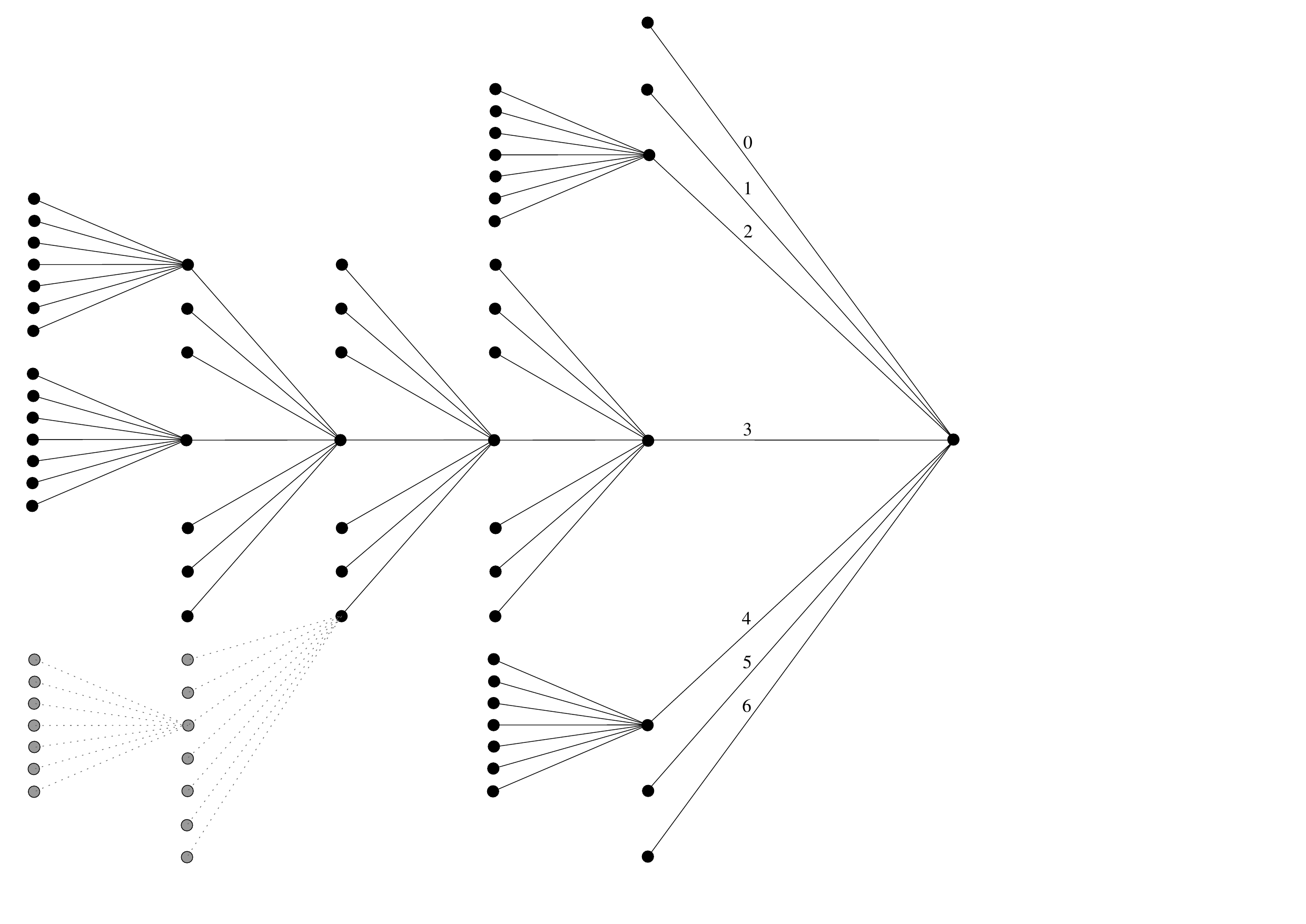}
\vspace{-0.30in}
\caption{The tree shown {\em without} the two dotted branches
is the MAP model $T_1^*$ obtained from 
$n=22900$ observations, with $D=260$. 
Its posterior probability 
$\pi(T_1^*|x)\approx 0.0174$
and its prior $\pi(T_1^*)\approx 5.7 \times 10^{-11}$.
The whole tree shown is the fifth {\em a posteriori}
most likely model $T^*_5$, and $T_2^*$, $T_3^*$ and
$T_4^*$ were found to be small variations around
$T_1^*$ and $T_5^*$, all with depth 5. The corresponding 
posterior odds $\pi(T_1^*|x)/\pi(T_i^*|x)$, for $i=2,3,4,5$,
are $1.094$, $1.367$, $1.496$ and $2.467$, respectively.}
\label{fig:SPXtree}
\end{figure}

The shape of the MAP model $T_1^*$ contains 
significant information.
Since its maximal
depth is 5, in order to determine the distribution
of the next sample we never
have to look more than five days back -- corresponding to
a week of trading days. The smaller the changes
in the most recent S\&P values, the further back we need to look
in order to predict tomorrow's value. For example,
if the difference between yesterday and today is
larger than $\pm2\%$, we need look no further than
yesterday; if it is between $1\%$ and $2\%$, we need
to look at the day before yesterday as well;
and if it is smaller than $\pm1\%$, we need
to look even further back, but no more than a week.

The sum of the posterior probabilities
of the top~$k=5$ models is less than $6.5\%$.
But with $n=22900$ data points, $m=7$ and $D=260$,
the complexity of \BCTk~becomes
prohibitive for large values of $k$.
In order to explore $\pi(T|x)$ further,
we ran the RW sampler with $T^{(0)}=T_1^*$ for $N=10^6$
iterations.
The acceptance rate was
$\approx 44.9\%$, and a total of 356531
different models were visited. 
The MCMC frequencies of the 25 most visited
models shown in Figure~\ref{fig:SPXtop25}
indicate
that the sampler has converged 
after $N=10^6$ iterations.

\begin{figure}[ht!]
\includegraphics[width=2.7in]{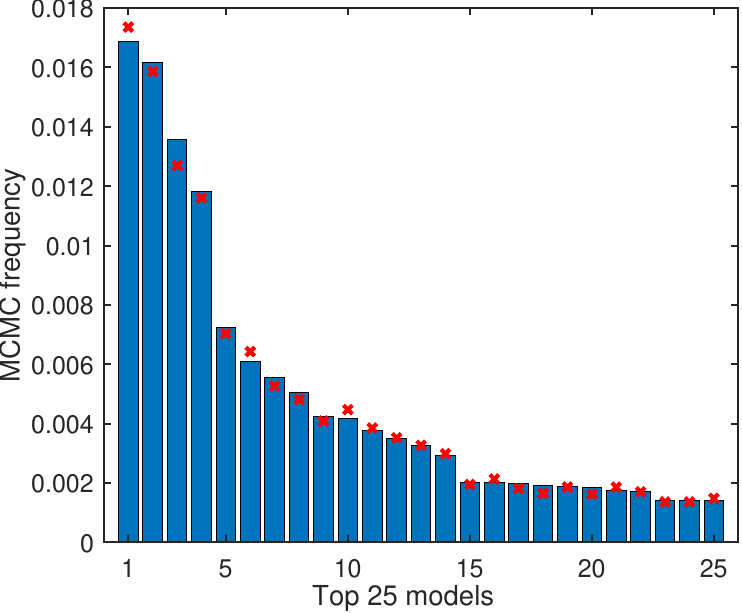}

\vspace*{-3.65in}

\hspace{2.68in}
\includegraphics[width=3.9in]{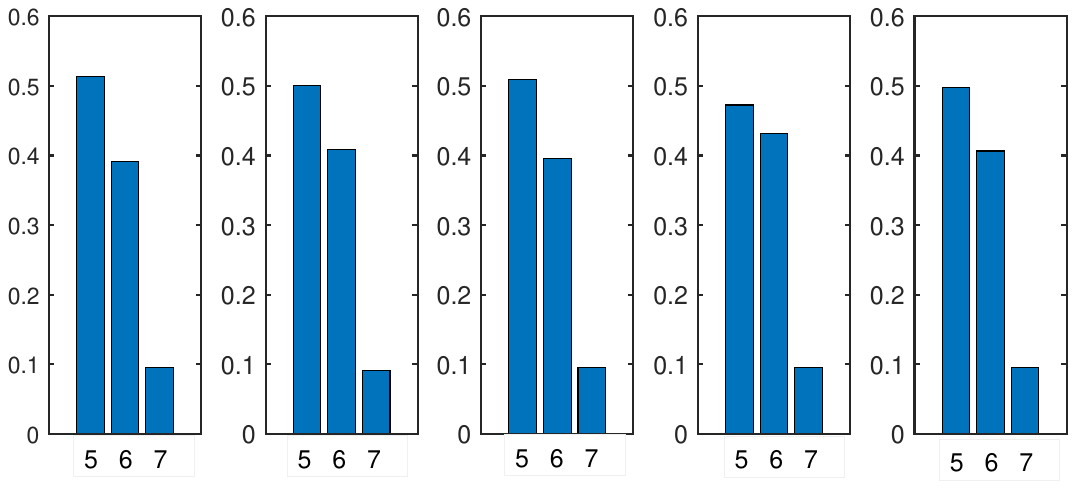}

\vspace*{-1.9in}
\caption{Left: MCMC 
empirical frequencies of the 25 most frequently 
visited models, after $N=10^6$ iterations.
The corresponding true posterior probabilities 
are marked with a red `x'.
Right: 
Markov order estimation:
The five histograms 
show the empirical frequencies of the depths 
of the 1,000 most visited models after $10^6$ 
MCMC iterations, in five independent repetitions of the 
same experiment.
}
\label{fig:SPXtop25}
\end{figure}

The MCMC output can also be used for 
Markov order estimation,
by providing an approximation to
the posterior distribution on model depth.
The empirical distributions of the model
depths obtained in five repetitions of the same experiment
are shown in~Figure~\ref{fig:SPXtop25}.
}\end{example}

\begin{example}[A bimodal posterior]
\label{ex:bimodal}
{\em
Consider a 3rd order chain $\{X_n\}$ 
on the alphabet $A=\{0,1,2,3,4,5\}$ with the property that
each $X_n$ depends on the past $(X_{n-1},X_{n-2},X_{n-3})$ 
only via $X_{n-3}$. Specifically, suppose that,
for $i,j,a,b\in A$,
$$\Pr(X_n=j|X_{n-1}=a,X_{n-2}=b,X_{n-3}=i)=Q_{ij},$$ 
where the transition matrix $Q=(Q_{ij})$ is 
given in Section~\ref{s:explicit} of the supplementary 
material.

The model of $\{X_n\}$ viewed as a variable-memory
chain is clearly the complete tree 
of depth~3, but 
the dependence of each $X_n$ on its 
past is only meaningful if it can extend at least
three time steps back: The first and second most
recent symbols are independent of $X_n$.

Therefore, it is not surprising that the MAP model
identified by the \BCTk~algorithm 
(with $k=5$, based on $n=1850$ samples) 
is simply the root, $T_1^*=\Lambda=\{\lambda\}$, 
with $T_2^*$ a complex tree of depth 3 
(with 54 leaves at depth 3) being a close second;
their posterior probabilities are
$\pi(T_1^*|x)\approx 0.3512$
and $\pi(T_1^*|x)\approx 0.2373$, respectively.
The next three {\em a posteriori} most likely 
models $T_3^*,T_4^*,T_5^*$ were found to be small
variations of $T_2^*$, also of depth 3.

%

Running the RW sampler with 
$T^{(0)}=T_1^*$, 
we found that it never visited any of the other 
top $5$ models after $10^6$ iterations,
and similarly starting at $T_2^*$
it never visited $T_1^*$. Although
$T^*_2$ can 
theoretically be reached from $T_1^*$ 
in just 12 MCMC steps, most models between them
have extremely small posterior probabilities;
e.g., the complete tree of depth one, $T_c(1)$, which 
must necessarily be visited in order to move
between $T_1^*$ and $T_2^*$,
has $\pi(T_c(1)|x)\approx 3.2\times 10^{-19}$.

In contrast, the jump sampler with 
jump
parameter $p=1/2$, 
made frequent jumps between
$T_1^*$ and $\{T_2^*,T_3^*,T_4^*,T_5^*\}$.
Starting with $T^{(0)}=T_1^*$, after 
$N=10^5$ MCMC iterations it
appears to have explored the bulk of the
posterior distribution, having
visited all of the significant
parts of its support.
The empirical frequencies of 
the top~5 models were very close to their
actual posterior probabilities 
(Figure~\ref{fig:bimod2}),
the total number
of unique models visited was 39, and
the sum of their posterior probabilities
was $\approx99.8\%$.

\begin{figure}[ht!]
\begin{center}
\includegraphics[width=5.5in]{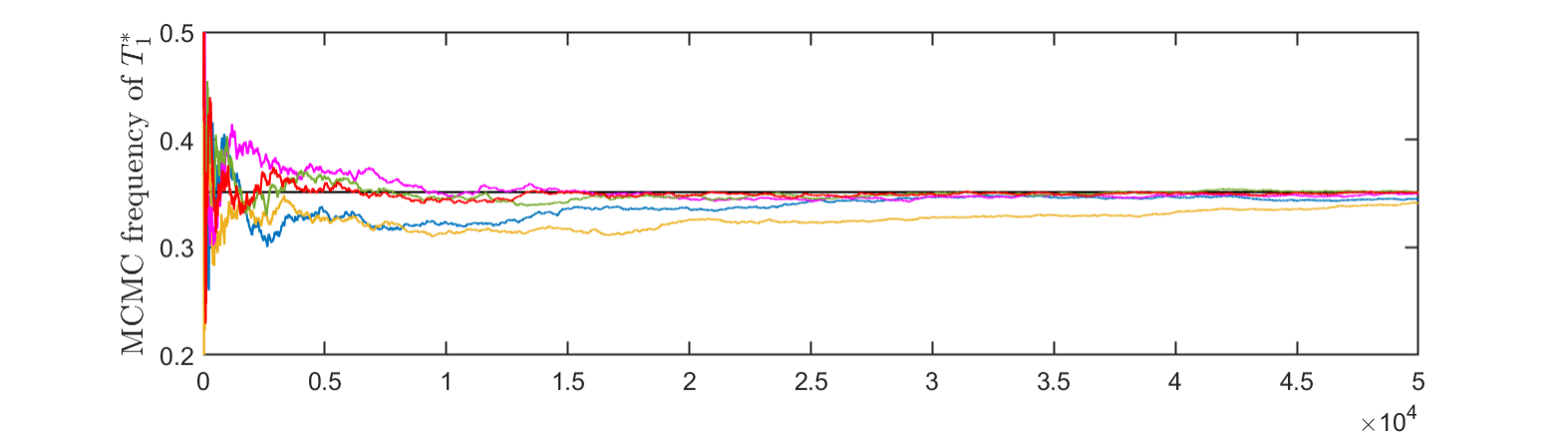}

\includegraphics[width=5.5in]{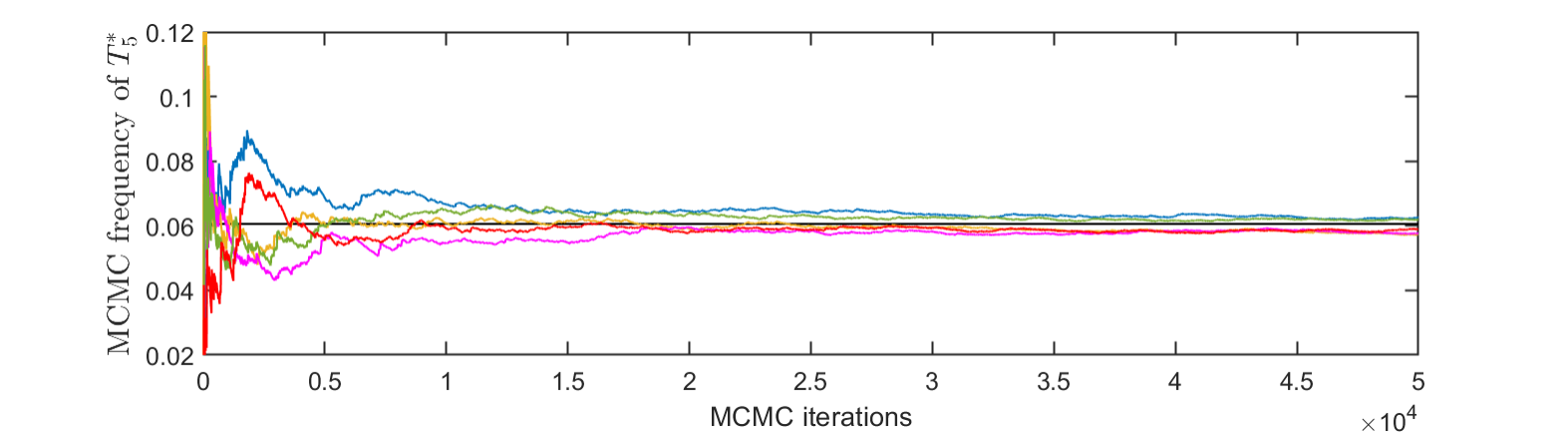}
\end{center}
\vspace*{-0.2in}
\caption{Results of the jump sampler.
The top plot shows 
the empirical frequency with which 
$T_1^*$ appears in the MCMC sample,
as a function of the number of MCMC
iterations.
The five graphs plotted correspond 
to five independent repetitions of the experiment
with $50,000$ iterations;
the horizontal line is the limiting frequency,
$\pi(T_1^*|x)$.  The bottom plot
shows corresponding results for $T_5^*$.
}
\label{fig:bimod2}
\end{figure}


In terms of Markov order estimation,
although the MAP model $T_1^*$ has depth~0, 
the mode of the posterior distribution of
the depth parameter is~3:
The $N=10^5$ MCMC samples obtained
above consist entirely of models having
depths either~0 or~3, with
corresponding MCMC frequencies
of approximately 35.13\% and 64.87\%,
respectively. 

More generally, for any statistic $F(\theta,T)$, 
we can obtain MCMC samples 
$\{(\theta^{(t)},T^{(t)})\}$ 
using the joint sampler of
Section~\ref{s:MH},
and use the values
$\{F(\theta^{(t)},T^{(t)})\}$ 
to provide an approximation to the posterior 
$\pi(F|x)$. Standard Bayesian methods
\citep{geisser:book,bernardo-smith:book,gelman:book}
can then be applied to provide
point estimates, credible sets, and other
relevant information.

For example, suppose we wish to estimate the parameter
$\theta_s(j)=\Pr(X_0=j|X_{-D}^{-1}=s),$
for the specific context $s=020$ and $j=5$.
The maximum likelihood estimate (MLE)
is
$\hat{\theta}^{\rm MLE}_s(j)=a_s(j)/M_s,$
and a commonly used Bayesian counterpart to the MLE,
$\hat{\theta}^{\rm MAP}_s(j)$,
is the mode 
of the full conditional density 
$\pi(\theta_s(j)|x,T)$ in~(\ref{eq:full-cond})
of $\theta_s(j)$ given 
the data $x$ and the
MAP model $T=T_1^*$.
Another Bayesian alternative
to the MLE is the posterior mean,
which can be approximated as,
$$\hat{\theta}^{\rm MCMC}_s(j)
:=\frac{1}{N}\sum_{t=1}^N\theta^{(t)}_s(j),$$
and an estimator
with smaller
variance can also be obtained via 
Rao-Blackwellization~\citep{gelfand-smith:90},
using~(\ref{eq:full-cond}):
$$\hat{\theta}^{\rm RB}_s(j)
:=\frac{1}{N}\sum_{t=1}^N
E\big(\theta_s(j)\big|x,T^{(t)}\big).$$
In this example, we obtain the values:

\vspace{-0.05in}

  \begin{center}
    \begin{tabular}{|c|c|c|c||c|}
		\hline
	& & & & \vspace*{-0.13in}\\
	$\displaystyle{\hat{\theta}^{\rm MLE}_s(j)}$ &
	$\hat{\theta}^{\rm MAP}_s(j)$ &
	$\hat{\theta}^{\rm MCMC}_s(j)$ &
	$\hat{\theta}^{\rm RB}_s(j)$ &
	true $\theta_s(j)$\\
		\hline
	0.1290 & 0.1871 & 0.1512 & 0.1516 & 0.15 \\
     		\hline
     \end{tabular}
  \end{center}


\noindent
The estimates
$\hat{\theta}^{\rm MCMC}_s(j)$ and
$\hat{\theta}^{\rm RB}_s(j)$ 
are based on $N=10^5$ MCMC samples
obtained by the jump version of
the joint sampler,
with jump
parameter $p=1/2$.
From the same samples
we get an estimate of the posterior
distribution of $\theta_{020}(5)$,
shown in Figure~\ref{fig:thetaH}.

\begin{figure}[ht!]
\includegraphics[width=2.8in]{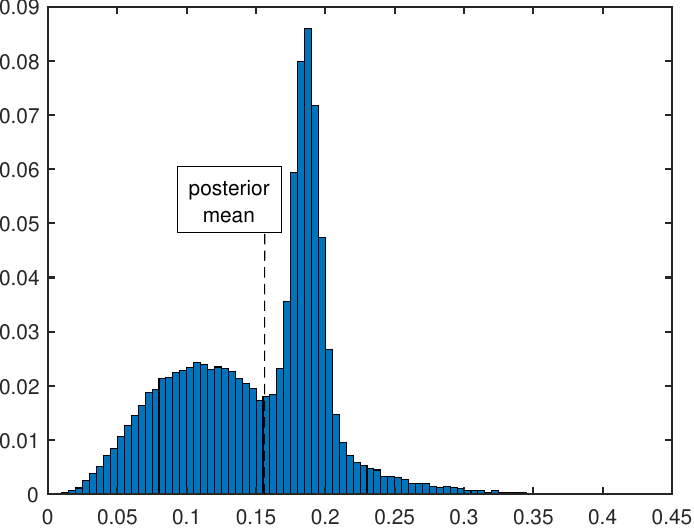}

\vspace{-2.1in}

\hspace{2.8in}
\includegraphics[width=3.55in]{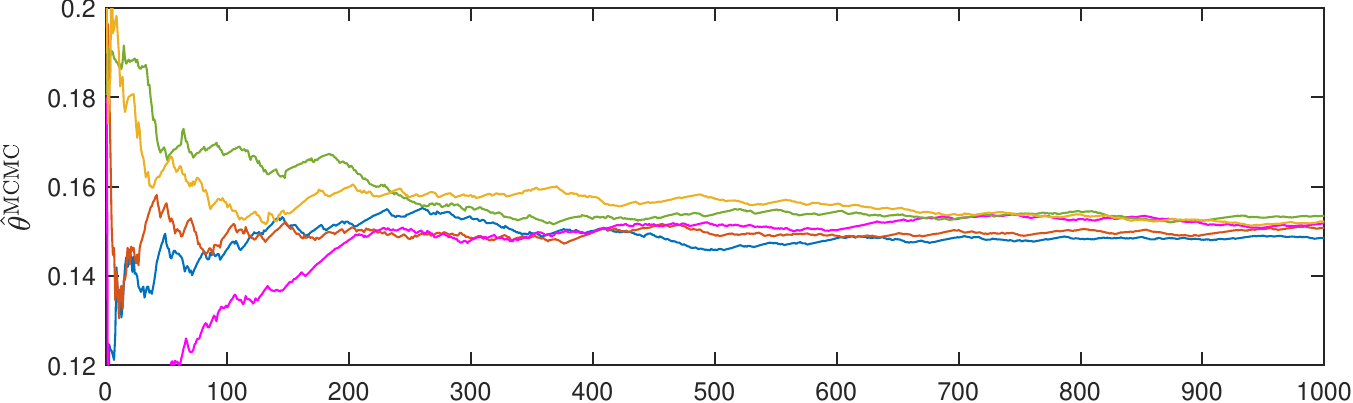}

\hspace{2.8in}
\includegraphics[width=3.55in]{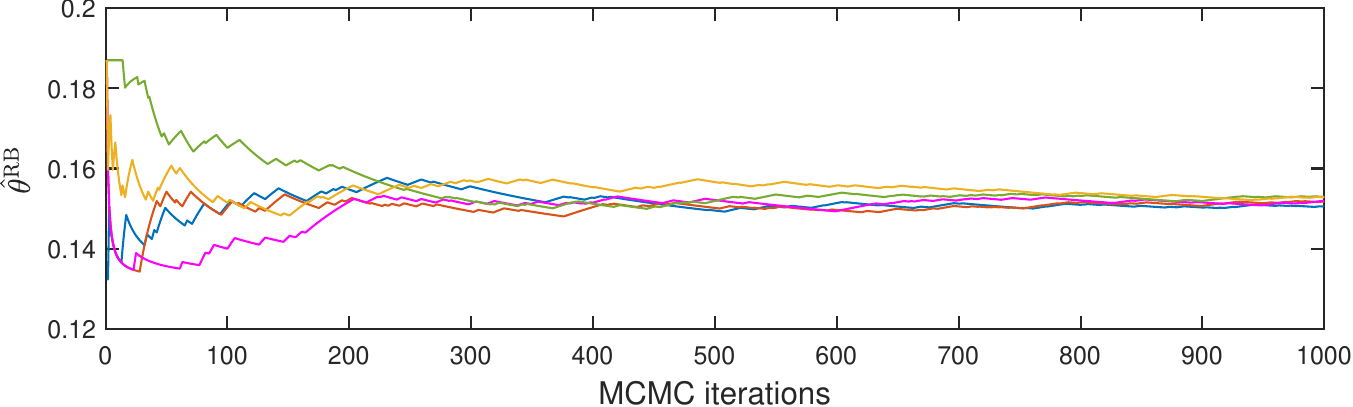}
\caption{Left: Histogram of the $N=10^5$ MCMC 
samples $\theta^{(t)}_s(j)$,
with $D=3$, $\beta=0.95$, $s=020$ and $j=5$.
Right:
Corresponding MCMC estimates.
The top plot shows 
the ergodic averages
$\hat{\theta}^{\rm MCMC}_s(j)$ 
as a function of the number of MCMC
iterations.
The five graphs plotted correspond 
to five independent repetitions
with only $N=1,000$ MCMC iterations each.
The bottom plot
shows corresponding results for
$\hat{\theta}^{\rm RB}_s(j)$.
}
\label{fig:thetaH}
\end{figure}

Note that the values of the four estimates above
are quite different, which is 
common, especially for
small data sizes $n$. On the other hand, the fact that the
Bayesian estimates are more accurate
than the MLE is by no
means universal. 

Finally, in Figure~\ref{fig:thetaH}
we also show the results of five different MCMC experiments,
indicating that the ergodic averages in the estimators
$\hat{\theta}^{\rm MCMC}_s(j)$ and
$\hat{\theta}^{\rm RB}_s(j)$ converge
quickly,
and that the variance of 
$\hat{\theta}^{\rm RB}_s(j)$ appears to be smaller,
as expected.
}\end{example}

\newpage

\section{Prediction}
\label{s:prediction}

Given
a {\em training sequence} $\mbox{$x_1^t=(x_1,x_2,\ldots,x_t)$}$
of length $t$
from a discrete time series $x$ with values in the alphabet
$A=\{0,1,\ldots,m-1\}$, 
we wish to sequentially predict the next $n$ values of the
{\em test sequence}\, $x_{t+1}^{t+n} =(x_{t+1},x_{t+2},\ldots,x_{t+n})$.
At each step $i=1,2,\ldots,n$,
given the past samples $x_1^{t+i-1}$,
the prediction of the next sample
$x_{t+i}$ is expressed as a conditional distribution
$\Qhati(z|x_1^{t+i-1})$, $z\in A$,
and the performance of a predictor
$\clQ=\{\Qhati\;;\;i=1,2,\ldots, n\}$ 
is evaluated by 
the normalised, cumulative log-loss.

In the remainder of this section we describe the four 
prediction methods that
will be used below, 
together with the natural predictor induced by the
\BCT\ framework discussed
in Section~\ref{s:sequential}.
In Sections~\ref{s:PredBasic} and~\ref{s:PredReal} we report 
the results obtained by all five methods on simulated 
and real data sets,
and we discuss them in Section~\ref{s:Pconclusions}.

A common approach to prediction is to
first use the training data to learn a model and 
associated parameters, and then use the conditional 
distributions of this model as predictors.
Indeed, this is 
exactly the form of the predictors proposed by all  
methods, other than \BCT,
described below. 

\medskip

\noindent
{\bf Bayesian context trees.}
As outlined in Section~\ref{s:sequential},
the canonical predictor within the \BCT\ framework 
is the one based on the posterior predictive 
distribution, $\hat{Q}_i(x_{t+i}|x_1^{t+i-1})=
P^*_D(x_{i+1}|x_1^i)$, cf.~(\ref{eq:defn}),
which can easily be computed sequentially
via the \CTW\ algorithm. 
By averaging over all models, $P^*_D(x_{i+1}|x_1^i)$
incorporates the model uncertainty
and it
avoids the problem of model selection 
by replacing it with model averaging.

When the data $x$ are generated by a
stationary, irreducible variable-memory
chain, in the special case $\beta=1/2$
it is shown by \cite{weissman-et-al:di}
that the \BCT\ predictor is consistent
with probability one, asymptotically in the size of the
test data, for any finite training sequence.
Moreover, as the following discussion indicates,
the \BCT\ predictor essentially achieves 
the optimal minimax regret with respect to log-loss.
Theorem~\ref{thm:redundancy} gives a 
nonasymptotic lower bound, for the special 
case of binary chains and $\beta=1/2$; it was 
established in~\cite{ctw-4:93, willems-shtarkov-tjalkens:95}.

\begin{theorem} 
\label{thm:redundancy}
For any binary, variable-memory chain
model $T\in\clT(D)$ 
and associated parameters $\theta=\{\theta_s\;;\;s\in T\}$, 
for any data sequence $x_1^n$
of arbitrary length $n$,
and any initial context $x_{-D+1}^0$,
the prior predictive likelihood for 
$\beta=1/2$ satisfies,
\be
\log P_D^*(x_1^n|x_{-D+1}^0)
\geq
\log P(x_1^n|x_{-D+1}^0,\theta,T)
	-
	\frac{|T|}{2}
	\log n + C,
\label{eq:LB}
\ee
with an explicit constant $C=C(T)$,
independent of $n$ and $\theta$.
\end{theorem}

The lower bound~(\ref{eq:LB}) is asymptotically
tight up to and including the $\log n$ term,
both in expectation and for individual sequences: Corresponding
upper bounds follow from the
fundamental ``converse'' theorem by
\cite{rissanen:84,rissanen:86b},
and from the general results by \cite{weinbergeretal:94}.
These results clearly indicate
that the \BCT\ predictor indeed essentially achieves 
the optimal minimax regret
\citep{barron:14}.

\medskip

\noindent
{\bf Variable length Markov chains.}
For prediction using variable-length Markov chains
we use the prediction function in the {\sf R} 
package {\sf VLMC}.
This selects a VLMC model
based on the training data, and then predicts
future samples by the maximum likelihood parameters
associated with this model.
Here we only show results 
for the best-BIC-VLMC model, as it 
was found to be the most effective choice in practice;
see also the relevant comments in Section~\ref{s:simulated}.
When the data $x$ are generated by a stationary
and ergodic variable-memory chain, the results
of~\cite{buhlmann:99} imply that the VLMC predictor 
is asymptotically consistent,
but for consistency it is the size
of the training data that needs to grow to infinity.
Further methodology
in connection with VLMC prediction 
is developed in~\cite{buhlmann:00}.

\medskip

\noindent
{\bf Mixture transition distribution.}
The MTD approach to
prediction is again based on first
selecting a model and then performing
prediction using an approximation to the 
maximum likelihood parameters associated with that model.
Among the four MTD versions discussed
in Section~\ref{s:compare},
we only report results obtained by the best-BIC-MTDg,
since the
performance of the other three methods
was either identical or inferior.

\medskip

\noindent
{\bf Sparse Markov chains.}
As described in the 
Introduction, SMCs are a generalisation of the 
\BCT\ models in $\clT(D)$: Each SMC model consists
of a partition of the set of all contexts of depth $D$
into different states, 
such that all contexts in the same state
induce the same
conditional distribution on the next symbol
of the underlying chain. 
In~\cite{jaaskinen:14},
a class of priors for SMCs
are defined, and 
algorithms for selecting approximate MAP
versions of models and parameters are
developed in the subsequent work~\cite{xiong:16}.
In our experiments, prediction is performed
based on the resulting model and
parameters obtained using
the code publicly available
at \texttt{\url{https://www.helsinki.fi/bsg/filer/SMCD.zip}}.


\medskip

\noindent
{\bf Conditional tensor factorisation.}
The CTF models of \cite{sarkar:16},
described in the Introduction,
use conditional tensor factorisation to 
represent the full transition 
probability tensor of a higher-order chain.
In our experiments 
we use the CTF software publicly available 
at \texttt{\url{https://github.com/david-dunson/bnphomc}}.
Prediction is again performed in two steps.
First a model is selected via the results of an 
MCMC sampler on model space, and appropriate
parameters are chosen by an approximation 
to their posterior mean via Gibbs sampling.
The induced predictive distributions are
then obtained from the selected model and
parameters. 

\medskip

Throughout our experiments, we take the maximal 
depth to be $D=10$ for \BCT, MTD, SMC and CTF.
Following standard practice \citep{begleiter:04}
in most cases we split the data
50-50 into a training set and a test set.
One difficulty that occasionally arises with the VLMC and MTD 
predictors is that
they use maximum likelihood parameter estimates
for their models, which 
in some cases means that they assign zero conditional
probability to certain symbols, and which in turn results
in poor performance and an infinite log-loss. 



Finally we note that different versions of the \CTW\ algorithm
have been used for prediction in earlier work,
including \cite{ron-et-al:96,begleiter:04,dimitrakakis:2010}.


\subsection{Simulated data}
\label{s:PredBasic}

\begin{figure}[ht!]
\centerline{
\includegraphics[width=3.2in]{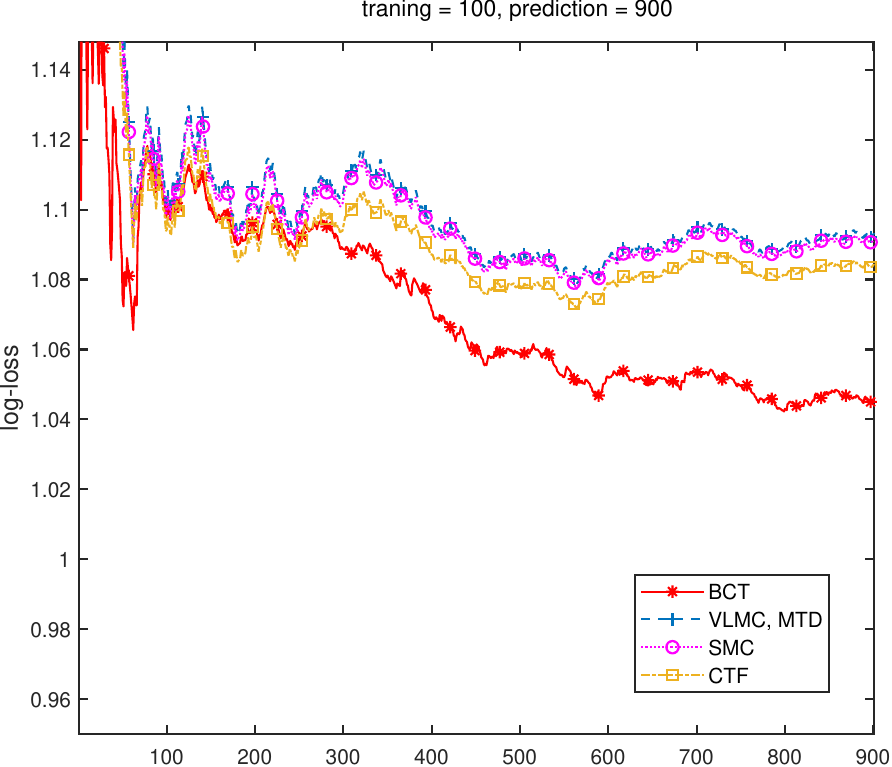}
\includegraphics[width=3.2in]{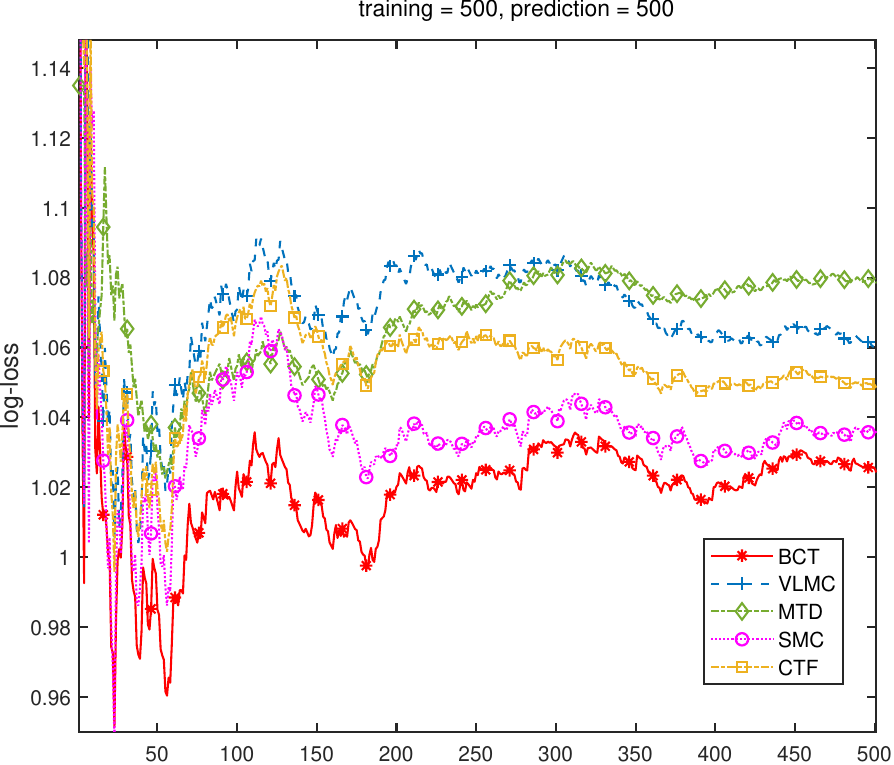}
}
\vspace{0.05in}
\centerline{
\includegraphics[width=3.2in]{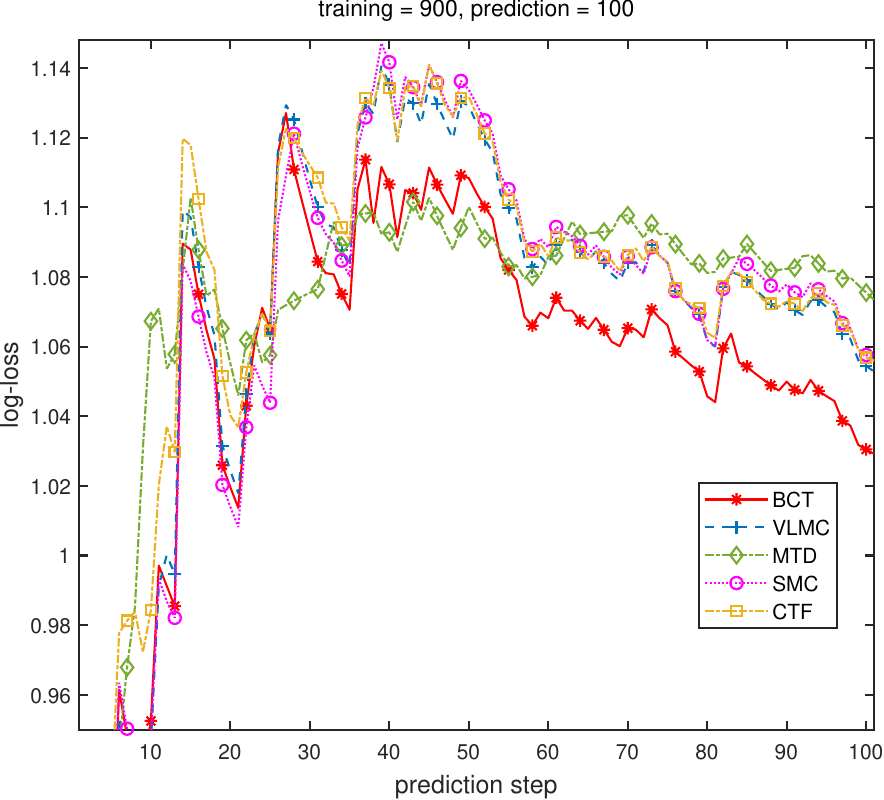}
\includegraphics[width=3.2in]{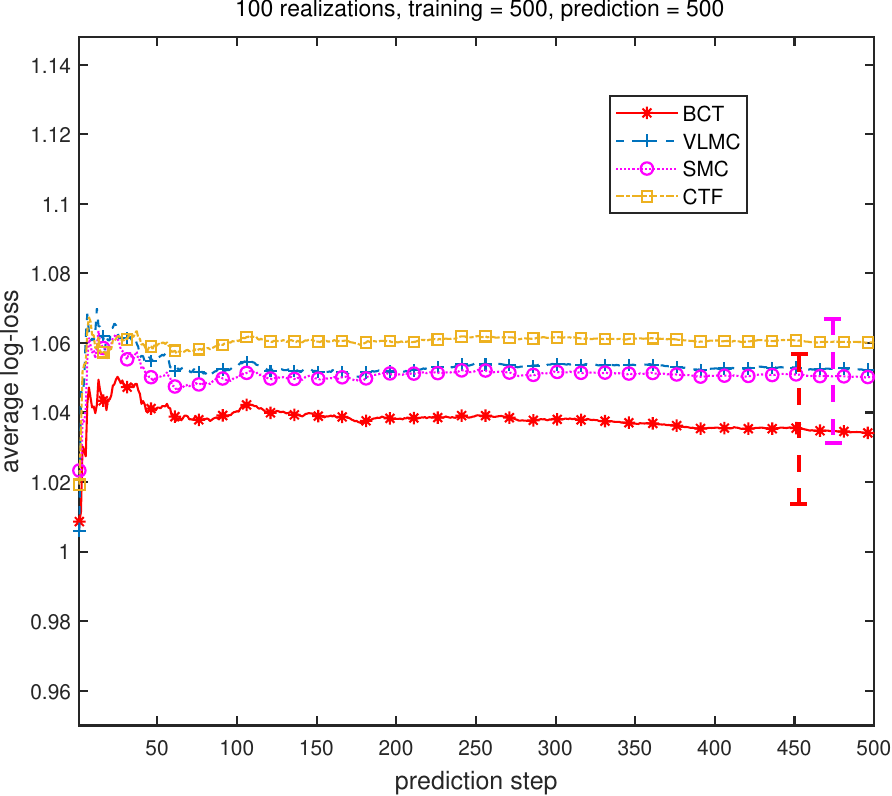}
}
\vspace{-0.05in}
\caption{Prediction results on simulated data.
First three plots: 
Log-loss achieved by each of the five methods, as a
function of the size of the test data.
Bottom right: Log-loss achieved by \BCT, VLMC, SMC and CTF 
on $n=500$ test samples with $t=500$ training samples,
averaged over 100 independent repetitions of the same 
experiment. One-standard-deviation error bars 
are also plotted for \BCT\ and SMC near the end
of the test data. The vertical (log-loss) axis has
the same range in all four plots.}
\label{fig:spred1}
\end{figure}

The experiments here are based on 1,000 
samples simulated
from 
the 5th order chain with alphabet size $m=3$
in Section~\ref{s:simulated}.
Figure~\ref{fig:spred1}
shows the log-loss achieved by all five
predictors 
as a function of the number of predicted samples
in three different cases.

The results in the first plot are
based on $t=100$ training samples and 
$n=900$ test samples. The
\BCT\ is seen to perform consistently 
better than the other four methods,
and the difference between 
\BCT\ and the second best method, CTF,
increases as more data get predicted,
reaching a significant $\approx 3.7\%$ by the end.
This is in part due to the fact that
the \BCT\ predictor continues to get updated
past the training stage, whereas the models
employed by the other four methods based
on only $100$ training samples are too 
simple to be effective.
SMC, VLMC and MTD all produce the empty model
corresponding to i.i.d.\ data, with
VLMC and MTD also having the same parameters,
inducing the exact same predictors (hence
shown as a single graph).
CTF, on the other hand, uses a first order Markov model,
leading to slightly better performance. 

The second plot shows the log-loss obtained with
$t=500$ training samples and $n=500$ test samples.
The results are similar, however, 
the difference between \BCT\ and the other methods is 
now smaller.
Since the training set is larger,
the other four methods have a better chance to capture
more of the structure present in the data.
The VLMC model is a simple tree of depth two,
and the CTF model also corresponds to a second order
chain. MTD again produces an i.i.d.\ model,
and SMC, which is the closest to \BCT,
with a log-loss difference of $\approx 2.5\%$,
identifies a second order model with five states.

With $t=900$ training and $n=100$ test samples, 
\BCT\ again appears to have the best performance,
although there are larger fluctuations due to the smaller
test data size.
The results 
of SMC, VLMC and CTF are almost identical,
and significantly better than MTD.
SMC uses a second order model with 4 states,
the CTF model also has order~2, MTD selects
the i.i.d.\ model, and VLMC, which is
closest to \BCT\ by the end (by a 2.2\% difference),
produces a tree model of depth 3 with 5 leaves.

\newpage

Finally, we examine 
the log-loss of these predictors averaged over 
100 independent realisations simulated from this 
chain, each time with a 50-50 split between 
training and test data. The last plot in
Figure~\ref{fig:spred1} 
shows the results obtained by all methods 
except MTD, which was based
on an i.i.d.\ model every time.
The average performance of \BCT\ 
is better
than that of the other methods;
after the first 10 samples or so, the log-loss of the
\BCT\ stays consistently approximately 1.6\% lower than
that of the second best method, SMC.

\subsection{Real data}
\label{s:PredReal}

\noindent
{\bf SARS-CoV-2 gene.}
Here we examine the spike (S) gene, in positions
21,563--25,384 of the SARS-CoV-2 
genome \citep{wu:20}
described in Section~\ref{s:realdata}.
The importance of this gene is that it codes for the surface 
glycoprotein whose function was identified in~\cite{yan:20,lan:20}
as critical, in that it binds onto the Angiotensin Converting 
Enzyme 2 (ACE2) receptor on human epithelial cells,
giving the virus access to the cell and thus facilitating the 
Covid-19 disease. 

\begin{figure}[ht!]
\centerline{
\includegraphics[width=3.2in]{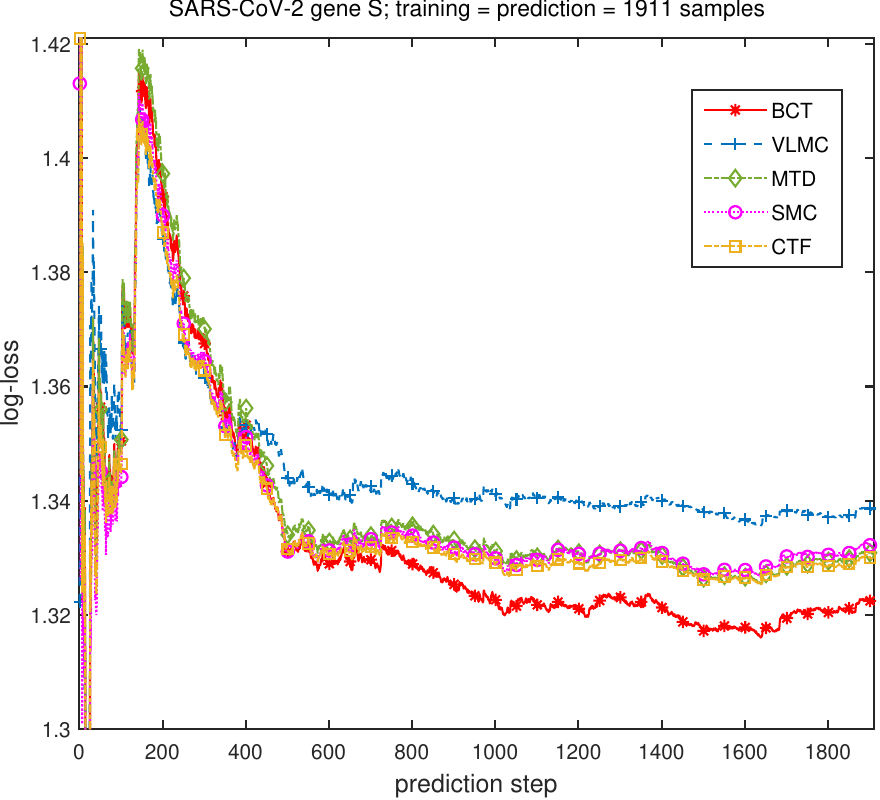}
\includegraphics[width=3.2in]{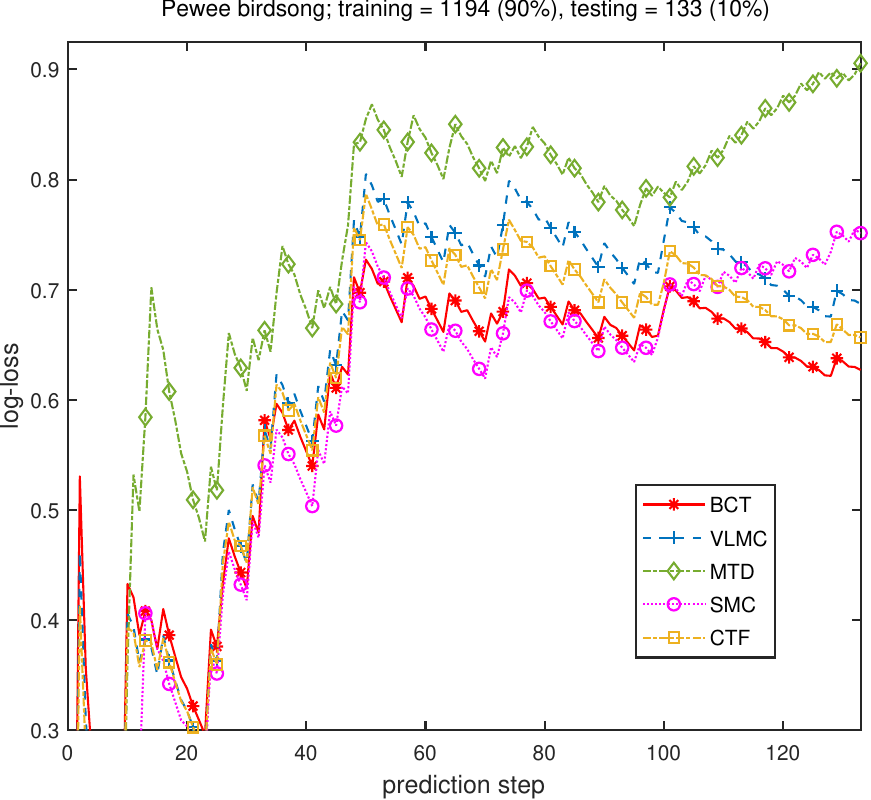}
}
\caption{Prediction results on real data. Left:
SARS-CoV-2 gene (S),
with $t=1911$ training
samples and $n=1911$ test samples.
Right:
Pewee birdsong data, with $t=1194$ training
samples and $n=133$ test samples,
corresponding to a 90\%-training/10\%-testing split.
}
\label{fig:Preal}
\end{figure}

The data, consisting of a
3,822 bp-long gene sequence,
was again split 50-50
into a training set and test set.
Figure~\ref{fig:Preal} shows the prediction 
results obtained
by all five methods.
\BCT\ consistently achieves the smallest log-loss.
On the training data,
the MAP tree produced by \BCT\ corresponds to the
full first order Markov model
with posterior probability of $\approx 98\%$, 
while on the entire data set the MAP tree has depth~2.
The MAP model $T_1^*$ now has 
$\pi(T_1^*|x)\approx 49.5\%$, while the first order 
chain which appears as $T_2^*$ 
has $\pi(T_2^*|x)\approx48\%$. 
This ability of \BCT\ to perform sequential updates 
and adaptive model averaging explains, in part,
its superior performance.

The second best method, CTF, uses a first order
Markov model, as does MTD, which achieves
near-identical performance.
SMC uses a first order model that 
groups $\{G,T\}$ into a single state,
and VLMC similarly produces a tree of depth 1
that groups $\{A,G,T\}$ into a single state.

\medskip


\noindent
{\bf Pewee birdsong.}
The data,
discussed in Section~\ref{s:realdata},
consists of 1,327 samples in a three-letter
alphabet.
When half of it was used for training and the other half
for testing,
VLMC and MTD both had an infinite log-loss after the 420th 
symbol in the test sequence; this
was mentioned earlier as a possible
drawback of methods that use maximum likelihood
estimates for the parameters.
Among the methods that give a finite log-loss,
\BCT\ was found to have the best performance.
SMC produced a second order model with five states,
and by the end of the test data its log-loss
was larger than that of the \BCT\
by 13.2\%. CTF used a fourth order model for
prediction, resulting in a log-loss larger
than that of \BCT\ by 1.2\%.

In order to illustrate 
the relative performance of all five
methods, in Figure~\ref{fig:Preal}
we report the prediction results on the last 
10\% of the samples, after the first 90\%
have been used for training.
The \BCT\ has smallest log-loss overall,
approximately 4.8\% lower than that
of the second best method, CTF, at the end of the
experiment.
The CTF and MTD predictors both use a fourth order model,
SMC finds a second order model with 5 states,
and VLMC uses a simple tree model of depth 4.

\subsection{Discussion}
\label{s:Pconclusions}

The \BCT\ predictor was found to have
consistently better performance than the other four methods considered,
achieving a log-loss that was between 1.2\% and 4.8\% better than
that of the second best method in each case.
This is due, in part, to the fact that 
the \BCT\ predictor 
is based on averaging over all models
and parameters with respect to their posterior distribution,
and that it allows for sequential updates beyond the
training stage.

The method that performed the closest to \BCT\ in most 
cases was CTF, which usually identified the same Markov order 
as the other methods. The VLMC and MTD predictors were 
found to be consistently and significantly less effective 
than \BCT, and as noted earlier they sometimes assigned
zero probability to the occurrence of certain symbols
in the data, resulting in an infinite log-loss.
Another difficulty with CTF, VLMC and MTD is that
they required much more computational effort
in the training stage.

The SMC predictor was found in most cases to have performance similar to VLMC.
As the context tree models in $\clT(D)$ are a subset of
all SMCs, the fact that \BCT\ performs better 
than SMC highlights the 
power of the full Bayesian
\BCT\ framework.
In contrast, SMC uses a single
(and often poor) estimate of its MAP model,
leading to much less satisfactory results.

\newpage

\section{Concluding remarks}
\label{s:discussion}

The Bayesian context trees framework introduced in this work
was found to be effective for several core statistical tasks
in the analysis of discrete time series. Variable-memory
Markov chains are a rich class of models that 
capture important aspects of the higher-order temporal structure
naturally present in many types of real-world data, and the associated
exact inference algorithms in Sections~\ref{s:MMLA}--\ref{s:kMAPT}
provide efficient tools for modelling, estimation 
and prediction. These algorithms, along with the
variable-dimensional MCMC samplers of Section~\ref{s:MH},
also facilitate accurate posterior computations
that offer a quantitative measure of uncertainty
for the estimated models and parameters. 
The resulting methods were found to 
outperform several of the most commonly used
approaches on both simulated and real-world data.

The exact nature of the algorithms in 
Sections~\ref{s:MMLA}--\ref{s:kMAPT},
particularly the ability to compute the prior predictive
likelihood, opens the door to numerous other applications
that are the object of ongoing work,
including anomaly detection, segmentation, and change-point detection.
Similarly, the 
low complexity 
of the
algorithms 
opens the door 
to their effective use in problems involving `big data'.
As has been noted in before connection with VLMCs, the main 
practical limitation in the 
application of these tools is when the alphabet
size is large.

While our focus has been exclusively on discrete-valued
time series,
the ideas and techniques introduced
in this work can be generalised to 
real- or vector-valued observations.
Important directions of ongoing 
relevant research include the extension 
of the exact inference algorithms 
to various types of tree-structured
autoregressive models,
and the development of corresponding tools
for inference with hidden 
Bayesian context trees.

\newpage

{\small
\section*{Acknowledgments}

\addcontentsline{toc}{section}{Acknowledgments}

We gratefully acknowledge interesting
conversations with 
Petros Dellaportas,
Marina Doufexis,
Chris Gioran, 
Carl Rasmussen, 
Stephen Souras,
and
Frans Willems.
We are also grateful to 
Georgia Gregoriou for providing us 
with the spike train data examined
in Section~\ref{s:realdata},
and to the Associate Editor and
anonymous reviewers for numerous
helpful suggestions.
}


\renewcommand{\baselinestretch}{0} 

{\small
\bibliographystyle{chicago}
\addcontentsline{toc}{section}{References}

\def\cprime{$'$}

}


\newpage

\renewcommand{\appendixname}{Supplementary Material}
\renewcommand{\appendixtocname}{Supplementary Material}
\renewcommand{\appendixpagename}{Supplementary Material}
\addappheadtotoc

\begin{appendices}

\appendixpage

\section{Proofs of Lemma~\ref{lem:prior} and Theorems~\ref{thm:MMLA}
and~\ref{thm:MAPT}}
\label{s:proofs}

Throughout this section, for the sake of clarity of notation,
we simply write $\pi_D(T)$ for the model prior
$\pi_D(T;\beta)$, without explicit reference to $\beta$.

\medskip

\noindent
{\em Proof of Lemma~\ref{lem:prior}.}
The proof is by induction. Note first
that for $D=0,1$ the result is trivial
to check. Also observe that we can write
any tree $T$ which does 
not contain only the root node~$\lambda$, 
as the union $T=\cup_j T_j$ of a collection 
of $m$ subtrees $T_0,T_1,\ldots,T_{m-1}$.
Clearly we will then have,
\be
|T|=\sum_j|T_j|,\;\;\mbox{and}
\;\;L_D(T)=\sum_j L_{D-1} (T_j).
\label{eq:Tunion}
\ee

For the inductive step,
suppose that the result holds for
all depths less than or equal to 
some $d\leq D-1$. We will show that
it holds for $d+1$ as well.
Let $\Lambda=\{\lambda\}$ denote the tree
that consists only of the root node
$\lambda$. Using~(\ref{eq:Tunion}),
we have,
\be
\sum_{T\in\clT(d+1)} \pi_{d+1}(T) 
&=&\pi_{d+1}(\Lambda)+\sum_{T\in\clT(d+1),T\neq\Lambda}
	\alpha^{|T|-1}
	\beta^{|T|-L_{d+1}(T)}
	\nonumber\\
&=&\beta+\sum_{T_0,T_1,\ldots,T_{m-1}\in\clT(d)}
	\alpha^{\sum_j|T_j|-1}
	\beta^{\sum_j|T_j|-\sum_jL_d(T_j)}
	\label{eq:step1}\\
&=&\beta+\alpha^{m-1}
	\sum_{T_0,T_1,\ldots,T_{m-1}\in\clT(d)}
	\prod_j
	\alpha^{|T_j|-1}
	\beta^{|T_j|-L_d(T_j)}
	\nonumber\\
&=&\beta+\alpha^{m-1}
	\prod_j
	\sum_{T_j\in\clT(d)}
	\pi_d(T_j)
	\nonumber\\
&=&\beta+\alpha^{m-1} = 1,
	\label{eq:step2}
\ee
where~(\ref{eq:step1}) follows by~(\ref{eq:Tunion}),
and~(\ref{eq:step2}) follows from the inductive 
hypothesis and the assumption that $\beta=1-\alpha^{m-1}$.
\qed

\medskip

\noindent
Before giving the proof of Theorem~\ref{thm:MMLA},
we note for later use the following simple properties
of the prior $\pi_D(T)$. Recall that we write
$\Lambda=\{\lambda\}$ for the tree consisting of
only the root node $\lambda$.

\begin{lemma}
\label{lem:unions}
\begin{itemize}
\item[$(i)$]
If $T\in\clT(D)$, $T\neq\Lambda$, is
expressed as the union $T=\cup_jT_j$ of the
subtrees $T_j\in\clT(D-1)$,
then,
\be
\pi_{D}(T) = a^{m-1} \prod_{j=0}^{m-1} \pi_{D-1}(T_j),
\label{eq:prior-ind}
\ee
where $\alpha^{m-1}=1-\beta$.
\item[$(ii)$]
If $T\in\clT(D)$, $t\in T$ is at depth $d<D$, 
$S\in \clT(D-d)$ is nonempty, and $T\cup S$ consists 
of the tree $T$ with $S$ added as a subtree rooted
at $t$, then,
$$\pi_D(T\cup S) = \beta^{-1}\pi_D(T)\pi_{D-d}(S).$$
\end{itemize}
\end{lemma}

\medskip

\noindent
{\em Proof.}
The result of~$(i)$ immediately follows 
from the observation~(\ref{eq:Tunion})
in the proof of Lemma~\ref{lem:prior}.
Similarly,~$(ii)$ follows from the simple
observations that
$|T\cup S|=|T|+|S|-1$ and
$L_D(T\cup S)=L_D(T)+L_{D-d}(S)$,
together with the definition of the prior.
\qed

\medskip

\noindent
{\em Proof of Theorem~\ref{thm:MMLA}.}
First we note that, without loss of generality,
we may assume that the tree $\TMAX$ is
the complete tree of depth $D$; if some
node $s$ of the complete tree is
not in $\TMAX$, we simply assume 
that it has an all-zero count vector $a_s$.

The proof is again by induction. 
We adopt the notation of the proof of~Lemma~\ref{lem:prior}
and observe that, 
in view of Lemma~\ref{lem:integrate}, 
it suffices to show that,
\be
P_{w,\lambda} = 
\sum_{T \in \clT(D)} 
\pi_{D}(T)\prod_{s \in T} P_{e} (a_s).
\label{eq:target}
\ee
We claim that the following more general 
statement holds: For any node $s$ at depth
$d$ with $0\leq d\leq D$, we have,
\be
P_{w,s} = \sum_{U \in \clT(D-d)} \pi_{D-d}(U) 
\prod_{u \in U} P_{e} (a_{su}),
\label{eq:claim}
\ee
where $su$ denotes the concatenation of contexts
$s$ and $u$.
Clearly~(\ref{eq:claim}) implies~(\ref{eq:target}) upon
taking $s=\lambda$, 
and~(\ref{eq:claim}) is trivially true for nodes
$s$ at depth $D$, since it reduces to the 
fact that $P_{w,s}=P_{e,s}$ for leaves $s$,
by definition.

Suppose~(\ref{eq:claim}) holds for all nodes $s$
at depth $d$ for some fixed $0 < d \leq D$. 
Let $s$ be a node at depth $d-1$;
then, by the inductive hypothesis,
\ben
P_{w,s} 
&=& \beta P_e(a_s) + (1-\beta)\prod_{j=0}^{m-1} P_{w,sj} \\
&=& \beta P_e(a_s) + (1-\beta)\prod_{j=0}^{m-1}\left 
	[\sum_{T_j \in \clT(D-d)} 
	\pi_{D-d}(T_j) \prod_{t \in T_j}P_e(a_{sjt}) \right],
\een
where $sjt$ denotes the concatenation of context
$s$, then symbol $j$, then context $t$, in that order.
Therefore,
\ben
P_{w,s} 
&=& \beta P_e(a_s) + (1-\beta) \sum_{T_0,T_1,\ldots,T_{m-1}\in\clT(D-d)}
	\prod_{j=0}^{m-1} \left[ \pi_{D-d}(T_{j})
	\prod_{t \in T_j}P_{e}(a_{sjt}) \right ] \\
&=& \beta P_e(a_s) + \frac{1-\beta}{\alpha^{m-1}}
	\sum_{T_0,T_1,\ldots,T_{m-1}\in \clT(D-d)} 
	\pi_{D-d+1}(\cup_jT_j) 
	\left[ 
	\prod_{j=0}^{m-1} 
	\prod_{t \in T_j} P_e(a_{sjt}) \right],
\een
where 
for the last step we have used~(\ref{eq:prior-ind})
from Lemma~\ref{lem:unions}.
Concatenating every symbol $j$ with every leaf of the 
corresponding tree $T_j$, we end up with all the leaves
of the larger tree $\cup_jT_j$. Therefore,
\ben
P_{w,s} 
&=& \beta P_e(a_s) + \frac{1-\beta}{\alpha^{m-1}}
	\sum_{T_0,T_1,\ldots,T_{m-1}\in \clT(D-d)} 
	\pi_{D-d+1}(\cup_jT_j) 
	\prod_{t \in \cup_jT_j} P_e(a_{st}),
\een
and since $1-\beta=\alpha^{m-1}$ and
$\pi_d(\Lambda)=\beta$ for all $d\geq 1$,
\ben
P_{w,s} 
&=& \pi_{D-d+1}(\Lambda)P_e(a_s)
	+ \sum_{T_0,T_1,\ldots,T_{m-1}\in \clT(D-d)} 
	\pi_{D-d+1}(\cup_jT_j) 
	\prod_{t \in \cup_jT_j} P_e(a_{st})\\
&=& \pi_{D-d+1}(\Lambda)P_e(a_s)
	+ \sum_{T\in \clT(D-d+1),T\neq\Lambda} 
	\pi_{D-d+1}(T) 
	\prod_{t \in T} P_e(a_{st})\\
&=& 
	\sum_{T\in \clT(D-d+1)} 
	\pi_{D-d+1}(T) 
	\prod_{t \in T} P_e(a_{st}).
\een
This establishes~(\ref{eq:claim}) for all nodes
$s$ at depth $d-1$, completing the inductive
step and the proof of the theorem.
\qed

\medskip

\noindent
{\em Proof of Theorem~\ref{thm:MAPT}.}
Again we observe that, without loss of generality,
we may assume that the tree $\TMAX$ is
the complete tree of depth $D$; if some
node $s$ of the complete tree is
not in $\TMAX$, we simply assume 
that it has an all-zero count vector $a_s$.
It is easy to see that, for $\beta\geq 1/2$,
these assumptions
are equivalent to the ones in the description
of the algorithm, giving the same initial values
to all leaves of $\TMAX$.

The proof is once again by induction,
and we adopt the same notation as in
the proofs of~Lemma~\ref{lem:prior}
and Theorem~\ref{thm:MMLA}.
First we will prove that,
\be
P_{m,\lambda} = 
\max_{T \in \clT(D)} 
P(x,T),
\label{eq:pre-targetM}
\ee
which,
in view of Lemma~\ref{lem:integrate}, 
is equivalent to,
\be
P_{m,\lambda} = 
\max_{T \in \clT(D)} 
\pi_{D}(T)\prod_{s \in T} P_{e} (a_s).
\label{eq:targetM}
\ee
As in the proof of Theorem~\ref{thm:MMLA},
we claim that the following more general 
statement holds: For any node $s$ at depth
$d$ with $0\leq d\leq D$, we have,
\be
P_{m,s} = \max_{U \in \clT(D-d)} \pi_{D-d}(U) 
\prod_{u \in U} P_{e} (a_{su}),
\label{eq:claimM}
\ee
where $su$ denotes the concatenation of contexts
$s$ and $u$.
Taking $s=\lambda$ in~(\ref{eq:claimM}) 
gives~(\ref{eq:targetM}),
and~(\ref{eq:claimM}) is trivially true for nodes
$s$ at depth $D$, since it reduces to the 
fact that $P_{m,s}=P_{e,s}$ for leaves $s$,
by definition.

For the inductive step, we assume that~(\ref{eq:claimM}) 
holds for all nodes $s$ at depth $d$ for some fixed $0 < d \leq D$,
and consider a node $s$ at depth $d-1$.
By the inductive hypothesis we have,
\ben
P_{m,s} 
&=& \max\left\{\beta P_e(a_s), (1-\beta)\prod_{j=0}^{m-1} P_{m,sj}\right\} \\
&=& \max\left\{
	\beta P_e(a_s) , (1-\beta)\prod_{j=0}^{m-1}\left 
	[\max_{T_j \in \clT(D-d)} 
	\pi_{D-d}(T_j) \prod_{t \in T_j}P_e(a_{sjt}) \right]
	\right\}\\
&=& \max\left\{
	\beta P_e(a_s) , (1-\beta) \max_{T_0,T_1,\ldots,T_{m-1}\in\clT(D-d)}
	\prod_{j=0}^{m-1} \left[ \pi_{D-d}(T_{j})
	\prod_{t \in T_j}P_{e}(a_{sjt}) \right ]
	\right\} \\
&=& 	\max\left\{
	\beta P_e(a_s) , \frac{1-\beta}{\alpha^{m-1}}
	\max_{T_0,T_1,\ldots,T_{m-1}\in \clT(D-d)} 
	\pi_{D-d+1}(\cup_jT_j) 
	\left[ 
	\prod_{j=0}^{m-1} 
	\prod_{t \in T_j} P_e(a_{sjt}) \right]
	\right\},
\een
where the last step follows by~(\ref{eq:prior-ind})
from Lemma~\ref{lem:unions}.
Arguing as in the proof of Theorem~\ref{thm:MMLA},
\ben
P_{m,s} 
&=& 
	\max\left\{
	\pi_{D-d+1}(\Lambda)P_e(a_s)
	, \max_{T_0,T_1,\ldots,T_{m-1}\in \clT(D-d)} 
	\pi_{D-d+1}(\cup_jT_j) 
	\prod_{t \in \cup_jT_j} P_e(a_{st})
	\right\}\\
&=& 
	\max\left\{
	\pi_{D-d+1}(\Lambda)P_e(a_s)
	, \max_{T\in \clT(D-d+1),T\neq\Lambda} 
	\pi_{D-d+1}(T) 
	\prod_{t \in T} P_e(a_{st})
	\right\}\\
&=& 
	\max_{T\in \clT(D-d+1)} 
	\pi_{D-d+1}(T) 
	\prod_{t \in T} P_e(a_{st}).
\een
This establishes~(\ref{eq:claimM}) for all nodes
$s$ at depth $d-1$, completing the inductive
step and hence also proving~(\ref{eq:pre-targetM})
and~(\ref{eq:targetM}).

To complete the proof of the theorem, 
it now suffices to show that,
\be
P_{m,\lambda} = 
P(x,T^*_1),
\label{eq:pre-targetM2}
\ee
because this is exactly~(\ref{eq:max-root}),
and combined with~(\ref{eq:pre-targetM})
it implies,
$$
\max_{T \in \clT(D)} 
P(x,T)
=
P(x,T^*_1),$$
which, after dividing both sides
by the prior predictive likelihood, $P^*_D(x)$,
gives~(\ref{eq:thm-MAPT}).
By Lemma~\ref{lem:integrate},~(\ref{eq:pre-targetM2}) is
equivalent to,
\be
P_{m,\lambda} = 
\pi_{D}(T^*_1)\prod_{s \in T^*_1} P_{e} (a_{s}),
\label{eq:targetM2}
\ee
and, once again,
we will establish the following more general 
statement: For any node $s$ at depth
$d$ with $0\leq d\leq D$, we have,
\be
P_{m,s} = 
\max\Big\{\beta P_e(a_s),
(1-\beta)\prod_{j=0}^{m-1} P_{m,sj}\Big\}
=
\pi_{D-d}(T(s)) 
\prod_{t \in T(s)} P_{e} (a_{st}),
\label{eq:claimM2}
\ee
where $T(s)$ is the tree that the
\BCT~algorithm would produce if it
started its step~$(iii)$ at node $s$.
Taking $s=\lambda$ in~(\ref{eq:claimM2}) 
gives~(\ref{eq:targetM2}),
and~(\ref{eq:claimM2}) is again trivially true 
for leaves $s$ at depth $D$, by the 
definition of the maximal probabilities
$P_{m,s}$.

Finally, for the inductive step,
suppose~(\ref{eq:claimM2}) holds for
all nodes at depth $0 < d \leq D$,
and let $s$ be a node at 
depth $d-1$. We consider two 
separate cases: 
$(i)$~If the maximum in~(\ref{eq:claimM2}) 
is achieved by the first term, then 
$P_{m,s} = \beta P_{e,s}$ and $T(s)$ consists
of $s$ only, so that~(\ref{eq:claimM2}) holds trivially;
$(ii)$~If the maximum in~(\ref{eq:claimM2}) 
is achieved by the second term, then
$T(s)=\cup_jT(sj)$, and
using~(\ref{eq:prior-ind}) 
from Lemma~\ref{lem:unions}
and the inductive hypothesis,
we obtain:
\ben
P_{m,s} 
&=& (1-\beta) \prod_{j=0}^{m-1} P_{m,sj} \\
&=& (1-\beta) \prod_{j=0}^{m-1}\left[\pi_{D-d}(T(sj))
	\prod_{t \in T(sj)}P_e(a_{sjt})\right] \\
&=& 
	\pi_{D-d+1}(\cup_j T(sj)) 
	\prod_{j=0}^{m-1}\prod_{t \in T(sj)} P_{e} (a_{sjt}) \\
&=& 
	\pi_{D-d+1}(\cup_j T(sj)) 
	\prod_{t \in \cup_jT(sj)} P_{e} (a_{st}) \\
&=& \pi_{D-d+1}(T(s)) 
\prod_{t \in T(s)} P_e (a_{st}).
\een
This establishes~(\ref{eq:claimM2}) and 
completes the proof of the theorem.
\qed

\section{The actual {\hmath $k$}-{\sf BCT} algorithm}
\label{s:SMkBCT}

Here we describe a practical version
of the algorithm, which only differs
from the idealised version of Section~\ref{s:kMAPT}
in the initialisation step of the maximal
probabilities and the position vectors at 
the leaves of depth $d<D$. After describing
the algorithm, we also make some comments
on its implementation complexity.

\begin{enumerate}
\item[$(i)$]
Initially, perform two preprocessing steps:
\begin{enumerate}
\item[$(ia)$]
Execute the first two 
steps $(i)$ and $(ii)$ of
the idealised \BCTk~algorithm on the complete
$m$-ary tree of depth $D$, with all the
count vectors $a_s$ assumed to be equal 
to zero, $a_s=(0,0,\ldots,0)$ for all $s$.
Since all nodes at the same depth are identical,
this process can be carried out effectively, 
by performing the relevant computations only
at a single node (instead of all $m^d$ nodes)
for each depth $0\leq d\leq D$.
\item[$(ib)$]
Build the tree $\TMAX$ from the contexts
of $x_{-D+1}^n$ and 
compute the count vectors $a_s$ 
and the probabilities $P_{e,s}=P_e(a_s)$ 
at all nodes $s$ of $\TMAX$,
as in steps~$(i)$--$(iii)$ of \CTW.
\end{enumerate}
\item[$(ii)$]
Starting at the leaves 
and proceeding towards the
root, at each
node $s$ we compute a list of $k$
maximal probabilities $P_{m,s}^{(i)}$
and $k$ position vectors 
$c^{(i)}_s=(c^{(i)}_s(0),c^{(i)}_s(1),\ldots,c^{(i)}_s(m-1))$,
for $i=1,2,\ldots,k$, recursively as follows.
\begin{enumerate}
\item[$(iia)$]
At each leaf $s$ at depth $D$, 
with a nonzero count vector $a_s$, 
let $P^{(1)}_{m,s}=P_{e,s}$
and $c_s^{(1)}=(0,0,\ldots,0)$.
For $i=2,3,\ldots,k$, we leave $P^{(i)}$
and $c_s^{(i)}$ undefined.
\item[$(iib)$]
Similarly, at each leaf $s$ at depth $D$, 
with an all-zero count vector $a_s$, 
let $P^{(1)}_{m,s}=P_{e,s}=1$,
$c_s^{(1)}=(0,0,\ldots,0)$, and
for $i=2,3,\ldots,k$, leave $P^{(i)}$
and $c_s^{(i)}$ undefined.
\item[$(iic)$]
At each leaf $s$ at depth $d<D$,
we let the list of the maximal probabilities 
$P_{m,s}^{(i)}$ and position vectors 
$c_s^{(i)}$ of $s$ be those that
are computed for a node at depth $d$ in
the preprocessing stage $(ia)$.
\item[$(iid)$]
Continue with steps $(iib)$ and $(iic)$ 
as in the idealised version of \BCTk.
\end{enumerate}
\item[$(iii)$]
We perform the same steps as described
in $(iiia)$--$(iiid)$ of the idealised
version, with the following addition:
\begin{enumerate}
\item[$(iiie)$]
While examining a node $s$ 
at depth $0<d<D$ in step
$(iiib)$ or $(iiic)$ of the idealised
algorithm, we may reach a point where 
the algorithm dictates that we examine 
its $m$ children, 
when these children are {\em not} 
included in the tree $\TMAX$. In that
case, we add them to $\TMAX$, and 
we define their corresponding maximal
probabilities and position vectors
according to the initialisation 
described in step $(iib)$ or $(iic)$ 
above, depending on whether $d=D-1$
or $d<D-1$, respectively.
\end{enumerate}
\item [$(iv)$]
As in the idealised version,
output the $k$ resulting trees $T^*_i$
and the $k$ maximal probabilities at the root,
$P^{(i)}_{m,\lambda}$, $i=1,2,\ldots,k$.
\end{enumerate}

\noindent
{\bf Complexity of \BCTk.}
As discussed in 
Sections~\ref{s:summary}
and~\ref{s:sequential} of the main text,
the complexity of the \BCTk\ algorithm in its most
naive implementation is $O(nmD\times k^m)$. The factor $k^m$ comes 
from step $(iic)$ of the idealized algorithm, where for every internal 
node $s$,
all possible combinations from the $m$ lists of probabilities $P_{m,sj}$
are computed. However, the fact that these lists $P_{m,sj}$ are already 
ordered means that the top-$k$ combinations can be found more efficiently, 
without performing an exhaustive search over all combinations. 
In particular, a best-first search can be employed to find only the 
top-$k$ combinations: A priority queue can be used to maintain all 
possible candidates and efficiently pick the next best. This reduces 
the complexity to $O \left( nmD \times  km\log (km) \right )$.

\section{Proof of Theorem~\ref{thm:MAPT-k}}
\label{s:proofs2}

\noindent
Before giving the proof of
Theorem~\ref{thm:MAPT-k}, we observe 
that it is easy (though somewhat tedious)
to verify that the two versions of the
\BCTk~algorithm described in Section~\ref{s:SMkBCT}
are equivalent, in that they produce identical
results as long as $\beta\geq 1/2$. 
Therefore, for the sake of simplicity,
the proof below is given only for the
idealised \BCTk~algorithm.

We adopt the same notation as in Section~\ref{s:proofs}.

\medskip

\noindent
{\em Proof of Theorem~\ref{thm:MAPT-k}.}
The proof is again by induction,
and we adopt the same notation as in
the proof of Theorem~\ref{thm:MAPT}.
First we will prove that, for all $1\leq i\leq k$,
\be
P^{(i)}_{m,\lambda} = 
\maxi_{T \in \clT(D)} 
P(x,T),
\label{eq:pre-targetM-k}
\ee
which,
in view of Lemma~\ref{lem:integrate}, 
is equivalent to,
\be
P^{(i)}_{m,\lambda} = 
\maxi_{T \in \clT(D)} 
\pi_{D}(T)\prod_{s \in T} P_{e} (a_s).
\label{eq:targetM-k}
\ee
As in the proof of Theorem~\ref{thm:MAPT},
we claim that the following more general 
statement holds: For all $i$ and any node $s$ at depth
$d$ with $0\leq d\leq D$, we have,
\be
P^{(i)}_{m,s} = \maxi_{U \in \clT(D-d)} \pi_{D-d}(U) 
\prod_{u \in U} P_{e} (a_{su}),
\label{eq:claimM-k}
\ee
where $su$ denotes the concatenation of contexts
$s$ and $u$.
Taking $s=\lambda$ in~(\ref{eq:claimM-k}) 
gives~(\ref{eq:targetM-k}),
and~(\ref{eq:claimM-k}) is trivially true for nodes
$s$ at depth $D$, since at depth $D$ we only 
consider $i=1$ and then
$P^{(1)}_{m,s}=P_{e,s}$ for leaves $s$,
by definition.

\newpage

For the inductive step, we assume that~(\ref{eq:claimM-k}) 
holds for all $i$ and all nodes $s$ at depth $d$ for some 
fixed $0 < d \leq D$,
and consider a node $s$ at depth $d-1$.
By the inductive hypothesis we have,
in the notation of~(\ref{eq:probimax}),
that
$P^{(i)}_{m,s}$ is equal to,
\begin{align*}
 & \maxi\left[
	\bigcup_{i_0=1}^{k_0}\bigcup_{i_1=1}^{k_1}\cdots
	\bigcup_{i_{m-1}=1}^{k_{m-1}}
	\left\{ (1-\beta)\prod_{j=0}^{m-1} P^{(i_j)}_{m,sj}\right\}
	\bigcup\, \big\{\beta P_e(a_s)\big\}
	\right] \\
=& \maxi\left[
	\bigcup_{i_0=1}^{k_0}\bigcup_{i_1=1}^{k_1}\cdots
	\bigcup_{i_{m-1}=1}^{k_{m-1}}
	\left\{(1-\beta)\prod_{j=0}^{m-1}\left 
	[\maxij_{T_j \in \clT(D-d)} 
	\pi_{D-d}(T_j) \prod_{t \in T_j}P_e(a_{sjt}) \right]
	\right\}
	\bigcup\, \big\{\beta P_e(a_s)\big\}
	\right]\\
=& \maxi\left[
	\bigcup_{i_0=1}^{k_0}\bigcup_{i_1=1}^{k_1}\cdots
	\bigcup_{i_{m-1}=1}^{k_{m-1}}
	\right.\\
&
	\hspace{0.6in}
	\left\{(1-\beta) 
	\maxiz_{T_0\in\clT(D-d)}
	\maxio_{T_1\in\clT(D-d)}
	\cdots
	\maxim_{T_{m-1}\in\clT(D-d)}
	\prod_{j=0}^{m-1} \left[ \pi_{D-d}(T_{j})
	\prod_{t \in T_j}P_{e}(a_{sjt}) \right ] \right\}\\
&	
	\hspace{0.9in}
	\bigcup\, \big\{\beta P_e(a_s)\big\}
	\Bigg] \\
=& \maxi\left[
	\bigcup_{i_0=1}^{k_0}\bigcup_{i_1=1}^{k_1}\cdots
	\bigcup_{i_{m-1}=1}^{k_{m-1}}
	\right.\\
&
	\hspace{0.6in}
	\left\{
	\maxiz_{T_0\in\clT(D-d)}
	\maxio_{T_1\in\clT(D-d)}
	\cdots
	\maxim_{T_{m-1}\in\clT(D-d)}
	\pi_{D-d+1}(\cup_jT_j) 
	\left[ 
	\prod_{j=0}^{m-1} 
	\prod_{t \in T_j} P_e(a_{sjt}) \right]
	\right\}\\
& 	
	\hspace{0.9in}
	\bigcup\, \big\{\beta P_e(a_s)\big\}
	\Bigg],
\end{align*}
where the last step follows by~(\ref{eq:prior-ind})
from Lemma~\ref{lem:unions}.
Therefore,
$P^{(i)}_{m,s}$ is equal to,
\begin{align*}
& 
	\maxi\left[
	\bigcup_{i_0=1}^{k_0}\bigcup_{i_1=1}^{k_1}\cdots
	\bigcup_{i_{m-1}=1}^{k_{m-1}}
	\left\{
	\maxiz_{T_0\in\clT(D-d)}
	\cdots
	\maxim_{T_{m-1}\in\clT(D-d)}
	\pi_{D-d+1}(\cup_jT_j) 
	\prod_{t \in \cup_jT_j} P_e(a_{st})
	\right\}
	\right.\\
&	\hspace{0.9in}
	\bigcup\big\{
	\pi_{D-d+1}(\Lambda)P_e(a_s)\big\}
	\Bigg]\\
=& 
	\maxi\left[
	\bigcup_{i'=1}^k
	\left\{
	\maxip_{T\in \clT(D-d+1),T\neq\Lambda} 
	\pi_{D-d+1}(T) 
	\prod_{t \in T} P_e(a_{st})
	\right\}
	\bigcup\big\{
	\pi_{D-d+1}(\Lambda)P_e(a_s)\big\}
	\right]\\
=& 
	\maxi_{T\in \clT(D-d+1)} 
	\pi_{D-d+1}(T) 
	\prod_{t \in T} P_e(a_{st}).
\end{align*}
This establishes~(\ref{eq:claimM-k}) for all $i$
and all nodes
$s$ at depth $d-1$, completing the inductive
step and hence also proving~(\ref{eq:pre-targetM-k})
and~(\ref{eq:targetM-k}).

To complete the proof of the theorem, 
it now suffices to show that, for all
$1\leq i\leq k$,
\be
P^{(i)}_{m,\lambda} = 
P(x,T^*_i),
\label{eq:pre-targetM2-k}
\ee
because, combined with~(\ref{eq:pre-targetM-k})
this implies,
$$
\maxi_{T \in \clT(D)} 
P(x,T)
=
P(x,T^*_i),$$
which, after dividing both sides
by the prior predictive likelihood,
gives~(\ref{eq:thm-MAPT-k}).
By Lemma~\ref{lem:integrate},~(\ref{eq:pre-targetM2-k}) is
equivalent to,
\be
P^{(i)}_{m,\lambda} = 
\pi_{D}(T^*_i)\prod_{s \in T^*_i} P_{e} (a_{st}),
\label{eq:targetM2-k}
\ee
and, once again,
we will establish the following more general 
statement: For all $i$ and any node $s$ at depth
$d$ with $0\leq d\leq D$, we have,
\be
P^{(i)}_{m,s} 
&= &
 \maxi\left[
	\bigcup_{i_0=1}^{k_0}\bigcup_{i_1=1}^{k_1}\cdots
	\bigcup_{i_{m-1}=1}^{k_{m-1}}
	\left\{ (1-\beta)\prod_{j=0}^{m-1} P^{(i_j)}_{m,sj}\right\}
	\bigcup\, \big\{\beta P_e(a_s)\big\}
	\right] 
	\label{eq:claimM2-k-pre}
	\\
&=&
\pi_{D-d}(T^{(i)}(s)) 
\prod_{t \in T^{(i)}(s)} P_{e} (a_{st}),
\label{eq:claimM2-k}
\ee
where $T^{(i)}(s)$ is the $i$th tree that 
\BCTk~would produce if it
started its step~$(iii)$ at node $s$.
Taking $s=\lambda$ in~(\ref{eq:claimM2-k}) 
gives~(\ref{eq:targetM2-k}),
and~(\ref{eq:claimM2-k}) is again trivially true 
for leaves $s$ at depth $D$, by the 
definition of the maximal probabilities
$P_{m,s}$.

Finally, for the inductive step
assume~(\ref{eq:claimM2-k}) holds for
all nodes at depth $0 < d \leq D$,
and let $s$ be a node at 
at depth $d-1$. We consider two 
separate cases: 
$(i)$~If the maximum in~(\ref{eq:claimM2-k-pre})
is achieved by the 
last term, then 
$P^{(i)}_{m,s} = \beta P_{e,s}$ and $T^{(i)}(s)$ consists
of $s$ only, so that~(\ref{eq:claimM2-k}) holds trivially;
$(ii)$~If the maximum in~(\ref{eq:claimM2-k-pre}) 
is achieved by the collection of indices
$(i_0,i_1,\ldots,i_{m-1})$, then
$T^{(i)}(s)=\cup_jT^{(i_j)}(sj)$, and
using~(\ref{eq:prior-ind}) from Lemma~\ref{lem:unions}
and the inductive hypothesis,
we obtain:
\ben
P^{(i)}_{m,s} 
&=& (1-\beta) \prod_{j=0}^{m-1} P^{(i_j)}_{m,sj} (a_{sj}) \\
&=& (1-\beta) \prod_{j=0}^{m-1}\left[\pi_{D-d}(T^{(i_j)}(sj))
	\prod_{t \in T^{(i_j)}(sj)}P_e(a_{sjt})\right] \\
&=& 
	\pi_{D-d+1}(\cup_j T^{(i_j)}(sj)) 
	\prod_{j=0}^{m-1}\prod_{t \in T^{(i_j)}(sj)} P_{e} (a_{sjt}) \\
&=& 
	\pi_{D-d+1}(\cup_j T^{(i_j)}(sj)) 
	\prod_{t \in \cup_jT^{(i_j)}(sj)} P_{e} (a_{st}) \\
&=& 
	\pi_{D-d+1}(T^{(i)}(s)) 
	\prod_{t \in T^{(i)}(s)} P_e (a_{st}).
\een
This establishes~(\ref{eq:claimM2-k}) and 
completes the proof of the theorem.
\qed

\section{Arbitrary Dirichlet parameters}
\label{s:dirichlet}

In some cases it may be desirable to consider a more general 
Dirichlet prior $\pi(\theta|T)$ on the parameters $\theta$
associated with a given model $T\in\clT(D)$.
Instead of the Dir$(1/2,1/2,\ldots,1/2)$ distribution
defined in~(\ref{eq:theta-prior}) we may place
a different, general Dir$(\gamma_s(0),\gamma_s(1),\ldots,\gamma_s(m-1))$
prior
on each context $s\in T$.

For an arbitrary 
collection of hyperparameters
$\gamma=\{\gamma_s
=(\gamma_s(0),\ldots,\gamma_s(m-1))\;;\;s\in T\},$
with each $\gamma_s(j)>0$,
let $\pi(\theta|T)
=\prod_{s\in T}\pi(\theta_s)$, with,
$$\pi(\theta_s) =
\pi(\theta_s(0),\theta_s(1),\ldots,\theta_s(m-1))
=
\frac{\Gamma(M'_s)}{\prod_{j=0}^{m-1}\Gamma(\gamma_s(j))}  
\prod_{j=0}^{m-1} \theta_s(j)^{\gamma_s(j) - 1}
\propto  \prod_{j=0}^{m-1} \theta_s(j)^{\gamma_s(j) - 1},
$$
where $M'_s:=\sum_{j=0}^{m-1}\gamma_s(j)$, $s\in T$.

As in Section~\ref{s:mml}, the marginal
likelihood $P(x|T)$ of a data string $x$ given a model $T\in\clT(D)$ 
can be computed explicitly. Lemma~\ref{lem:integrate2} below is the
analog of Lemma~\ref{lem:integrate} in this case. 

\begin{lemma}
\label{lem:integrate2}
The marginal likelihood $P(x|T)$ 
of the observations
$x$ given a model $T$ is,
$$P(x|T)
=\int P(x,\theta|T)d\theta
=\int P(x|\theta,T)\pi(\theta|T)d\theta
=\prod_{s\in T}P_e(a_s,\gamma_s),$$
where the count vectors 
$a_s=(a_s(0),a_s(1),\ldots,a_s(m-1))$
are defined
in {\em (\ref{eq:vector-a})} as before
and 
the {\em estimated probabilities} $P_e(a_s,\gamma_s)$ are 
now defined by,
\be
P_{e}(a_s,\gamma_s):= 
\frac{\Gamma(M'_s)}{\Gamma(M_s+M'_s)}\prod_{j=0}^{m-1}
\frac{\Gamma(a_s(j)+\gamma_s(j))}{\Gamma(\gamma_s(j))},
\label{eq:Peg}
\ee
where $M_s:=a_s(0)+a_s(1)+\cdots+a_s(m-1)$ and
$M'_s:=\gamma_s(0)+\gamma_s(1)+\cdots+\gamma_s(m-1)$,
again with the convention that any empty 
product is taken to be equal to 1.
\end{lemma}

Now it is straightforward to modify the 
\CTW, \BCT~and \BCTk~algorithms,
simply by replacing the estimated probabilities
$P_e(a_s)$ of~(\ref{eq:Pe}) by $P_e(a_s,\gamma_s)$
as in~(\ref{eq:Peg}). A careful
inspection of the proofs of 
Theorems~\ref{thm:MMLA},~\ref{thm:MAPT} and~\ref{thm:MAPT-k} 
shows that their results remain valid
in this more general case.

\begin{corollary}
Suppose that, for each model
$T\in\clT(D)$, the prior $\pi(\theta|T)$ on the parameters 
$\theta=\{\theta_s\;;\;s\in T\}$ is the product of 
{\em Dir}$(\gamma_s(0),\gamma_s(1),\ldots,\gamma_s(m-1))$
distributions, for an arbitrary collection
of hyperparameters $\gamma$. If 
the estimated probabilities
$P_e(a_s)$ of~{\em (\ref{eq:Pe})} are replaced by $P_e(a_s,\gamma_s)$
as in~{\em (\ref{eq:Peg})} in the \CTW, \BCT~and \BCTk~algorithms,
then the results of 
Theorems~\ref{thm:MMLA},~\ref{thm:MAPT} and~\ref{thm:MAPT-k} 
remain valid exactly as stated.
\end{corollary}

Similarly, all of the additional results in Section~\ref{s:additional}
remain valid as stated there, with the exception of the 
expression for the full conditional density in~(\ref{eq:full-cond}).
In the case of a general prior $\pi(\theta|T)$ in terms
of the hyperparameters $\gamma$, the same computation shows
that here:
\begin{equation*}
\pi (\theta |T,x) 
=  \prod _{s \in T} \text{Dir}\big (a_s(0) + \gamma_s(0), \ldots , 
	a_s(m-1) + \gamma_s (m-1)\big ).
\end{equation*}

Finally, we note that sequential updates,
as outlined in Section~\ref{s:sequential},
can be carried out for the \CTW, \BCT, 
and $k$-\BCT\ algorithms in this more general
setting as well.
The only change is in the computation required 
to update $P_e(a_s)$ in step~$(iii')$,
where, in the general case,
$P_e(a_s, \gamma_s)$ is updated by
multiplying its earlier value by,
\begin{equation*}
\frac{a_s(j) + \gamma _s (j) -1}{M_s ' + M_s -1},
\end{equation*}
using the updated values of $a_s$ and $M_s$. 

\section{Explicit computations and numerical values}
\label{s:explicit}

In this section we give more details on some expressions
and explicit numerical values 
that are omitted in the main text.

\medskip

\noindent
{\bf RW sampler.} 
The ratios $r(T,T')$ in the acceptance
probabilities of the RW sampler in Section~\ref{s:MH} 
are given by, 
$$
r(T,T'):=
	\frac{\pi(T'|x)}{\pi(T|x)}\times
\begin{cases}
	1/2,
	&\mbox{in case (a)};\\
	m^{D-1}/2,	
	&\mbox{in case (b)};\\
	(|T|-L_D(T))/N_D(T'),
	&\mbox{in case (c), if $T'\neq T_c(D)$}; \\
	2m^{-D+1},
	&\mbox{in case (c), if $T'= T_c(D)$}; \\
	N_D(T)/(|T'|-L_D(T')),
	&\mbox{in case (d), if $T'\neq \Lambda$}; \\
	2,
	&\mbox{in case (d), if $T'=\Lambda$},
\end{cases}
$$
where 
$N_D(T)$ denotes the number of internal
nodes in a tree $T$ having only $m$
descendants,
and the posterior odds $\pi(T'|x)/\pi(T|x)$ 
can be computed via~(\ref{eq:mpost}).

\medskip

\noindent
{\bf Jump sampler.}
The jump sampler of Section~\ref{s:MH}
incorporates
the RW sampler's 
proposal distribution 
as one of its steps:
Let $q(T'|T)$ denote the proposal 
probabilities of the RW sampler, 
so that $q(T'|T)$ equals 1, $m^{-D+1}$, 
$1/[2(|T|-L_D(T))]$ and $1/[2N_D(T)]$, in each of its 
cases $(ia)$, $(ib)$, $(ic)$ and $(id)$, respectively.
We say that two models $T,T'$ are {\em neighbours}
if they differ by exactly one branch of $m$ children.

The ratios $r(T,T')$ in the acceptance
probabilities of the jump sampler
are given by, 
$$
r(T,T'):=
	\frac{\pi(T'|x)}{\pi(T|x)}\times
	\begin{cases}
	\frac{(1-p)q(T|T')
	+pk^{-1}
	\IND\{T\in\clT^*\}
	}
	{(1-p)q(T'|T)
	+pk^{-1}
	\IND\{T'\in\clT^*\}
	},
	&\;\;\;\mbox{if $T,T'$ are neighbours},\\
	\IND\{T\in\clT^*\},
	&\;\;\;\mbox{if $T,T'$ are not neighbours},
	\end{cases}
$$
where $\IND\{\cdot\}$ denotes the indicator
function of the event $\{\cdot\}$.

\medskip

\noindent
{\bf Parameter values.} The transition matrix $Q=(Q_{ij})$ 
in Example~\ref{ex:bimodal} is:
\[ 
\left( \begin{array}{cccccc}
0.5 &  0.2 &  0.1  &  0    &  0.05 &   0.15\\
0.4 &  0   &  0.4  &  0.2  &  0    &   0\\
0.3 &  0.1 &  0.23 &  0.12 &  0.05 &   0.2\\
0.05&  0.1 &  0.05 &  0.05 &  0.03 &   0.72\\
0   &  0   &  1    &  0    &  0    &   0\\
0.1 &  0.2 &  0.3  &  0.2  &  0.05 &   0.15\\
\end{array} \right).
\]
The parameter vector $\theta=(\theta_s)$ of the 5th order
chain in Section~\ref{s:simulated} is given by:
\ben
&& 
\hspace{-0.18in} 
	\theta_1=(0.4,0.4,0.2),\; \theta_2=(0.2,0.4,0.4),\\
&& 
\hspace{-0.18in} 
	\theta_{00}=(0.4,0.2,0.4),\; \theta_{01}=(0.3,0.6,0.1),\\
&&
\hspace{-0.18in} 
	\theta_{022}=(0.5,0.3,0.2),\\
&& 
\hspace{-0.18in} 
	\theta_{0212}=(0.1,0.3,0.6),\;
	\theta_{0211}=(0.05,0.25,0.7),\;  \theta_{0210}=(0.35,0.55,0.1),\\
&& 
	\theta_{0202}=(0.1,0.2,0.7),\;
	\theta_{0201}=(0.8,0.05,0.15),\\
&& 
\hspace{-0.18in} 
	\theta_{02002}=(0.7,0.2,0.1),\; \theta_{02001}=(0.1,0.1,0.8),\;
	\theta_{02000}=(0.3,0.45,0.25).
\een

\section{Model selection examples}
\label{s:compare2}

\subsection{Simulated data}
\label{s:scores2}

\begin{figure}[ht!]
\centerline{\includegraphics[width=2.4in]{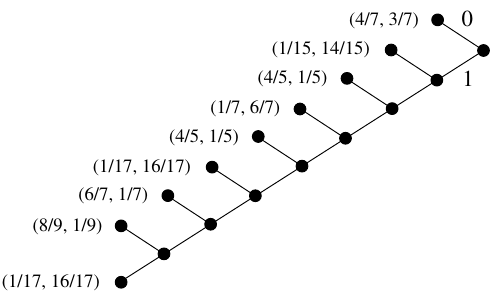}}
\caption{The model and parameters of the
renewal-like binary chain.}
\label{fig:scores3}
\end{figure}

\noindent
{\bf Renewal process.}
The binary variable-memory chain $\{X_n\}$ with
model and parameters shown in Figure~\ref{fig:scores3}
was examined in the VLMC work of \cite[p.~303]{buhlmann:00}.
Note that the distribution of the next symbol
produced by the chain only depends on how far
in the past the most recent ``0'' appeared.

\noindent
{$\bullet\;\;n=200$.} With $n=200$ samples $x$,
the \BCT~algorithm with $D=10$ and $\beta=1-2^{-m+1}=1/2$
produces a MAP model
$T_1^*$ which is the same as the true model
pruned at depth~4; its prior probability
is $\pi(T_1^*)\approx0.002$ and its posterior 
$\pi(T_1^*|x)\approx 0.0725$.
The next four of the top $k=5$ models
produced by the \BCTk~algorithm are small
variations of $T_1^*$, of maximal depths 4 or 5.
The total posterior probability of the top 5 models
is $\approx 0.1782$. 

The best-BIC-VLMC gives the same model as 
the \BCT~algorithm (with good AIC and BIC scores), 
while the
default-VLMC and best-AIC-VLMC both produce 
a tree of depth 8. It is the same as the true model
up to depth 4, but it also includes the depth-8 branch
corresponding to the context $s=01111001$,
which does not appear in the true model.
It has a marginally better AIC score than
$T_1^*$ (by $\approx 1\%$), 
but a much worse BIC score, somewhat
overfitting the data. 

The best-BIC and best-AIC versions of MTD both give $D=2$;
and the best-BIC and best-AIC versions of MTDg both give $D=4$. 
In all four cases, the resulting AIC and BIC scores
are not competitive with those of the \BCT~and 
the best-BIC-VLMC algorithms. 

\medskip

\begin{figure}[ht!]
\centerline{\includegraphics[width=6.0in]{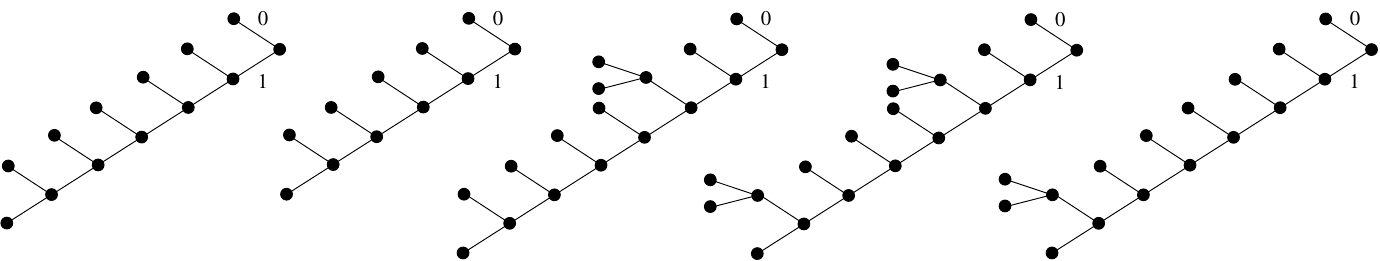}}
\caption{The top $k=5$ models obtained by \BCTk~with $n=1,000$
samples from the renewal-like chain.
The prior probability of $T_1^*$ is $\pi(T_1^*)\approx1.2\times 10^{-4}$
and its posterior is $\pi(T_1^*|x)\approx0.0799$. The posterior
odds $\pi(T_1^*|x)/\pi(T_i^*|x)$ of the next four models 
are approximately 1.25, 1.98 2.48 and 4,
for $i=2,3,4,5$, respectively.
The total posterior probability of the top 5 models
is approximately $0.2336$.}
\label{fig:scores4}
\end{figure}

\noindent
{$\bullet\;\;n=1,000$.} The results of the \BCTk~algorithm
(with $k=5$, $D=10$ and $\beta=1-2^{-m+1}=1/2$)
on a sample of length $n=1,000$,
reproduce more of the 
true underlying structure;
see Figure~\ref{fig:scores4}.
The MAP tree $T_1^*$ is the 
true model pruned at depth $6$, 
and it has at least as good
BIC and AIC scores as all other methods.

Despite the larger data length, the
best-BIC-VLMC again gives the depth-4 version
of the true model, while 
the default-VLMC and best-AIC-VLMC 
produce larger trees of depth 11, very similar between
them but very different from 
the true model and the MAP tree -- all of their leaves
except one are not present in the true model.
The long branches of depth 8-11 are clear examples 
of overfitting, resulting in poor BIC scores.
Despite this,
even the AIC score of the best-AIC-VLMC tree is 
within less than $0.1\%$ of the AIC score of the MAP model.

The best-AIC-MTD and best-BIC-MTD both give $D=5$
as the optimal depth, while the 
best-AIC-MTDg and best-BIC-MTDg both give $D=4$. 
Once again, their scores are not competitive 
with those of the MAP model and the VLMC results.

\medskip

\begin{figure}[ht!]
\centerline{\includegraphics[width=5.0in]{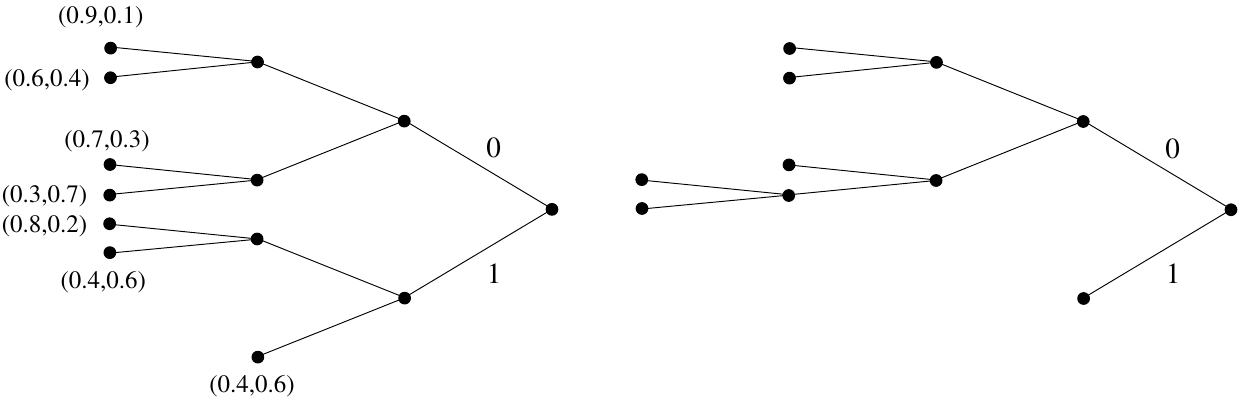}}

\vspace{-0.2in}

\hspace{2.2in} $(a)$ \hspace{1.9in} $(b)$
\caption{$(a)$ The model from which 
the data are generated for the third order binary chain.
$(b)$ The MAP model
produced by the \BCT~algorithm based on $n=200$ samples.}
\label{fig:scores5}
\end{figure}

\noindent
{\bf A third order binary chain.} 
Here we examine data generated from the 
3rd order variable-memory chain in 
example~V4 of the MTD paper \cite[p.~353]{berchtold:02}.
The model and associated parameters are shown
in Figure~\ref{fig:scores5}~(a).

\begin{figure}[ht!]
\centerline{\includegraphics[width=6.2in]{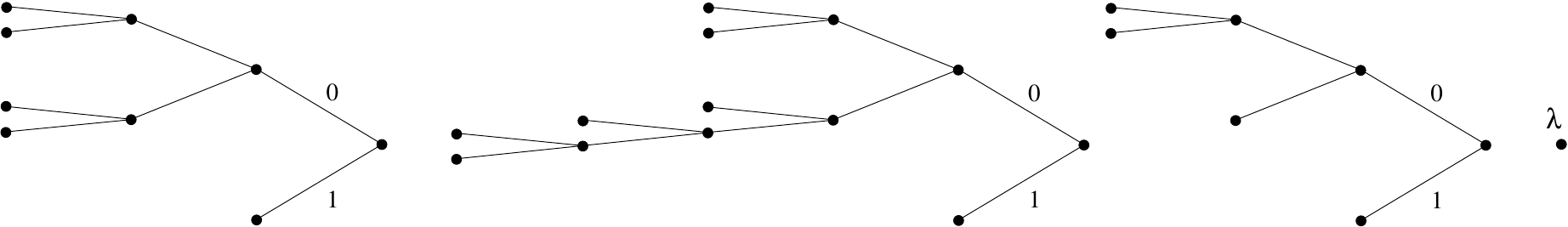}}
\caption{The four {\em a posteriori} most likely models $T_i^*$
after $T_1^*$ obtained by the \BCTk~algorithm on
$n=200$ samples from the third order binary chain.
Notice that $T_5^*$ is empty tree $\Lambda$ consisting
of just the root node $\lambda$.
The posterior odds $\pi(T_1^*|x)/\pi(T_i^*|x)$ for
$i=2,3,4,5$ are approximately 
1.27, 1.33, 2.53 and 2.78,
respectively,
and the sum of the posteriors of the top 5 models
is $\approx 0.12$.}
\label{fig:scores6}
\end{figure}

\noindent
{$\bullet\;\;n=200$.} The MAP model $T_1^*$ obtained 
by \BCTk~(with $D=10$, $\beta=1-2^{-m+1}=1/2$ and $k=5$) from
$n=200$ simulated samples of this chain
is shown in Figure~\ref{fig:scores5}~(b); its 
prior probability is $\pi(T_1^*)\approx 4.9\times 10^{-4}$,
and its posterior $\pi(T_1^*|x)\approx 0.0363$.
The 
top part of the true tree is correctly identified, the lower part is 
cropped at 1, and there is an extra branch of depth 4 that is not in 
the true model. The next four {\em a posteriori} most likely models are
shown in Figure~\ref{fig:scores6}.

\begin{figure}[ht!]
\includegraphics[width=6.1in]{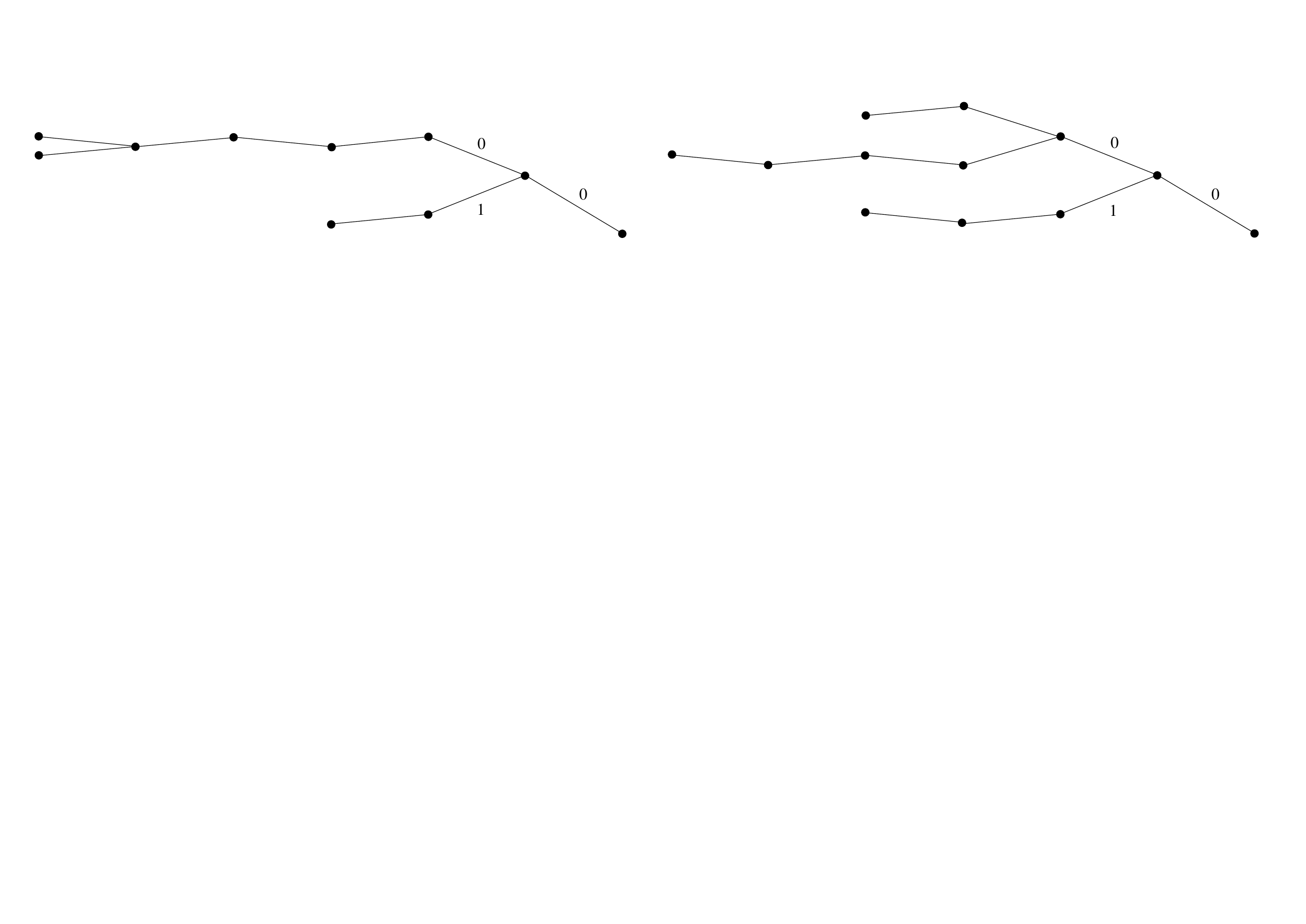}
\vspace{-3.1in}
\caption{The models produced by the default-VLMC (left)
and the best-AIC-VLMC (right) from $n=200$ samples of
the third order binary chain.}
\label{fig:scores7}
\end{figure}

The result of the best-BIC-VLMC is the same as $T_2^*$. Both $T_1^*$ and
$T_2^*$ have very good (and very similar) AIC and BIC scores.
The results of the default-VLMC and the best-AIC VLMC are 
shown in Figure~\ref{fig:scores7}.
They both have depth 6, and they bear little resemblance to 
the true model. Their AIC scores are good but not significantly better 
than the scores of $T_1^*$ and $T_2^*$. Their BIC scores are rather poor, 
once again suggesting that the results of default-VLMC and
best-AIC-VLMC 
overfit the data.

The best-AIC-MTD and best-BIC-MTD both give $D=3$,
corresponding to a 3rd order chain which is not
terribly different from the true model.
Its BIC score is about 0.4\% 
better than that of the MAP and the best-BIC-VLMC models,
while its AIC is slightly worse than the other two.
The best-BIC-MTDg gave $D = 0$,
and the best-AIC-MTDg gave $D = 3$. In both cases of MTDg, 
the corresponding scores were not competitive with 
those of the other methods.  


\newpage

\begin{figure}[ht!]
\centerline{\includegraphics[width=5.0in]{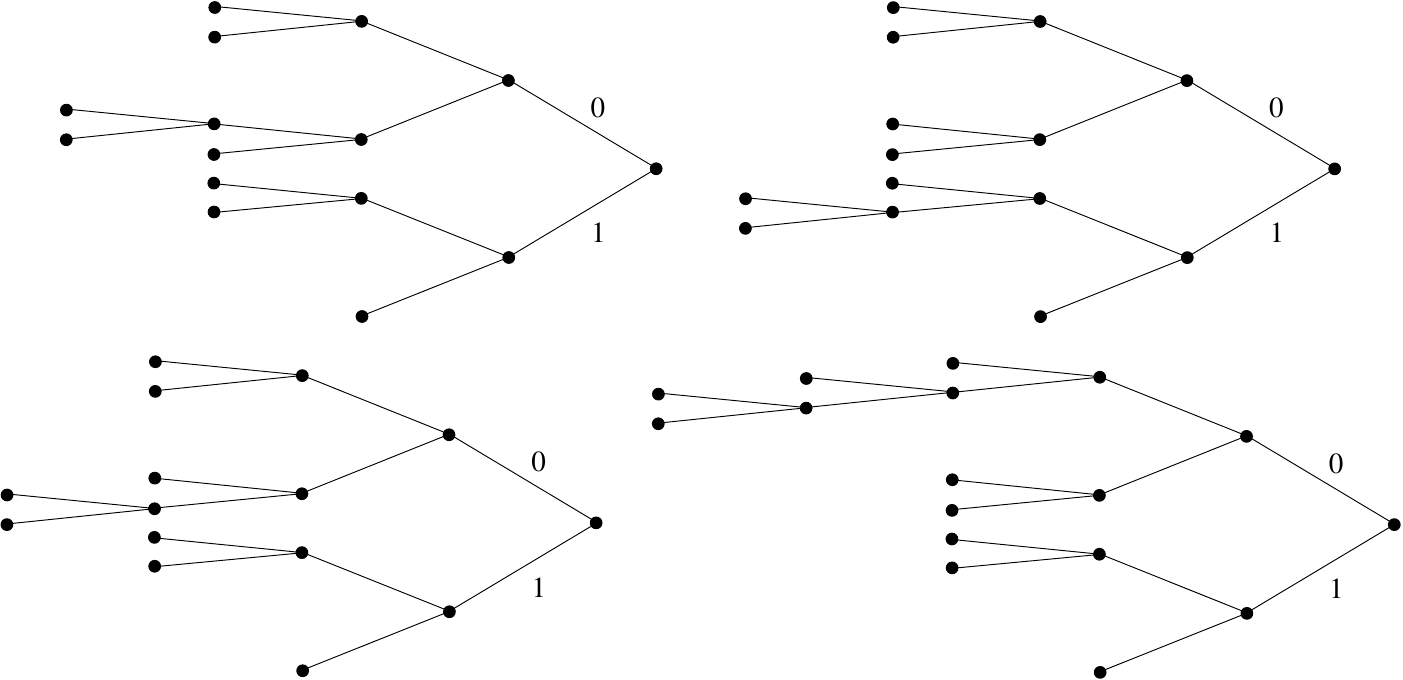}}
\caption{The four {\em a posteriori} most likely models $T_i^*$
after $T_1^*$ obtained by the \BCTk~algorithm from 
$n=1,000$ samples of the third order binary chain.
Their posterior odds $\pi(T_1^*|x)/\pi(T_i^*|x)$ for
$i=2,3,4,5$ are approximately 
2.52, 3.96, 4.66 and 5.47, respectively,
and the sum of the posteriors of the top 5 models
is $\approx 0.2179$.}
\label{fig:scores8}
\end{figure}

\noindent
{$\bullet\;\;n=1,000$.} The MAP tree $T_1^*$ produced by 
\BCTk~on $n=1,000$ samples 
(with $D=10$, $\beta=1-2^{-m+1}=1/2$ and $k=5$) 
is the true underlying model; its prior
probability is $\pi(T_1^*)\approx 1.2\times 10^{-4}$ and its 
posterior $\pi(T_1^*|x)\approx 0.1065$. The next four
{\em a posteriori} most likely models are shown in 
Figure~\ref{fig:scores8}.

The best-BIC-VLMC produces the true model as well, 
while the default-VLMC and best-AIC-VLMC produce very 
large trees of depth 18, which are not at all similar 
to the true model. They have good AIC scores 
(although not much better than the MAP and best-BIC-VLMC models), 
but their BIC scores are significantly worse.

The best-AIC-MTD and best-BIC-MTD give $D = 3$,
which corresponds to a model with a BIC
about 0.2\% higher than that of
the true underlying tree.
The best-AIC-MTDg and best-BIC-MTDg also
give $D = 3$, but the scores of the corresponding
models are worse than those produced by the other 
methods.  


\newpage

\subsection{Real data}
\label{s:FB}

{\bf Financial data.}
We consider the tick-by-tick price changes of the Facebook
stock price during a six-and-a-half-hour-long trading period 
on October 3, 2016. The price change is recorded every time
there is a trade; if the price goes down during the $i$th
trade we set $x_i=0$, if it stays the same we set $x_i=1$,
and if it goes up $x_i=2$. This produces a
ternary time series of length $n = 50,745$. 

\begin{figure}[ht!]
\centerline{\includegraphics[width=2.3in]{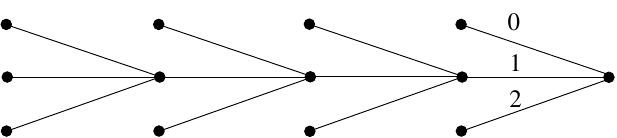}}
\caption{The MAP model for the Facebook stock
price data.}
\label{fig:FB}
\end{figure}

The MAP model $T_1^*$ produced by \BCTk~with $D=20$,
$\beta=1-2^{-m+1}=3/4$ and $k=5$ is shown in Figure~\ref{fig:FB}. 
Its prior is
$\pi(T_1^*)\approx 2.9\times 10^{-4}$, and
its posterior is $\pi(T_1^*|x)\approx 0.416$.
Note that $T_1^*$ admits an interpretation
very similar to that of the models obtained
for the other financial data set we considered 
in Example~\ref{ex:SP} of Section~\ref{s:MCMC}:
In order to 
determine the distribution of the next value,
look as far back as necessary until you see a price
change, or until you have looked four places back.

\begin{figure}[ht!]
\centerline{\includegraphics[width=4.0in]{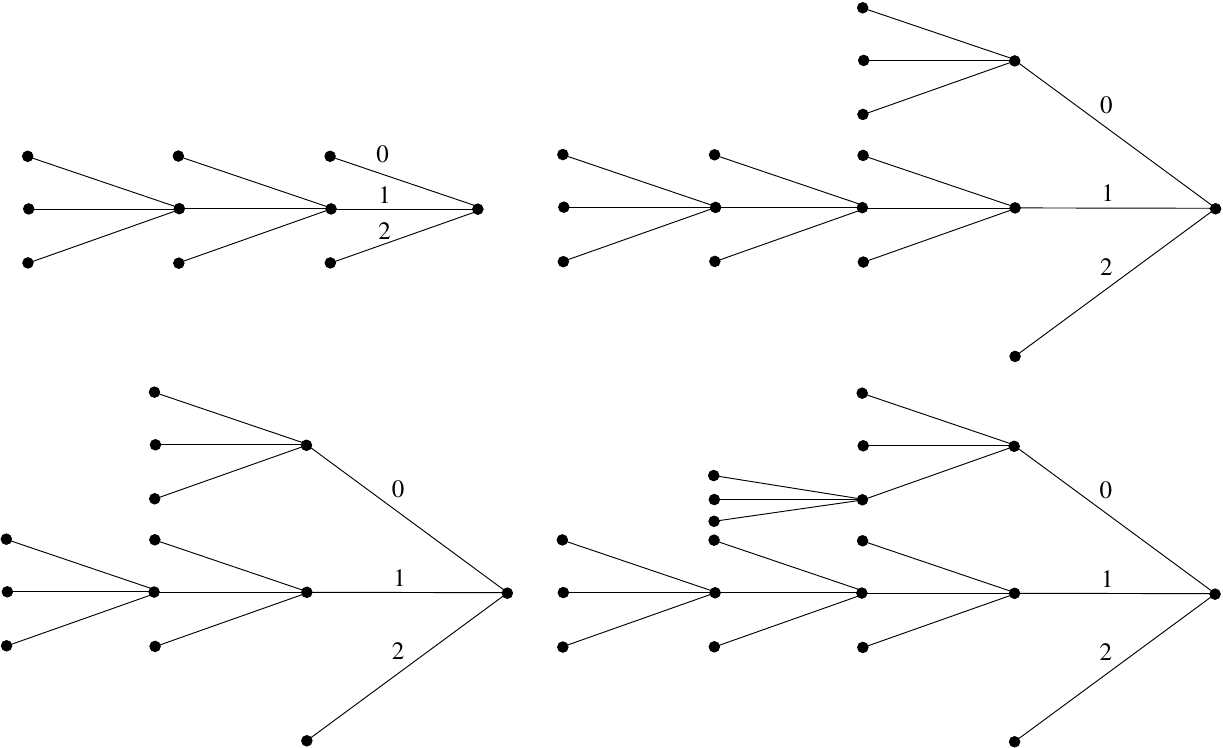}}
\caption{The four {\em a posteriori} most likely models
$T_2^*,T_3^*,T_4^*,T_5^*$ after $T_1^*$
for the Facebook stock price data.
Their posterior
odds $\pi(T_1^*|x)/\pi(T_i^*|x)$ are
$1.017, 5.074,5.161,72.8$,
for $i=2,3,4,5$, respectively, and the sum of the posterior
probabilities of all five models is $\approx 0.993$.}
\label{fig:FB4}
\end{figure}

The second {\em a posteriori} most likely model $T_2^*$ 
(show on the top left of Figure~\ref{fig:FB4}) is the
same at $T_1^*$ but pruned at depth 3.
Its posterior
probability is $\pi(T_2^*|x)\approx 0.409$,
and its form 
together with the fact that the sum of the posteriors
of the top two models is over $82\%$ reinforces the
above interpretation of the dependence structure in the
data. The next three {\em a posteriori} most likely models
are also shown in Figure~\ref{fig:FB4}.

VLMC produces a model which is almost identical to 
the MAP tree, the only difference being that VLMC gives
the same parameter vectors $\theta_s$ to contexts 
$s=1111$ and 1112.
Their AIC and BIC scores are also essentially identical.

Finally, MTD gives a model of depth $D = 4$,
with AIC and BIC scores only slightly worse than 
those of the MAP model,
and MTDg gives $D=2$ with the
corresponding model having poor AIC and BIC scores.

\end{appendices}

\end{document}